\documentclass[preprint,superscriptaddress,nofootinbib,amsmath,amssymb,aps] {revtex4-1}
\usepackage{graphicx}
\usepackage{wrapfig}
\usepackage{rotating}
\usepackage{epstopdf}
\usepackage{amsmath}
\usepackage{amsfonts}
\usepackage{mathtools}
\usepackage{adjustbox}
\usepackage{amssymb}
\usepackage{dcolumn}
\usepackage{bm}
\usepackage{color}
\usepackage[linktoc=all]{hyperref}
\usepackage{cleveref}
\usepackage[section]{placeins}
\usepackage[freestanding,sicmds]{hepunits}
\usepackage[normalem]{ulem} 

\numberwithin{equation}{section}
\newcommand{\lag}{\mathcal{L}}
\newcommand{\ud}{\,\mathrm{d}}
\newcommand{\sigann}{\,\langle\sigma_{\text{ann}} v\rangle}
\newcommand{\sla}{ \rlap{\hspace{0.2em}/}}
\newcommand{\g}{\,\mathrm{g}}
\newcommand{\cm}{\,\mathrm{cm}}
\newcommand{\Earth}{\,\mathrm{Earth}}

\begin{document}
\count\footins = 1000

\preprint{Imperial/TP/2021/MC/03}

\title{
Cosmology and Signals of Light Pseudo-Dirac Dark Matter}

\author{Mariana Carrillo Gonz\'alez} 
\affiliation{Theoretical Physics, Blackett Laboratory, Imperial College, London, SW7 2AZ, U.K}

\author{Natalia Toro}
\affiliation{SLAC National Accelerator Laboratory, 2575 Sand Hill Road, Menlo Park, CA, USA}

\begin{abstract}
In this paper, we analyze the cosmological evolution, allowed parameter space, and observational prospects for a dark sector consisting of thermally produced pseudo-Dirac fermions with a small mass splitting, coupled to the Standard Model through a dark photon.  This scenario is particularly notable in the context of sub-GeV dark matter, where the mass-off-diagonal leading interaction limits applicability of both CMB energy injection constraints and standard direct detection searches. We present the first general study of the thermal history of pseudo-Dirac DM with splittings from 100 eV to MeV, focusing on the depletion of the heavier ``excited'' state abundance via scatterings and decays, and on the distinctive signals arising from its small surviving abundance. We analyze CMB energy injection bounds on both DM annihilation and decay, accelerator-based probes, and new line-like direct-detection signals from the excited DM down-scattering on either nuclei or electrons, as well as future search prospects in each channel.  We also comment on the relevance of this signal to the few-keV Xenon1T electron excess and on possible diurnal modulation of this signal, and introduce a signal-strength parametrization to facilitate the comparison of future experimental results to theoretical expectations.

\end{abstract}
\maketitle

\tableofcontents
\section{Introduction}
Although Dark Matter (DM) comprises 80\% of the Universe's matter density, its particle constituents and origin remain elusive.  A compelling possibility for the origin of DM is that it could arise as a thermal relic, which achieves chemical equilibrium with ordinary matter in the hot early Universe, then falls out of equilibrium as the Universe cools to a temperature and density where DM particles fail to annihilate efficiently \cite{Kolb:1990vq}.  This thermal relic scenario and, more specifically, the possibility that DM annihilates through the weak interactions of the Standard Model has motivated a large program of searches for weak-scale dark matter. 

Dark Matter particles with a mass of a few GeV or lower can also arise as thermal relics if their annihilation is dominated by a new force with GeV-scale mediator \cite{Boehm:2002yz,Pospelov:2007mp, ArkaniHamed:2008qn} rather than by the weak interactions \cite{Lee:1977ua} or TeV-scale new physics.  This possibility falls simultaneously in the blind spots of most direct detection experiments, which lack sensitivity to the low-energy nuclear recoils induced by (sub)GeV-scale DM, and of hadron accelerator probes that rely on DM interactions becoming strong at high energies \cite{Abercrombie:2015wmb, Abdallah:2015ter}. This situation has led to the proposal of new, complementary strategies to search for sub-GeV dark matter, based on either direct detection of DM-electron scattering \cite{Essig:2011nj, Essig:2015cda, Schutz:2016tid} or the production of DM particles in lower-energy, high-intensity accelerator experiments \cite{Izaguirre:2015eya,Battaglieri:2014qoa, Izaguirre:2014dua, Izaguirre:2013uxa, Izaguirre:2014bca, Izaguirre:2015yja, Izaguirre:2015zva, Diamond:2013oda, TheBelle:2015mwa, deNiverville:2012ij}.  These proposals offer substantial opportunities to test sub-GeV dark matter, with sufficient anticipated sensitivity to test several thermal DM hypotheses over a wide DM mass range.

In addition to the detection of DM scattering or production, a third class of observation sets important constraints on light DM: searches for signals of DM annihilation in the Galaxy, in other bound halos, and in the primordial early Universe.  The most powerful such constraint comes from observations of the cosmic microwave background (CMB), on which DM annihilation and decay can leave an imprint by slightly re-ionizing hydrogen after the time of recombination \cite{Adams:1998nr,Chen:2003gz,Padmanabhan:2005es,Chluba:2011hw, Slatyer:2015jla,Slatyer:2015kla, Slatyer:2009yq, Finkbeiner:2011dx}.   The resulting constraint is particularly powerful in models where the DM annihilation cross-section at the $\sim \rm{eV}$ temperature of last scattering is simply related to the annihilation cross-section during freeze-out, which is in turn fixed (for thermal relic DM) by the observed DM abundance.  Indeed, Planck data excludes dark matter lighter than $\sim10-50$ GeV (depending on the decay mode) that self-annihilates through $s$-wave processes. That simple constraint applies only if DM annihilation is unchanged between the era of relic freeze-out (at temperatures $T\sim 1/30 m_{DM}$) and recombination at $T\sim \rm{eV}$.  

Simple models readily violate this assumption. The most familiar example is $p$-wave annihilation, as in the case of scalar dark matter annihilating through a vector mediator into Standard Model particles, whose cross-section falls at low DM temperatures.  A second effect that can suppress annihilations at low temperatures is the reduced abundance at low temperatures of a DM species participating in the annihilation reaction.  Such a suppression is realized in the simplest technically natural model of GeV-scale thermal dark matter, consisting of a Dirac fermion DM annihilating to SM particles through a vector kinetically mixed with the photon.  This annihilation process is $s$-wave, but there is no symmetry in this model that forbids a splitting of the Dirac fermion into two mass eigenstates. When such a splitting is present, the leading vector coupling is mass-off-diagonal.  Therefore, the annihilation rate at late times is affected not only by the total dark matter density but also by the density of the heavier "excited" DM state, which can be exponentially suppressed at low temperatures/late times. This paper explores in detail the cosmology of this scenario. We explore how the depletion of the DM excited state weakens the constraints from the CMB on light dark matter, as well as detection prospects from both accelerator-based searches and direct detection. 

Recently, an excess in the electron recoil signal of the XENON1T detector was observed at energies between 2-3 keV \cite{Aprile:2020tmw}. It is possible that the scattering of an excited fermionic dark matter state off electrons can explain this excess when the mass splitting between the heavy and light DM states is of order of a few keV \cite{Bloch:2020uzh,Baryakhtar:2020rwy}. In order to understand more carefully if this possibility is not ruled out by previous observations, a precise calculation of the cosmological constraints becomes relevant.

The rest of this paper is organized as follows: in Section~\ref{sec:inel}, we introduce the inelastic vector-portal dark matter.  In Section~\ref{sec:tot}, we analyze the freeze-out of the \emph{total} dark matter abundance and show that, for a large region of the parameter space, we recover the results familiar from standard treatments of elastic dark matter freeze-out. In Section~\ref{sec:depopulation}, we consider the  abundance of the heavier (or ``excited'') mass eigenstate $\chi_h$ as a fraction of the dark matter, and its depletion through processes that convert it into the lighter mass eigenstate $\chi_l$,  with a focus on its abundance at recombination. We analyze separately the effects of DM scattering off SM matter, DM self-scattering, and decays ${\chi_h} f \rightarrow {\chi_l} + \rm{SM}$.  In Section \ref{sec:primordial}, we derive constraints in the plane of DM mass and mass splitting from Planck's CMB data by looking at residual annihilations and decays. Such constraints are obtained assuming thermal freeze-out through the vector portal and widely-used benchmarks for dark-sector couplings. Afterwards, we briefly review existing constraints and prospects from accelerator-based experiments in Section\ref{exp}. In Section \ref{directdetection}, we explore the direct detection landscape and analyze the resulting constraints from the absence of both nuclear and electron recoil signals in terrestrial experiments. We also analyze the signal sensitivity for various experimental searches and their constraining power for a thermal target. Finally, we discuss our results and conclude in Section~\ref{concl}. Details on the thermally averaged cross-sections, freeze-out calculations, and constraints for alternative benchmark scenarios can be found in the appendices. 

\section{Inelastic Vector-Portal Dark Matter} \label{sec:inel}
In this paper, we  analyze MeV- and GeV-scale dark matter that annihilates into Standard Model matter through a new interaction, or \emph{portal}, mediated by a light particle. The simplest technically natural case is a Dirac fermion dark matter ($\chi$) coupled to a vector boson that kinetically mixes \cite{Holdom:1985ag} with the photon,
\begin{equation}
\lag\supset \bar{\chi}(i\sla{D}-m_\chi)\chi+\frac{1}{2}m_A'^2 A'^{2}+\frac{1}{2}\epsilon_Y\,F^Y_{\mu\nu}F'^{\mu\nu},
\label{diracModel}\end{equation}
where the covariant derivative is given by $D_\mu\equiv\partial_\mu+ig_DA'_\mu$, $g_D$ is the dark coupling, $F^Y_{\mu\nu}$ is the hypercharge field strength tensor, and $F'^{\mu\nu}$ is the \textit{dark} field strength tensor corresponding to the new vector field $A'$. 
We define $\alpha_D\equiv g_D^2/4\pi$ in analogy with QED. The kinetic mixing parameter $\epsilon_Y$ is expected to be small since it can arise naturally at one- or two-loop level.  

In fact, no symmetry of this model forbids Majorana mass terms that would split the Dirac fermion $\chi$ into two Majorana mass eigenstates \cite{TuckerSmith:2001hy}.  We first summarize the basic features and DM phenomenology of this model in the absence of a Majorana splitting, then discuss the effects of this splitting on the DM physics. 

After electroweak symmetry breaking, the leading effect of the mixing operator is an induced kinetic mixing with the photon: 
\begin{equation}
\lag\supset\frac{1}{2}\epsilon\,F_{\mu\nu}F'^{\mu\nu},
\end{equation}
where $\epsilon\equiv\epsilon_Y\,\cos{\theta_w}$, $\theta_w$ is the weak mixing angle, and $F_{\mu\nu}$ is the QED strength field.  After diagonalizing the kinetic mixing terms, the SM matter originally charged under $U(1)_{\text{QED}}$ acquires a small $U(1)_D$ charge proportional to $\epsilon e$; this means that any Feynman diagram with a photon can be replaced by the \textit{dark} photon with the coupling rescaled by $\epsilon$. On the other hand, the DM fermions remain uncharged under $U(1)_{\text{QED}}$. 

When $m_{A'}\leq m_\chi$ (``secluded dark matter''), the dominant annihilation mode is through the $t$-channel process $\bar{\chi}\chi\rightarrow A'A'$ \cite{Pospelov:2007mp}. This channel is not suppressed by $\epsilon$ and can lead to DM underproduction unless $\alpha_D\ll\alpha$ \cite{Izaguirre:2014bca}. If, on the other hand, $m_{A'}> m_\chi$, then the leading annihilation process is $\bar{\chi}\chi \rightarrow f\bar f$ mediated through an $s$-channel $A'$  (additional special cases that arise when $m_{A'}$ is only slightly above $m_\chi$ are discussed in \cite{DAgnolo:2015ujb,Cline:2017tka,Fitzpatrick:2020vba,Fitzpatrick:2021cij}).  For MeV $\lesssim m_{\chi} \lesssim $ few GeV, this process can give rise to the correct relic density of dark matter for reasonable values of $\epsilon$ and moderate splittings between the $A'$ and $\chi$ masses.  However, both of the processes noted above are $s$-wave and therefore, for Dirac dark matter as in \eqref{diracModel}, constrained by CMB data over this entire mass range. 

A small Majorana mass term $\delta$ for the DM splits the Dirac DM into two near-degenerate mass eigenstates $\chi_l$ and $\chi_h$, with $m_{{\chi_h}}=m_{\chi_l}+\delta$. Moreover, the leading dark-photon coupling is mass-off-diagonal \cite{TuckerSmith:2001hy}.   As a result, although a small mass splitting has negligible effect on the DM freeze-out, it can drastically affect the late-time thermal history of DM annihilating through an $s$-channel dark photon and in turn the interpretation of the CMB constraint. We focus on the case $\delta \ll m_{\chi_l}$ (for discussion of the case $\delta \sim m_{\chi_l}$, see e.g. \cite{Izaguirre:2015zva}). In order to avoid under-production of DM, as in the previous case, we consider the scenario where $m_{A'}> m_{\chi_l}$. 

To present the interactions of this model more explicitly, we first decompose the Dirac fermion $\chi$ in two left-handed Weyl spinors $\chi$ and $\eta$ as $\chi = {\tiny \left( \begin{array}{c} \eta \\ \xi^\dagger \end{array} \right) }$ (following the conventions of \cite{Dreiner:2008tw}). The interactions and mass terms in the DM fermion sector become
\begin{align}
\lag\supset&-g_D\left(\eta^\dagger\bar{\sigma}^\mu A'_\mu\eta-\xi^\dagger\bar{\sigma}^\mu A'_\mu\xi\right)\nonumber\\
&-m_\chi\eta\xi - \frac{1}{2}m_\eta\eta\eta-\frac{1}{2}m_\xi\xi\xi+h.c., \label{introduceMajoMass}
\end{align}
where $m_\chi$ is the Dirac mass and $m_\eta$ and $m_\xi$ are the two Majorana masses allowed by symmetry.  The Majorana masses violate the dark $U(1)$ charge conservation but this is already broken in the model by the $A^\prime$ mass; although we do not discuss UV completions here explicitly, we note that such Majorana masses could readily arise in a Higgs-mechanism UV completion of the dark sector from $\xi$ and $\eta$ Yukawa couplings (or from higher-dimension couplings to Higgs fields, depending on the relative charges of the Higgs and dark fermions). These interactions split the spectrum into two nearly degenerate Majorana mass eigenstates, related to the Weyl fermions above as
\begin{align}
\chi_l=&\frac{1}{\sqrt{2}}\left(\eta+\left(\frac{m_\eta-m_\xi}{2m_\chi}+\sqrt{1+\frac{(m_\eta-m_\xi)^2}{4 m_\chi^2}}\right)\xi\right)\ ,\\
\chi_h=&\frac{1}{\sqrt{2}}\left(\eta+\left(\frac{m_\eta-m_\xi}{2m_\chi}-\sqrt{1+\frac{(m_\eta-m_\xi)^2}{4 m_\chi^2}}\right)\xi\right)\ ,
\end{align}
where the corresponding masses for the light $\chi_l$ and heavy $\chi_h$ states are given by 
\begin{align}
m_{\chi_l}&=\sqrt{m_\chi^2+(m_\eta-m_\xi)^2/4}-\delta/2 \ ,\\
m_{\chi_h}&=\sqrt{m_\chi^2+(m_\eta-m_\xi)^2/4}+\delta/2 \ ,
\end{align}
with $ \delta=m_\eta+m_\xi$. Switching to the mass eigenstate basis, the dark photon interactions take the form:
\begin{align}
\lag\supset\!\!-i g_D\!\!\frac{m_\chi}{\sqrt{m_\chi^2+(m_\eta-m_\xi)^2/4}} \!\!\left(\!{\chi_l}^\dagger\bar{\sigma}^\mu A'_\mu{\chi_h}\!-\!\chi_h^{\dagger}\bar{\sigma}^\mu A'_\mu{\chi_l}\!\right) \label{eq:inelastic-coupling} \\
-i g_D\frac{m_\eta-m_\xi}{\sqrt{m_\chi^2+(m_\eta-m_\xi)^2/4}} \left({\chi_h}^\dagger\bar{\sigma}^\mu A'_\mu{\chi_h}-\chi_l^{\dagger}\bar{\sigma}^\mu A'_\mu{\chi_l}\!\right). \label{eq:elastic-coupling}
\end{align}
Here we can see that there is an elastic and an inelastic contribution to the DM interactions. For small Majorana masses, the inelastic coupling is unsuppressed while elastic interactions are both velocity-suppressed and parametrically suppressed by the ratio of Majorana to Dirac masses. This means that the dominant $s$-channel annihilation mode in the early Universe is $\chi_h \chi_l \rightarrow \rm{SM}$, the rate of which can be suppressed at late times as the $\chi_h$ fractional abundance is depleted.  For simplicity, we take $m_\eta=m_\xi$, which corresponds to restoring the parity symmetry. In this case the elastic coupling vanishes exactly.  We will briefly revisit the case $m_\eta\neq m_\xi$ in Section \ref{sssec:hlll}. From now on, we analyze the freeze-out of the DM fermions taking into account only the inelastic part and show how the inelastic scatterings can weaken the CMB constraints.
\section{Boltzmann Equations and Thermal Benchmark} \label{sec:tot}
In this section, we analyze the requirements on the minimal model described above to give rise to the observed dark matter density. The number densities of the two dark matter species $\chi_l$ and $\chi_h$ are governed by Boltzmann equations that account for their annihilations as well as up- and down-scattering processes that convert $\chi_l$ into $\chi_h$ and vice versa. These processes follow from the inelastic interactions in Eq.~\eqref{eq:inelastic-coupling}.

\begin{figure}[!tb]
	\includegraphics[scale=0.7]{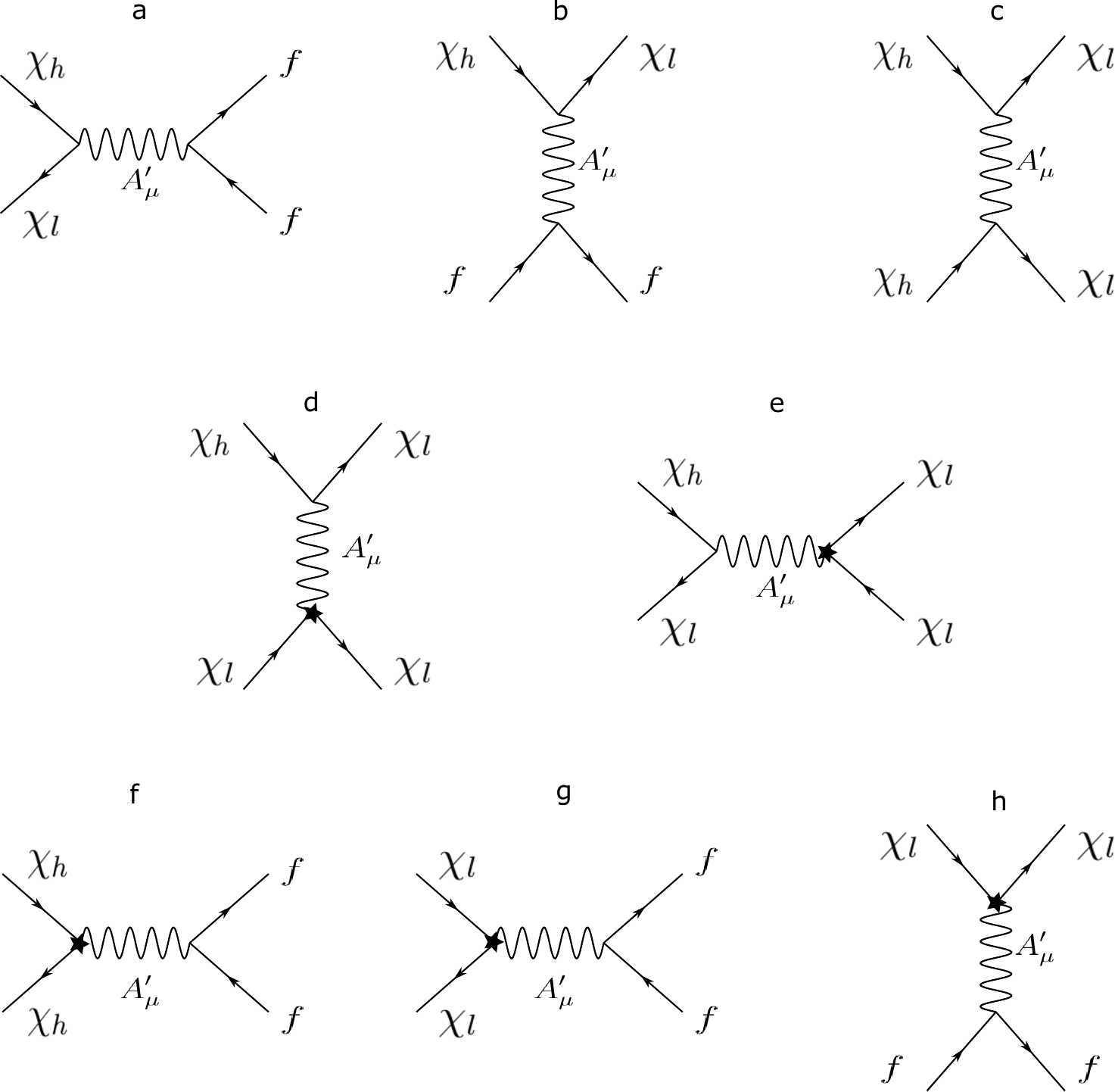}
	\caption{Feynman graphs for the different possible scattering processes, with time running from left to right. 
	Graphs a, b, and c in the top row are the dominant diagrams for DM annihilation, down-scattering off SM matter, and down-scattering off each other that contribute to the Boltzmann equations \eqref{bolx} and \eqref{bolxst} (graphs a and b are also responsible for observable signals --- CMB energy-injection and down-scattering direct detection, respectively).   These rely only on the leading \emph{inelastic} DM-dark-photon coupling \eqref{eq:inelastic-coupling}.  
	Graphs in the second and third lines include an elastic interaction from \eqref{eq:elastic-coupling}, denoted by a vertex with a star.  This coupling arises only in the parity-breaking case ($m_\eta \neq m_\xi$) and is suppressed by both $\delta/m$ and a velocity factor. Graphs d and e in the second line are the leading contributions to the semi-elastic processes discussed in Sec.~\ref{sssec:hlll} to the depletion of the excited state, which can sometimes contribute to depletion of $\chi_h$ despite these suppressions. The graphs in the bottom line are elastic contributions to DM annihilation (f and g) or scattering (h) that are mentioned in the text but negligibly small.
	\label{Fig:FeynmanGraphs} }
\end{figure}

First, we introduce the notation for the following analysis. We define the dimensionless variable $x\equiv\tfrac{m_{\chi_l}}{T}$ where $T$ stands for the standard model particles' temperature. Note that, as long as the dark matter-standard model interactions are in equilibrium the dark matter temperature will be equal to the standard model temperature. The thermally averaged cross section of the process with initial states $a$ $b$ is denoted $\langle\sigma_{a\,b}\,v\rangle$, and the thermally averaged decay rate is  $\langle\Gamma_{{\chi_h}}\rangle_i$. The final states of these processes are those arising from the inelastic interactions in Eq. \eqref{eq:inelastic-coupling}. $H(x)=H(m) x^{-2}$ is the Hubble parameter (from now on we will use the notation $H\equiv H(m)$), $Y_a(x)=\tfrac{n_a(s)}{s(x)}$ is the normalized number density of the species $a$ where $s(x)$ is the entropy density, and $Y_a^{\text{EQ}}(x)$ is the equilibrium number density of the species $a$. We also set the mass splitting between ground and excited state to be smaller than the DM mass. The Boltzmann equations for $\chi_l$ and $\chi_h$ are
\begin{align}
	Y_{\chi_l}'&=\frac{s}{H \; x^2} \Bigg[-\langle \sigma_{\text{ann}}\,v \rangle  \left(Y_{\chi_l} Y_{{\chi_h}}-Y^{\text{EQ}}_{\chi_l} Y^{\text{EQ}}_{{\chi_h}} \right) +\langle\sigma_{{\chi_h} f}\,v\rangle \left(Y_{\text{ferm}}
Y_{{\chi_h}}-\frac{Y^{\text{EQ}}_{{\chi_h}}}{Y^{\text{EQ}}_{{\chi_l}}}\;Y_{\text{ferm}}\,
Y_{{\chi_l}}\right) \nonumber \\
&+2\,\langle\sigma _{{\chi_h} {\chi_h}}\,v\rangle  \left(Y_{{\chi_h}}^2-\frac{(Y^{\text{EQ}}_{{\chi_h}})^2}{(Y^{\text{EQ}}_{{\chi_l}})^2}Y_{{\chi_l}}^2\right)\Bigg]  + \sum_i\frac{x}{H}\langle\Gamma_{{\chi_h}}\rangle_i \left(
Y_{{\chi_h}}-\frac{Y^{\text{EQ}}_{{\chi_h}}}{Y^{\text{EQ}}_{{\chi_l}}}\;\,
Y_{{\chi_l}}\right) , \label{bolx} 
\end{align}
\begin{align}
	Y_{{\chi_h}}'&=\frac{s}{H\; x^2}\Bigg[-\langle \sigma_{\text{ann}}\,v \rangle  \left(\!Y_{\chi_l} Y_{{\chi_h}}-Y^{\text{EQ}}_{\chi_l} \,Y^{\text{EQ}}_{{\chi_h}} \right) -\langle\sigma_{{\chi_h} f}\,v\rangle \left(Y_{\text{ferm}}
	Y_{{\chi_h}}-\frac{Y^{\text{EQ}}_{{\chi_h}}}{Y^{\text{EQ}}_{{\chi_l}}}\;Y_{\text{ferm}}\,
	Y_{{\chi_l}}\right) \nonumber \\
	&-2\,\langle\sigma _{{\chi_h} {\chi_h}}\,v\rangle  \left(Y_{{\chi_h}}^2-\frac{(Y^{\text{EQ}}_{{\chi_h}})^2}{(Y^{\text{EQ}}_{{\chi_l}})^2}Y_{{\chi_l}}^2\right)\Bigg] - \sum_i \frac{x}{H}\langle\Gamma_{{\chi_h}}\rangle_i \left(
	Y_{{\chi_h}}-\frac{Y^{\text{EQ}}_{{\chi_h}}}{Y^{\text{EQ}}_{{\chi_l}}}\;\,
	Y_{{\chi_l}}\right), \label{bolxst}
\end{align}
where prime denotes a derivative with respect to $x$. The thermally averaged cross section for the annihilation processes is $\sigann$, for which ${\chi_l}\,{\chi_h}\rightarrow e\,e$ (Fig.~\ref{Fig:FeynmanGraphs}-a) is a typical process and additional contributions arise from annihilation to heavier leptons and hadrons above their kinematic thresholds. The thermally averaged cross sections $\langle\sigma_{{\chi_h} f}\,v\rangle$ and $\langle\sigma _{{\chi_h} {\chi_h}}\,v\rangle $ correspond to the scatterings ${\chi_h}\,f\rightarrow {\chi_l}\,f$ (Fig.~\ref{Fig:FeynmanGraphs}-b) and ${\chi_h}\, {\chi_h}\rightarrow {\chi_l}\, {\chi_l}$ (Fig.~\ref{Fig:FeynmanGraphs}-c) respectively. For the scattering off fermions, we will only consider the dominant process which is the scattering off electrons. The sum over $i$ corresponds to different decay channels which will be explored in detail in the following section.

The dark matter annihilation to fermions changes $Y_\text{tot}=Y_{\chi_l}+Y_{\chi_h}$, but not the difference of abundances. Meanwhile, the scattering processes are the only ones that change the relative abundance of the ground and excited states. Each of these processes freezes out at a characteristic temperature where the expansion rate is approximately equal to the particles interaction rate. This happens when the relevant density, controlled by a Boltzmann factor or ratio of Boltzmann factors, becomes sufficiently small. In the annihilation case, this gives $T_\text{fo}\lesssim m_{\chi_l}$, for the scattering off electrons (positrons) $T_\text{fo}\lesssim m_e$, and for the scattering off dark matter $T_\text{fo}\lesssim \delta$. In this paper, we focus mostly on the limit $\delta< m_e< m_{\chi_l}$, which implies that the annihilation process freezes out earlier than the scatterings. Having the dark matter annihilations frozen-out gives a fixed $Y_\text{tot}$ which will let us analyze each scattering term on the RHS of the Boltzmann equation separately. It is important to notice that the limit $\delta< m_e< m_{\chi_l}$ will not always imply that the scattering off electrons freeze out earlier than the scattering off dark matter, since in these cases the cross sections could have a temperature dependence that comes into play. 

The equations \eqref{bolx} and \eqref{bolxst} can be summed to find a simpler equation for $Y_{\text{tot}}\equiv Y_{{\chi_l}}+Y_{{\chi_h}}$ given by 
\begin{equation}
Y_{\text{tot}}'\!=\!\frac{2\,s}{\,H\, x^2}\! \frac{e^{\frac{-\delta}{m_{\chi_l}}x}}{\left(1+e^{\frac{-\delta}{m_{\chi_l}}x}\right)^2} \!\left(\!-\langle \sigma_{\text{ann}} \rangle  \!\left(Y_{\text{tot}}^2-\left(Y^{\text{EQ}}_{\text{tot}}\right)^2 \right)\!\right) \ , \label{totall}
\end{equation}
where we have assumed that the scatterings keep the ratio $Y_{\chi_h}/Y_{\chi_l}=e^{\frac{-\delta}{m_{\chi_l}}x}$ in equilibrium and verified that this is valid in the parameter space of interest. The solution of this equation follows standard methods and is detailed in Appendix \ref{apfo}; for small $\delta$ the result recovers the standard result for elastic DM freeze-out. After this process freezes out, the total dark matter abundance will only change due to the expansion of the Universe. Given this, we can find an estimate for the annihilation cross section that gives rise to the observed dark matter density today. When $\delta\lesssim 0.1m_{\chi_l}$, the cross section is of order $\sim10^{-9}-10^{-8}\; \text{GeV}^{-2}$. On the other hand, if $\delta$ is comparable to $m_{\chi_l}$ the cross section required to obtain the observed dark matter abundance will be large and mass-dependent. These large cross sections are ruled out by accelerator-based experiments \cite{Izaguirre:2013uxa,Izaguirre:2014bca,Izaguirre:2015yja,Izaguirre:2015zva,Hook:2010tw,Curtin:2014cca,Essig:2013vha,Batell:2014mga,Alexander:2016aln,Aguilar-Arevalo:2017mqx,Lees:2017lec}, see Section \ref{exp} and Fig. \ref{eps}.

One can obtain an expression for the kinetic mixing $\epsilon$ by setting
\begin{equation}
	\sigann(m_{\chi_l},\delta,\alpha_D,m_{A'})\Big|_{T=T_{fo}}=\sigann^\text{t. relic}(m_{\chi_l},\delta)
		\label{epsilon}
	\end{equation}
where $\sigann^\text{t. relic}$ is the cross section required to produce the observed dark matter abundance, and $\sigann$ is the thermally averaged annihilation cross section in Eq.~\eqref{sigmaThAv}, which includes annihilation to electrons, heavier fermions and hadrons. Details on the calculation of this cross section can be found in Appendix \ref{App:CrossSec}. This is helpful since it allows us to reduce the parameter space in the dark matter-fermion scattering to only $m_{\chi_l}$, $\delta$, and $\alpha_D$.

\section{Depopulation of Excited State before Recombination} \label{sec:depopulation}
The dark matter excited state, $\chi_h$, can be depopulated through several different decay and scattering processes. In this section, we analyze the effect of each type of process on the final abundance $Y_{\chi_h}$.  Decays of $\chi_h$ into $\chi_l e^+e^-$ rapidly deplete the $\chi_h$ abundance when $\delta>2 m_e$; even when these are kinematically forbidden, radiative decays $\chi_h \rightarrow \chi_l +3\gamma$ can still be relevant over cosmological timescales.  In the absence of rapid decays, the $\chi_h$ abundance can still be depleted by inelastic down-scatterings of DM particles off each other and off Standard Model fermions, which convert $\chi_h$ into $\chi_l$.  Both processes must be considered as they experience different suppressions --- the scattering off SM fermions is suppressed by the kinetic mixing $\epsilon$, while the self-scattering is suppressed by the low DM abundance.  We find that DM self-scattering provides the stronger depletion for sub-GeV DM, while scattering off electrons is more important above a few GeV.  Finally, we identify the parameter region where semi-elastic scattering $\chi_h \chi_l \rightarrow \chi_l \chi_l$ can be important, in generalized models with parity-violating Majorana masses $m_\eta \neq m_\xi$. 

After the total DM abundance has frozen out, the heavy and light dark matter states can still maintain thermal contact via scattering and/or decays, and therefore their relative abundances satisfy a thermal equilibrium relation.  In this period $Y'_{\text{tot}}=0$ and $Y_{\chi_h}/Y_{\chi_l}=e^{-\frac{\delta}{T_\chi}}$, implying 
\begin{equation}
Y_{{\chi_l}}=Y_\text{tot}\frac{1}{1+e^{-\frac{\delta}{T_\chi}}}, \qquad Y_{{\chi_h}}=Y_\text{tot}\frac{e^{-\frac{\delta}{T_\chi}}}{1+e^{-\frac{\delta}{T_\chi}}} \ , \label{yyy}
\end{equation}
where $T_\chi$ is the DM temperature. Before $\chi f $-scattering freezes out, the dark matter and standard model are also in kinetic equilibrium, so $T_\chi=T$. After this process freezes out, the two temperatures decouple. These expressions for the pseudo-equilibrium abundance hold until the last inelastic DM scattering process freezes out (and assuming that DM annihilation freezes out before scattering). Thus, to find the dark matter abundances, we determine the freeze-out temperature for each process that contributes to the excited state depletion. We will compute this using the instantaneous freeze-out approximation for each process separately in the following subsections. Afterwards, we will analyze the regions of parameter space which are dominated by the different processes.

\subsection{Depopulation Through Decays}
 An important process to take into account if $\delta>2 m_e$ is the decay ${\chi_h}\rightarrow{\chi_l}+ 2 e$. When kinematically allowed, this decay will rapidly deplete the excited state. The decay rate is given by 
\begin{equation}
\Gamma_{{\chi_h}\rightarrow \chi_l+2 e}\simeq\frac{4 \alpha \alpha_D \epsilon^2 \delta^5}{15 \pi m_{A}^{'4}} \ ,
\end{equation} 
giving an excited state lifetime $\tau_{{\chi_h}}=\langle\Gamma_{{\chi_h}}\rangle^{-1}.$ This tree-level process has a lifetime much smaller than the age of the Universe at recombination. Thus it heavily depletes the excited state whenever $\delta >2 m_e$.

When $\delta<2 m_e$, the 2-electron decays are forbidden but $\chi_h$ can still undergo decays ${\chi_h}\rightarrow{\chi_l}+ 3 \gamma$ (via an electron loop) and ${\chi_h}\rightarrow{\chi_l}+ 2\nu$ (via $A^\prime-Z$ mixing). Their decay rates are given by\cite{Batell:2009vb}
\begin{align}
\Gamma_{{\chi_h}\rightarrow \chi_l+3\gamma}&\simeq\frac{17 \alpha^4 \alpha_D \epsilon^2 \delta^5}{2^73^65^3\pi^3 m_{A'}^4 } \frac{\delta^8}{m_e^8}\ , \label{Dphotons} \\
\Gamma_{{\chi_h}\rightarrow \chi_l+2\nu}&\simeq \frac{4 \sin^2{\theta_w^4}  \alpha_D \epsilon^2   \delta^5 }{315 \pi^3 \alpha  m _{A'}^4} G_F^2 \delta^4 \ ,
\end{align}
where $m_e$ is the electron mass and $G_F$ the Fermi coupling constant.  The decay to neutrinos is suppressed by $G_F \delta^4$ compared to the decay to electrons, while the decay to photons is suppressed by eight powers of $\delta/m_e$. For most of our parameter space, both lifetimes are larger than the age of the Universe, but for a large mass splitting close to the electron mass and sufficiently light DM, the decay to photons becomes relevant.  When these decay processes dominate, they lead to a relative abundance of the excited state at recombination of the form
\begin{equation}
Y^{PE}_{{\chi_h}}=Y_\text{tot}e^{-\Gamma_\text{decay} t_\text{rec}}  \ , \label{decays}
\end{equation}
where $t_\text{rec}$ is the age of the Universe at recombination and $\Gamma_\text{decay}$ the decay rate of the process. Note that the assumption of thermal DM fixes $\epsilon$ as a function of $m_\chi$, $m_{A^\prime}$, and $\delta$ as given by Eq.~\eqref{epsilon}.

\subsection{Depopulation Through Collisions}
In this subsection we analyze the scattering processes than can deplete the heavy state even when it is cosmologically long-lived. The two processes relevant when the Majorana masses respect parity ($m_\eta = m_\xi$) are scattering of $\chi_{h}$ off a Standard Model fermion (dominated by electrons) or off another $\chi_{h}$. We consider each of these in turn, then briefly examine the semi-elastic scattering process $\chi_h \chi_l \rightarrow \chi_l \chi_l$ that can further deplete very light DM if the two Majorana mass terms are unequal, allowing a small elastic DM interaction with the gauge boson.

\subsubsection{${\chi_h} f \rightarrow {\chi_l}f$ Scattering} \label{chif}
We start by considering the effects of the scattering off fermions. The dominant scattering off fermions is the one off electrons and positrons. Considering the processes ${\chi_h}\, e^\pm \rightarrow {\chi_l}\,e^\pm$ (Fig.~\ref{Fig:FeynmanGraphs}-b), we can get the freeze-out temperature and the abundance for the dark matter heavy and light states. The freeze-out temperature for these scatterings is obtained by assuming an instantaneous freeze-out and solving
\begin{eqnarray}
H(x_{{\chi} e^\pm}^{-2})\,=\,Y_e(x_{{\chi} e^\pm})\, s (x_{{\chi} e^\pm}^{-3})\, \langle\sigma_{{\chi_h}e^\pm}\,v\rangle, \label{foc}
\end{eqnarray}
where $H(x_{{\chi} e^\pm}^{-2})\equiv H(m_e)\,x_{{\chi} e^\pm}^{-2}$, $x_{{\chi} e^\pm}\equiv\frac{m_e}{T_{{\chi} e^\pm}}$, $Y_e=Y_{e^-}+Y_{e+}$, and $T_{{\chi} e^\pm}$ is the temperature at which both scattering processes freeze out. The freeze-out temperature for the scattering off electrons is the same as the scattering off positrons one since we assume \footnote{Assuming an electron asymmetry as large as the baryon one, $\eta_B\sim 10^{-10}$, the assumption $Y_{e^+}=Y_{e^-}$ is a good approximation for $T\geq 1/ (20 m_e)$.} $Y_{e^+}=Y_{e^-}$.

\begin{figure}[!h]
	\begin{center}
		\includegraphics[width=0.8\textwidth]{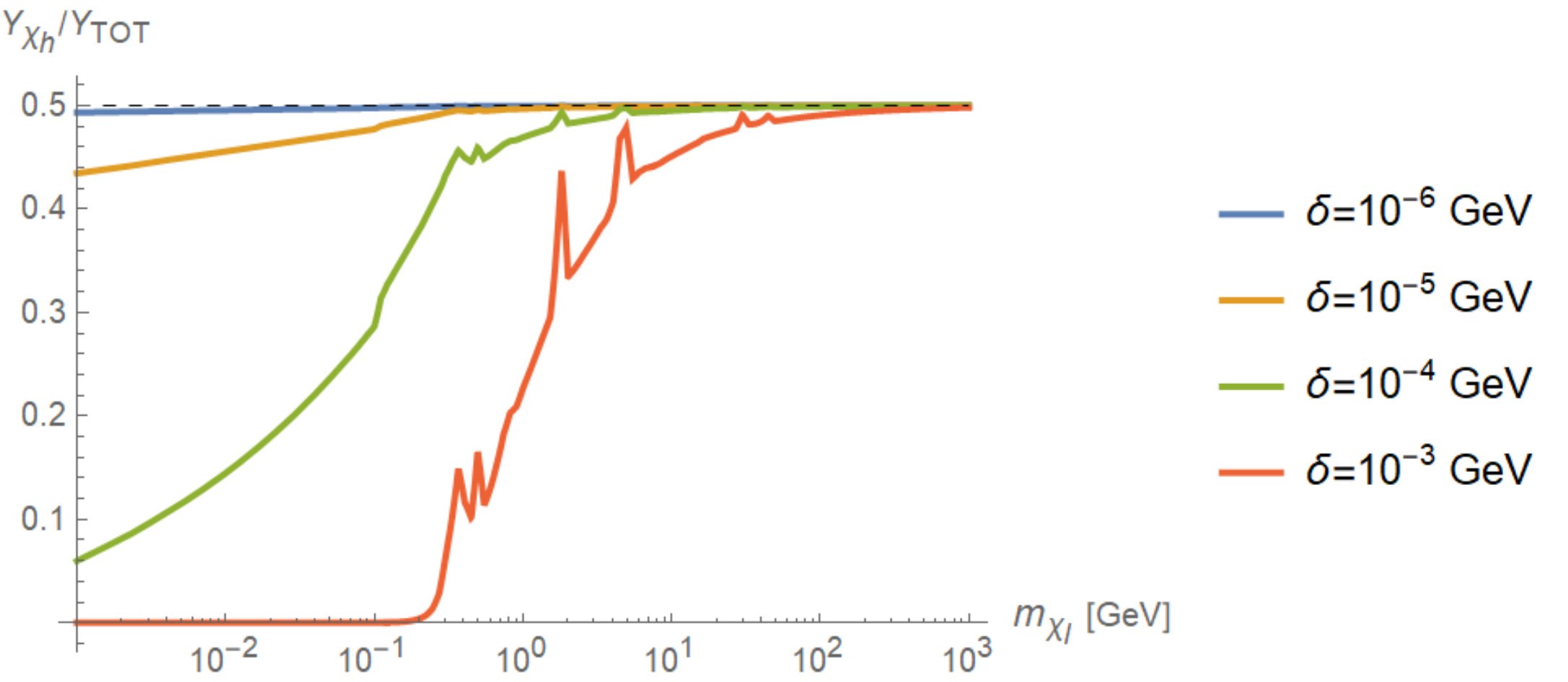}
	\end{center}
	\vspace{-0.5cm}
	\caption{This plot shows a comparison of the depletion of the excited state for different mass splittings and dark matter masses with $m_{A'}= 3 \, \left(m_{\chi_l}+\delta/2\right)$. We consider only the depletion produced by the scattering of dark matter off electrons and positrons (see Fig.~\ref{foscx} for the depletion by DM self-scattering). Note that we are considering an $\epsilon$ fixed for a thermal target. This value of $\epsilon$ depends on the total annihilation cross section which involves hadronic resonances giving rise to the observed peaks in the graph.} \label{fosce}
\end{figure}

Depending on the dark matter mass, this process can freeze out when the electron is non-relativistic ($Y_e\sim e^{-x}$) or when it is relativistic ($Y_e\sim constant$). The freeze-out happens when the electron is non-relativistic for $m_{\chi_l}\lesssim 1 \,  \text{GeV}$. In this case, the instantaneous freeze-out approximation gives:
\begin{equation}
x_{{\chi} e^\pm}=\log{\left[6\frac{\sqrt{10}}{(2\pi)^{3}}\frac{g}{g_*^{1/2}}\, M_{Pl}\, m_e\, \langle\sigma_{{\chi_h}e^\pm}v\rangle |_{x_{{\chi} e^\pm}}\,x_{{\chi} e^\pm}^{1/2}\right]}. \label{xxf}
\end{equation}
Here, we will use an analytical piece-wise approximation for $\langle\sigma_{{\chi_h}e^\pm}v\rangle $. The different regions in the approximation correspond to the cases where different particles are relativistic or non-relativistic, this is shown in detail in the Appendix \ref{csxf}, and the cross section's behavior is sketched in Fig. \ref{csxfT}. For larger masses, the freeze-out happens when the electron is relativistic which leads to:
\begin{equation}
x_{{\chi} e^\pm}\!=\!\left(\!\frac{63\,\sqrt{5}\,\zeta (3) }{4 \pi ^{7/2}  }\frac{g}{\sqrt{g_*(x_{{\chi} e^\pm})}}\frac{m_e^3 \,M_{\text{Pl}}}{m_{{\chi_l} }^2} \, \sigann) \!\right)^{\frac{1}{3}}\!\!.\!\! \label{xrel}
\end{equation}
In order to treat the relativistic to non-relativistic transition of the electron carefully, we have used the exact expression for the electron number density when solving for the freeze-out temperature.

When considering the effects of the dark matter-electron(positron) scattering, we find that for a large region of the parameter space of interest, the excited state and ground state abundance stay unchanged. From Fig. \ref{fosce}, we  can see that the region in which the dark matter abundance of the excited state is depleted significantly is for a mass splitting close to the electron mass and small DM masses. This can be explained by looking at the fact that, $x_{{\chi} e^\pm}=\mathcal{O}(10)$ for small DM masses, while for larger masses $x_{{\chi} e^\pm}=\mathcal{O}(10^{-1})-\mathcal{O}(1)$. This means that in order for this process to heavily deplete the excited state, whose abundance is determined by the Boltzmann factor $e^{-\frac{\delta}{m_e}x_{{\chi} e^\pm}}$, a mass splitting of the order of the electron mass is required.

\subsubsection{${\chi_h} {\chi_h}\rightarrow {\chi_l}\; {\chi_l}$ Scattering} \label{hhll}
Now, we consider the effects of the self-scattering of dark matter: ${\chi_h} {\chi_h}\rightarrow {\chi_l}{\chi_l}$ (Fig.~\ref{Fig:FeynmanGraphs}-c). Proceeding in a similar manner as the previous case, we obtain the freeze-out temperature by assuming an instantaneous freeze-out and solving $\Gamma_{{\chi_h} {\chi_h}}=H$. 

If $\chi \chi$-scattering freezes out before $\chi e$ scattering, then we would simply have 
\begin{equation}
x_{{\chi}{\chi}}^\text{early}=\log{\left[\frac{ \sqrt{\pi}}{3 \sqrt{5}}\frac{g_{*s}}{g_*^{1/2}} M_{Pl}Y_\text{tot}\, \delta\, \langle\sigma_{{\chi_h}\,{\chi_h}}v\rangle \frac{1}{x_{{\chi}{\chi}}}-1 \right]} \ .
	\vspace{0.3cm}
	\label{xX2}
\end{equation}
We include this case for completeness, but it is of limited interest since in this case it is $\chi e$ scattering rather than $\chi\chi$ scattering that determines the residual $\chi_h$ abundance at late times. 

If, instead, $\chi \chi$-scattering freezes out after $\chi e$ scattering, then the above equation must be modified to account for the kinetic decoupling between the dark matter and Standard Model sectors.  The $\chi_h\chi_h$ scattering cross-section and $\chi_h$ abundance are controlled by the DM temperature $T_\chi$, while Hubble expansion is controlled by the SM temperature $T$.  After $\chi e$ scattering freezes out, the dark matter temperature will redshift away as
\begin{equation}
T_{\chi}=T_{\chi e^\pm }\left(\frac{T \ g^{1/3}_{*s}(T)}{T_{\chi e^\pm}\ g^{1/3}_{*s}(T_{\chi e^\pm })}\right)^2 \ ,
\end{equation}
since, for the entire \mbox{parameter} space, the fermion-DM scattering always freezes out when the dark matter is non-relativistic. This leads to  
\begin{align}
x_{{\chi}{\chi}}^\text{late}=&\left( \frac{\delta}{T_{\chi e^\pm }}\!\right)^{\frac{1}{2}}\left(\frac{g^{1/3}_{*s}(T)}{g^{1/3}_{*s}(T_{\chi e^\pm })}\right) \left(\log{\left[\frac{ \sqrt{\pi}}{3 \sqrt{5}}\frac{g_{*s}}{g_*^{1/2}} M_{Pl} Y_\text{tot}\, \delta\, \langle\sigma_{{\chi_h}\,{\chi_h}}v\rangle \frac{1}{x_{{\chi}{\chi}}}-1 \right]}\right)^{\frac{1}{2}}  ,
\vspace{0.3cm}
\label{xX1}
\end{align} 
where $x_{{\chi}\,{\chi}}\equiv\frac{\delta}{T_{{\chi}\,{\chi}}}$ and $T_{{\chi}\,{\chi}}$ is the temperature at which the process ${\chi_h} {\chi_h}\rightarrow {\chi_l} {\chi_l}$ freezes out. 

As before, we need to know how the cross section scales in different regions; this is given by Eq. \eqref{crossecxx} and sketched in Fig. \ref{csxx}. Note that at leading order the Boltzmann factor driving the abundance of $\chi_h$ can be approximated as
\begin{align}
\left(\frac{\delta}{T_\chi}\right)^\text{late}\sim&\log\!{\!\left[\frac{ \sqrt{\pi}}{3 \sqrt{5}}\frac{g_{*s}}{g_*^{1/2}} M_{Pl}\,Y_\text{tot}\, \delta^{\frac{1}{2}}\,m_e^{\frac{1}{2}} \, \langle\sigma_{{\chi_h}\,{\chi_h}\rightarrow{\chi_l} \,{\chi_l}}v\rangle  \right]\!} \ , \label{BoltzChiChi} \\
\left(\frac{\delta}{T_\chi}\right)^\text{early}\sim&\log\!{\!\left[\frac{ \sqrt{\pi}}{3 \sqrt{5}}\frac{g_{*s}}{g_*^{1/2}} M_{Pl}\,Y_\text{tot}\, \delta\, \langle\sigma_{{\chi_h}\,{\chi_h}\rightarrow{\chi_l} \,{\chi_l}}v\rangle  \right]\!} \ .
\end{align}

\begin{figure}[!ht]
	\begin{center}
		\includegraphics[width=0.8\textwidth]{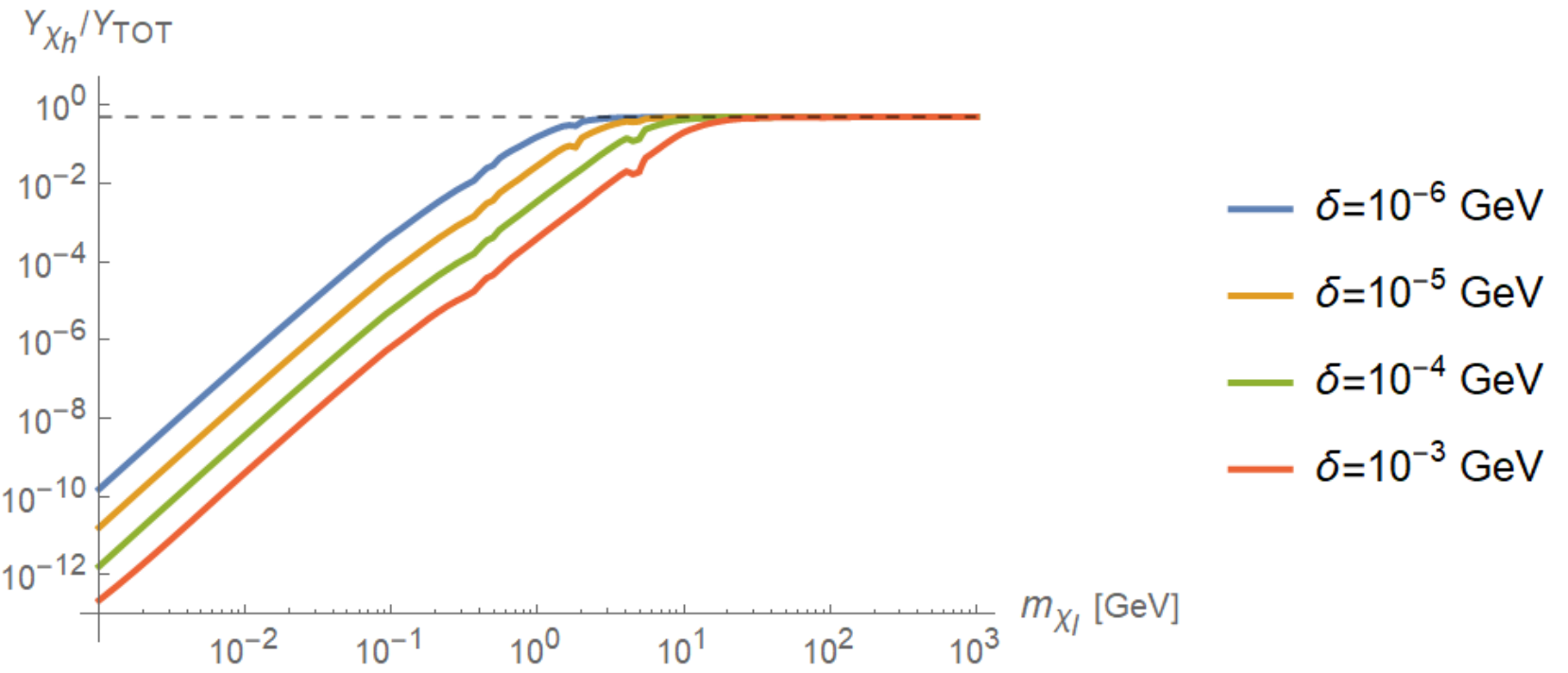} \\
	\end{center}
	\vspace{-0.3 cm}
	\caption{This plots shows the depletion of the heavy state for $m_{A'}=3 \left(m_{\chi_l}+\delta/2\right) ,  \ \alpha_D=0.5$. We consider only the depletion from $\chi\chi$ self-scattering (see Fig.~\ref{fosce} for the depletion by DM self-scattering). We see that the larger the mass splitting the more effective the depletion. Similarly, for a large ${\alpha_D^2}/{m_{A'}^4}$ the depletion is larger. The small peaks observed here arise due to the dependence on $T_{\chi e^\pm }$ which in turn is affected by hadronic resonances.} \label{foscx}
	\vspace{-0.3 cm}
\end{figure}

It is worth noting that for large DM masses this process freezes out early and does not significantly change  the abundance of the heavy state with respect to the light state. The value of the DM mass for which the depletion starts being effective depends on the dark coupling and dark gauge boson mass. Increasing the ratio ${\alpha_D^2}/{m_{A'}^4}$ causes the depletion to be more effective for heavier dark matter. At fixed ${\alpha_D^2}/{m_{A'}^4}$, the depletion of the excited state is more efficient for large mass splittings, see Fig. \ref{foscx}.

\subsubsection{Estimating the Effects of Semi-Elastic ${\chi_h} {\chi_l}\rightarrow {\chi_l}\; {\chi_l}$ Scattering}  \label{sssec:hlll}
If $m_\eta\neq m_\xi$, the elastic coupling of Eq. \eqref{eq:elastic-coupling} will also be present. In the limit of Majorana masses $m_{\eta,\xi} \ll m_\chi$, this interaction is suppressed relative to the inelastic one by $(m_\eta-m_\xi)/m_\chi$ (of order $\delta/m_\chi$ unless the two Majorana masses are fine-tuned against each other). Its effects are also velocity-suppressed in the non-relativistic limit.  The process ${\chi_h} {\chi_l}\rightarrow {\chi_l}\; {\chi_l}$, which we refer to as ``semi-elastic scattering'', involves one inelastic and one elastic coupling, and therefore its cross-section is suppressed relative to ${\chi_h} {\chi_h}\rightarrow {\chi_l}\; {\chi_l}$ by both kinematic and mass-ratio factors.   However, at low dark-sector temperatures the rate per $\chi_h$ depends only on  $Y_{\chi_l}$, which is slowly varying, and not on the Boltzmann-suppressed $Y_{\chi_h}$.  Therefore, this reaction can come to dominate after sufficiently efficient depletion of the $\chi_h$ abundance by other processes.  This can happen for sufficiently light DM mass and large enough splittings $\delta$.  

Setting $m_\xi=0$ for concreteness, so that $\delta = m_\eta$, we find that the scattering cross section scales as
\begin{equation}
	\langle\sigma \, v\rangle_{\chi_h\chi_l\rightarrow\chi_l\chi_l}=	\frac{64 \pi \, \alpha _D^2\,\delta^{2}}{ m_{A'}^4  }\left(\frac{\delta}{m_{{\chi_l} }}\right)^{3/2} \ .
\end{equation}
The region where the depletion of $\chi_h$ due to these scatterings is significant is seen in the purple triangle in the bottom right corner of the left-hand panel of Fig. \ref{regions}.  In this case, the depletion of the excited state is exponential, similar to the freeze-out of the scattering off relativistic electrons in subsection \ref{chif}, due to the scattering rate being controlled by the light state abundance which varies only polynomially after the total DM freeze-out.  However, because this reaction's rate depends on $m_\eta-m_\xi$, it is not fully determined from $m_\chi$ and $\delta$, and indeed can be absent if $m_\eta = m_\xi$ (as well as for scalar inelastic DM).  Therefore, we neglect this process in the rest of our analysis.  We will, however, continue to note the region where it would dominate over the $\chi_h \chi_h \rightarrow \chi_l \chi_l$ if $m_\xi=0$ and lead to substantial additional depletion of the DM abundance. In this triangular region of parameter space, our calculations of $\chi_h$ abundance and related constraints are robust only for the specific model with $m_\xi = m_\eta$, while for $m_\xi \neq n_\eta$ they should be considered as upper bounds on the abundance and resulting signals.  Outside the triangular region, our calculations are approximately valid even for more general models, so long as the Majorana masses are not fine-tuned (i.e. assuming $m_\xi \sim m_\eta \sim \delta$).

\subsection{Excited state abundance}
\begin{figure}[!h]
	\begin{center}
		\includegraphics[width=0.53\textwidth]{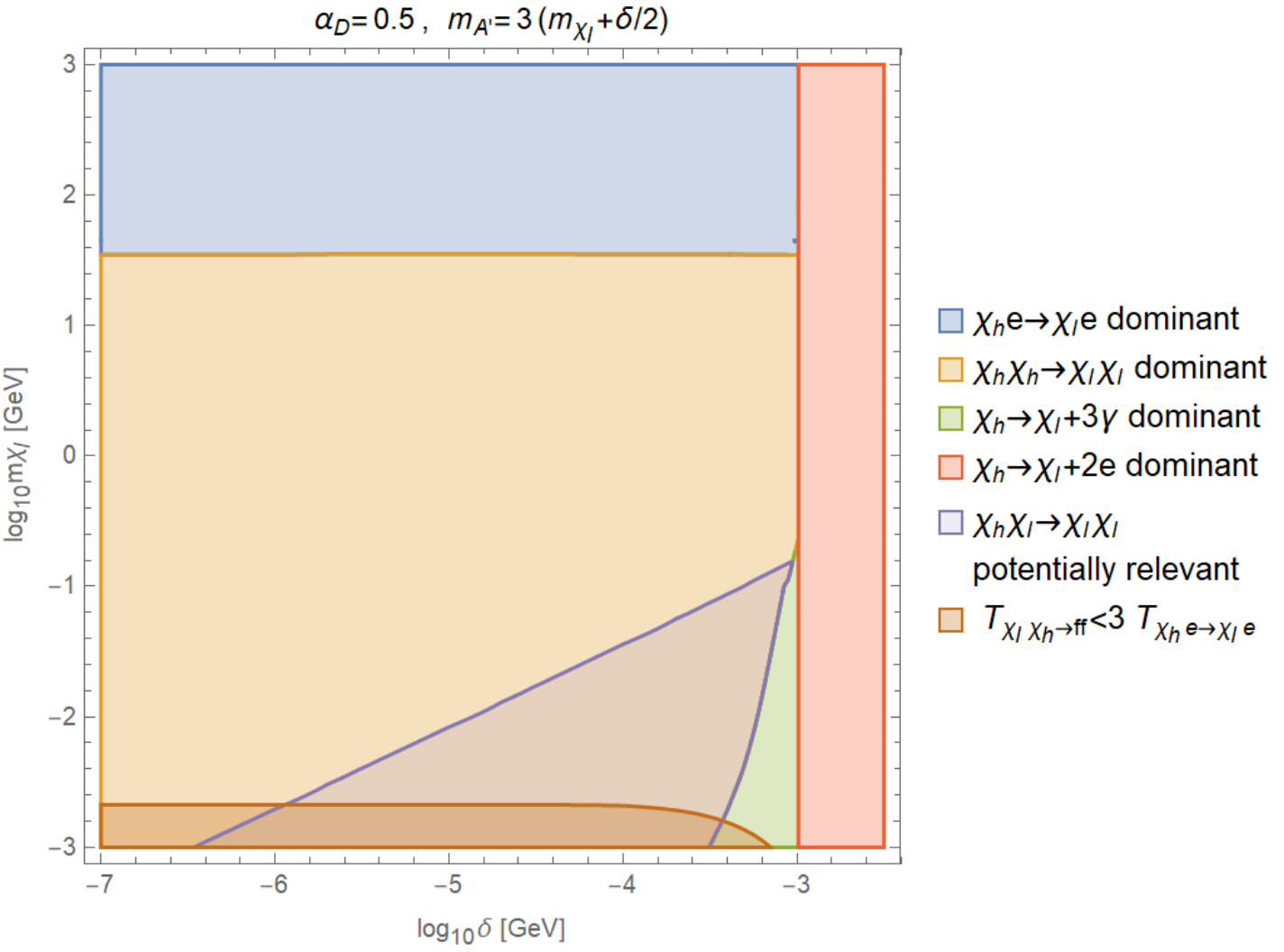}
		\includegraphics[width=0.46\textwidth]{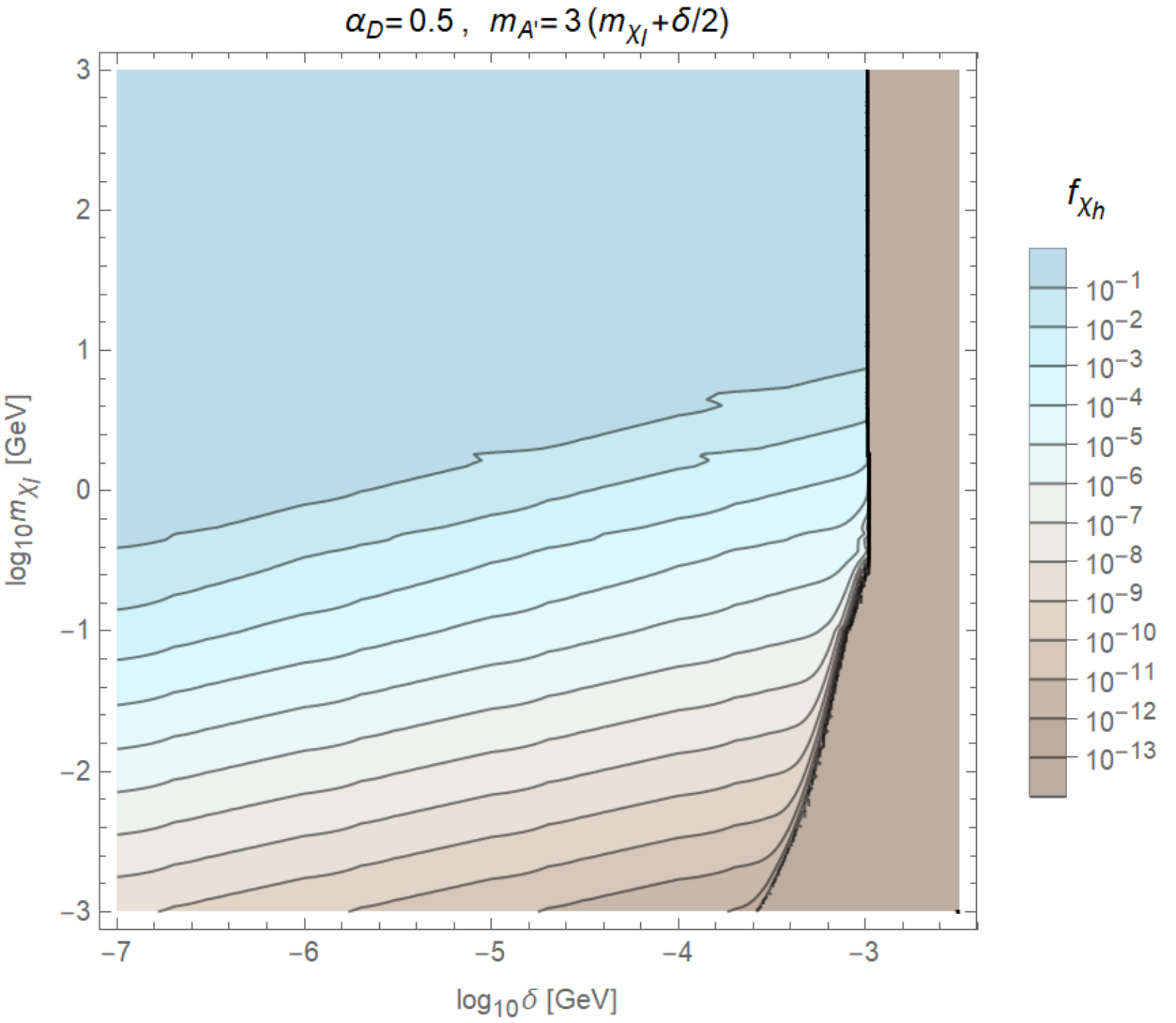}
	\end{center}
	\vspace{-0.5cm}
	\caption{The left-hand side shows the regions of the parameter space where the abundance of the excited sate at recombination is set by the decay to electrons, decay to photons, scattering off fermions, or scattering of dark matter. We also show the region where other processes may further deplete the abundance and the region where our approximations start to become less precise. The right-hand side shows the contours of the relative abundance of the dark matter excited state  at $T^\text{rec.}$, the recombination temperature. The abundance decreases sharply in regions where the decay processes dominate and at a slower rate when the scatterings are dominant. Analogous plots for additional dark-sector benchmarks are shown in Figures \ref{appFig:dominantReactions} and \ref{appFig:abundances}. \label{regions}}
\end{figure}

After analyzing each process that can lead to a significant depletion of the excited state, we assume that the excited abundance is set only by the dominant process in each region of our parameter space --- this should be accurate everywhere except perhaps near the boundaries between two regions. Since in the next section we are interested in analyzing the constraints from CMB data, here we focus on the abundances at recombination. Fig.~\ref{regions}(left) shows which process controls the $\chi_h$ abundance at recombination over the DM mass range $\MeV < m_\chi < \TeV$ and for splittings $10^{-7} \eV < \delta  < 3\, \MeV$. For $\delta>2m_e$ the decay to electrons and positrons is responsible for the largest depletion. For $\delta\lesssim 2m_e$, we have a small band where the decay to photons is the most relevant process. Note that decays also control the present-day $\chi_h$ abundance over a larger but similar-shaped triangular region. In the other regions, we can see that the scatterings dominate. For large dark matter masses, the largest depletion is obtained from the scattering off fermions. Lastly, for most of the analyzed parameter space, DM self-scattering is the most efficient process maintaining $\chi_h-\chi_l$ chemical equilibrium, and therefore controls the final $\chi_h$ abundance.   One should also note that there is a region where the ${\chi_h} {\chi_l}\rightarrow {\chi_l}\; {\chi_l}$ scattering could potentially deplete the excited state's abundance further. There is a small region at low DM masses where the temperature of kinetic decoupling from the SM, $T_{\chi_h e \rightarrow \chi_l e}$, is below the temperature of chemical decoupling $T_{\chi_h \chi_l \rightarrow f\bar f}$ by only a factor of 1.5-3. In this region, our use of the instantaneous freeze-out approximation may lead to $O(1)$ inaccuracies in the predicted abundance.  

In the right-hand side of Fig. \ref{regions}, we can observe the relative abundance of the excited state,
\begin{equation}
	f_{\chi_h}=\frac{Y_{\chi_h}}{Y_{\chi_h}+Y_{\chi_l}} \ .
\end{equation}
It is clear from this plot that in the regions where the decays dominate there is a sharp depletion of the excited state. This sharp depletion corresponds to the exponential suppression of the abundance in Eq. \eqref{decays}. Scatterings deplete the excited state at a slower rate since in this case the abundance is inversely proportional to the scattering cross-section. Nonetheless, for sub-GeV DM the depletion due to $\chi-\chi$ self-scattering can be considerable. 

\section{Primordial constraints and prospects} \label{sec:primordial}
Both annihilation and decay of DM can observably change the ionization history of the Universe during the era of recombination, which in turn alters the temperature and polarization power spectra of the CMB \cite{Chen:2003gz,Padmanabhan:2005es, Galli:2009zc, Padmanabhan:2005es, Finkbeiner:2011dx}.  Thus, the agreement of CMB data  with $\Lambda$CDM predictions can be used to set bounds on the energy injected per unit volume by DM annihilation or decay, with each case having a different time-dependence of energy injection and therefore giving rise to somewhat different constraints.  In our model, both the residual $\chi_h \chi_l \rightarrow f \bar f$ annihilation and the decays $\chi_h \rightarrow \chi_l + 3\gamma$ can be constrained by these bounds.  We will consider each case in turn. 
\subsection{Residual annihilations}
For DM annihilation, we follow the treatment of the Planck Collaboration \cite{Ade:2015xua,Aghanim:2018eyx} suitably modified to account for the fact that $\chi_h$ and $\chi_l$ can co-annihilate, but neither one can self-annihilate \footnote{There are subtle factors of two in the equations below. We use the normalization corresponding to Majorana DM, which is appropriate to our model and also assumed the Planck bound on $p_{ann}$ quoted below \cite{Ade:2015xua}.  In this context, the population-average annihilation rate for a mixture of $\chi_h$ and $\chi_l$ dark matter with fractions $f_h$ and $f_l$ is  $f_{\chi_h}^2 \langle \sigma_{hh} v \rangle + 2 \;f_{\chi_h}f_{\chi_l} \langle \sigma_{hl} v \rangle + f_{\chi_h}^2 \langle \sigma_{hh} v \rangle$. For our model ($m_\eta= m_\xi$), only the co-annihilation process is nonzero giving rise to the normalization given in \ref{pan}.  We also note that in the limit $f_{\chi_l} = f_{\chi_h} = 1/2$, \eqref{ez} recovers the standard expression for the energy release for annihilation by Dirac DM, as expected.}.  We write the rate of energy release by annihilating dark matter per unit volume as
\begin{equation}
	\left(\frac{\ud E(z)}{\ud t\, \ud V}\right)_\text{ann}= \,\rho_{\text{crit}}^2\,\Omega_{\text{CDM}}^2\,(1+z)^6 p_{\text{ann}}(z) \ , \label{ez}
\end{equation}
where $p_{\text{ann}}$ is defined as
\begin{equation}
	p_{\text{ann}}\equiv 2 \;f_{\chi_h}f_{\chi_l}\;\langle\sigma_{\text{ann}}v\rangle\frac{f(z)}{m_{\chi_l}} \ , \label{pan}
\end{equation}
with $f(z)$ the efficiency factor that measures how much of the injected power is actually deposited to the intergalactic medium (IGM).  Here, the dark matter densities are evaluated at recombination, which occurs after scattering processes have frozen out. As in the previous analysis, we have neglected the elastic annihilations in this equation. These elastic reactions (shown in Fig.~\ref{Fig:FeynmanGraphs} f and g) vanish in the parity-symmetric limit $m_\eta =m_\xi$; even for general Majorana masses they are both $p$-wave and suppressed by $(m_{\eta}-m_\xi)^2/m_\chi^2$, making them unobservable.

\begin{figure}[!b]
	\begin{center}
		\includegraphics[scale=1]{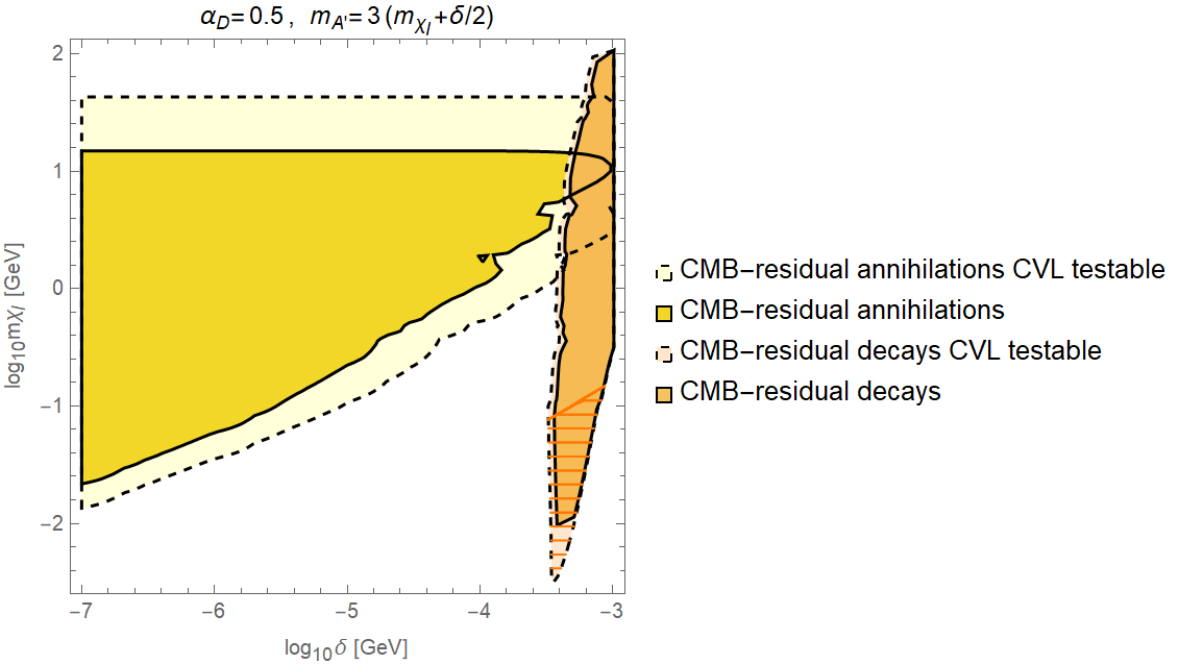}
	\end{center}
	\vspace{-0.3 cm}
	\caption{Primordial constraints for  $m_{A'} = 3 \,\left(m_{\chi_l}+\delta/2\right)$ and $\alpha_D=0.5$. The yellow zone in the plot corresponds to cross annihilation cross sections $\sigann>\sigann^\text{MAX}$, which is excluded by Planck 2018 data. The light yellow contour shows the maximum region that a cosmic variance limited experiment like Planck would be able to probe. The orange region corresponds to constraints arising from energy release due to decays, and correspondingly the light orange shows the CVL testable region. The striated orange area corresponds to the region where the exclusion can be weakened due to further depletion of the excited state from semi-elastic interactions.  Analogous plots for additional dark-sector benchmarks are shown in Figure \ref{appFig:cmb}. }
	\label{CMBbounds}
\end{figure}

By adding $p_{\text{ann}}$ as an additional parameter to the $\Lambda$CDM cosmology, the 2018 Planck Collaboration analysis \cite{Aghanim:2018eyx} obtains the upper limit 
\begin{equation}
	p_{\text{ann}}<p_{\text{ann}}^{\text{MAX}}\equiv2.74\times10^{-11}\, \text{GeV}^{-3} \ .
\end{equation} 
corresponding to the TT, TE, EE + low E + lensing + BAO data at $z = 600$ (redshift at which the effect of dark matter annihilation peaks \cite{Finkbeiner:2011dx}). We use the following prescription for the effective efficiency factor: we take $f(z)=f_\text{eff}^{3keV}$ as given in \cite{Slatyer:2015jla} for $m_{\chi_l}\leq 100\, \text{MeV}$, that is, for the region where the dark matter mass is below the muon threshold. This efficiency factor includes corrections to the simple prescription used by the Planck Collaboration \cite{Aghanim:2018eyx} to convert deposited power into perturbations to the ionization history. Above $2$ GeV, we are in a region dominated by modes other than electrons and positrons; following the results found in \cite{Slatyer:2015jla}, we approximate $f(z)=0.2$ for $m_{\chi_l}> 2\, \text{GeV}$. For the region in between, $100\; \text{MeV}< m_{\chi_l} \leq 2\; \text{GeV}$, we also have modes different than the electron that come into play and a more complicated structure in the ionization efficiency. Here, we approximate the efficiency factor as $f(z)=f_\text{eff}^{3keV}$, and we have verified that this amounts to a negligible effect in the resulting constraints.  
Using the prescription explained above, we can find the regions of our parameter space that are excluded by Planck's upper limit on the annihilation power; this is shown in Fig. \ref{CMBbounds}.  The contour of the constrained region is set by the efficiency of dark matter self-scattering, which depends on the ratio ${\alpha_D^2}/{m_{A'}^4}$.  As such, smaller values of $\alpha_D$ or larger $m_{A^\prime}$ compared to our benchmark model would enlarge the constrained region as shown in Fig.~\ref{appFig:cmb} in Appendix \ref{app:MoreBenchmarks}. Future CMB observations will improve on the Planck constraint, but the ultimate reach of this program is limited by cosmic variance.  The dashed pale-yellow region in Fig.~\ref{CMBbounds}  illustrates the ultimate limitation to CMB bounds on DM annihilation, based on the cosmic-variance-limited sensitivity $p_{ann}^{\text{MAX,CVL}} =9.21\times10^{-12}\, \text{GeV}^{-3}$ inferred from \cite{Ade:2015xua}.

To understand the scaling of the contours as we vary the coupling strength and the dark gauge boson mass, we proceed to find the temperature at which the process has to freeze out in order to saturate the limits imposed by Planck ($T_{\text{sat}}$). From Eq. \eqref{pan} we see that
\begin{equation}
	\frac{e^{-\delta/T_{\text{sat}}}}{(1+e^{-\delta/T_{\text{sat}}})^2}=\frac{p_{\text{ann}}^{\text{MAX}}\, m_{\chi_l}}{4\,\sigann^\text{MAX}\, f(z)},
\end{equation}
where we have used Eq. \eqref{yyy} for $Y_{{\chi_h}}\;\text{and}\;Y_{\chi_l}$. Using the freeze-out condition from Eq. \eqref{BoltzChiChi} and approximating the total abundance by Eq.~\eqref{ytot}, we can write 
\begin{equation}
	\frac{\delta}{T_{\text{sat}}}\sim\log{\left[\frac{32 \sqrt{2} \pi ^{3/2}}{3 \sqrt{5}}\frac{ \alpha_D^2 \delta\,  m_e^{\frac{1}{2}}
		}{\sigann R^4
			m_{\chi_l}^{7/2}}\right]},
\end{equation} 
where we have defined $R\equiv m_{A'}/m_{\chi_l}$ and used the fact that the freeze-out happens at $T\ll \delta$ when the cross-section in Eq. \eqref{crossecxx}  is constant. Given this we find
\begin{equation}
	\frac{\sigann^\text{MAX}}{\sigann}\propto \frac{p_{\text{ann}}^{\text{MAX}}}{\sigann^2}\frac{ \alpha_D^2}{R^4}\frac{\delta\,  m_{e}^{\frac{1}{2}}}{m_{\chi_l}^{5/2}} \ . \label{md}
\end{equation}
Solving for $m_{\chi_l}$ in Eq. \eqref{md} for the limiting situation $\sigann^\text{MAX}=\sigann$ we obtain
\begin{equation}
	\log{m_{{\chi_l} }}\sim\frac{2}{5}\log{\delta}+\frac{2}{5}\log{\left[C\frac{\alpha_D^2}{R^4} \right]},\label{scal}
\end{equation} 
where $C=p_{\text{ann}}^{\text{MAX}}m_e^{\frac{1}{2}}/\sigann^2$, which explains the shape of the yellow contour in Fig. \ref{CMBbounds}. We can also notice that smaller values of $\alpha_D/m_{A'}^2$ will lead to stronger constraints.

\subsection{Residual decays}\label{ssec:primordial-decays}
Similar to the above, the total energy release per unit volume due to $\chi_h$ decays is given by
\begin{equation}
	\left(\frac{\ud E(z)}{\ud t\, \ud V}\right)_\text{decay}= \,\rho_{\text{crit}}\,\Omega_{\text{CDM}}\,(1+z)^3 p_\text{dec} \ , \label{EnergyDecay}
\end{equation}
where 
\begin{equation}
	p_\text{dec}\equiv f_{\chi_h} \frac{\delta}{m_\chi} \frac{f_\text{dec}(z)}{\tau_{\chi_h}} \ , \label{pdec}
\end{equation}
with $f_\text{dec}$  the decay efficiency factor. Relative to standard formulas, our $p_\text{dec}$ is suppressed by the fraction $f_{\chi_h}$ of DM that can decay, and the fraction $\delta/m_{\chi}$ of the $\chi_h$ decay energy that is transferred to the SM bath (the remainder of the energy stays in the dark sector in the form of a non-relativistic $\chi_l$).  In addition, the efficiency factor $f_\text{dec}$ depends on the energies of decay products, which are typically set by $m_\chi$ but in our model are determined by $\delta$. We evaluate $f_{\chi_h}$ at $z=600$, which limits sensitivity to $\tau \gtrsim 10^{13}\,\s$ (a more careful treatment of short-lived species' decays is given in \cite{Slatyer:2016qyl}). In our case, only the decays to photons have long enough lifetimes to be relevant, so we will only consider the lifetime given by Eq.\eqref{Dphotons}. As in the annihilations case, one can derive a bound on $p_\text{dec}$ from CMB observations. This has been done in \cite{Slatyer:2016qyl} by using Planck 2015 data \cite{Ade:2015xua}, and can be translated into the following conservative constraint 
\begin{equation}
p_\text{dec}<p_\text{dec}^\text{MAX}= \frac{f_\text{dec}(z\sim300)}{\tau_{\chi_h}^\text{min}}\simeq \frac{1}{3 \times 10^{24} \text{ s}} \ , \label{pdecMAX}
\end{equation}
considering a decay to photons with a typical photon energy set by $\delta \sim 0.3-1 \,\MeV$, the range of $\delta$'s where this constraint is relevant. We consider the bound on the lifetime in Eq. \eqref{pdecMAX} as conservative approximation for our case considering the results in \cite{Slatyer:2016qyl}, where the bound on the lifetime is of order $10^{24}$ s for a large range of photon energies (DM masses). From this bound, we find that mass splittings of around $0.3-1$ MeV are mostly ruled out, see Fig. \ref{CMBbounds}. Since the excited state lifetime scales as $\delta^{13}$, the contours of the excluded region will not change drastically even if the bound on $p_\text{dec}$ changes by a few orders of magnitude. Similar to the annihilation case, we can understand the shape of the orange contour in Fig. \ref{CMBbounds} analytically. Combining Eqs.\eqref{pdec} and \eqref{pdecMAX}, together with the expressions for the excited state relative abundance and lifetime (Eq.\eqref{Dphotons}), we find
\begin{equation}
	\log{m_{{\chi_l} }}\sim -26\log{\delta}+2\log{\left[c \frac{\alpha_D^2}{R^4}\right]} \ .
\end{equation}  
where $c=m_e^{17/2}/\tau_{\chi_h}^\text{min}\sigann^2$.  This explains the steep slope on the left boundary of the orange contour, which arises due to the large dependence on $\delta$ of the excited state lifetime. Note that the slope changes above $10$ GeV since in that region the scattering off fermions dominates. The boundary on the right side of the orange contour is determined by the excited state abundance, so it is controlled by the exponential depletion of $\chi_{h}$. Given this, we can see that the CVL testable region \footnote{
Here we have estimated the $p_{dec}$ testable in a cosmic-variance-limited as $p_{dec}^{MAX,Planck2015}/3.8$, i.e. the same sensitivity improvement as would be achievable for annihilations \cite{Ade:2015xua}. This crude estimate is adequate given the rapid scaling of lifetime with our model parameters.} 
can only slightly improve the bounds on the long excited state lifetime regions.
\section{Constraints from Accelerator-Based Dark Matter Production}\label{exp}
\begin{figure}[!h]
 	\vspace{-0.5cm}
 	\begin{center}
 		\includegraphics[scale=.45]{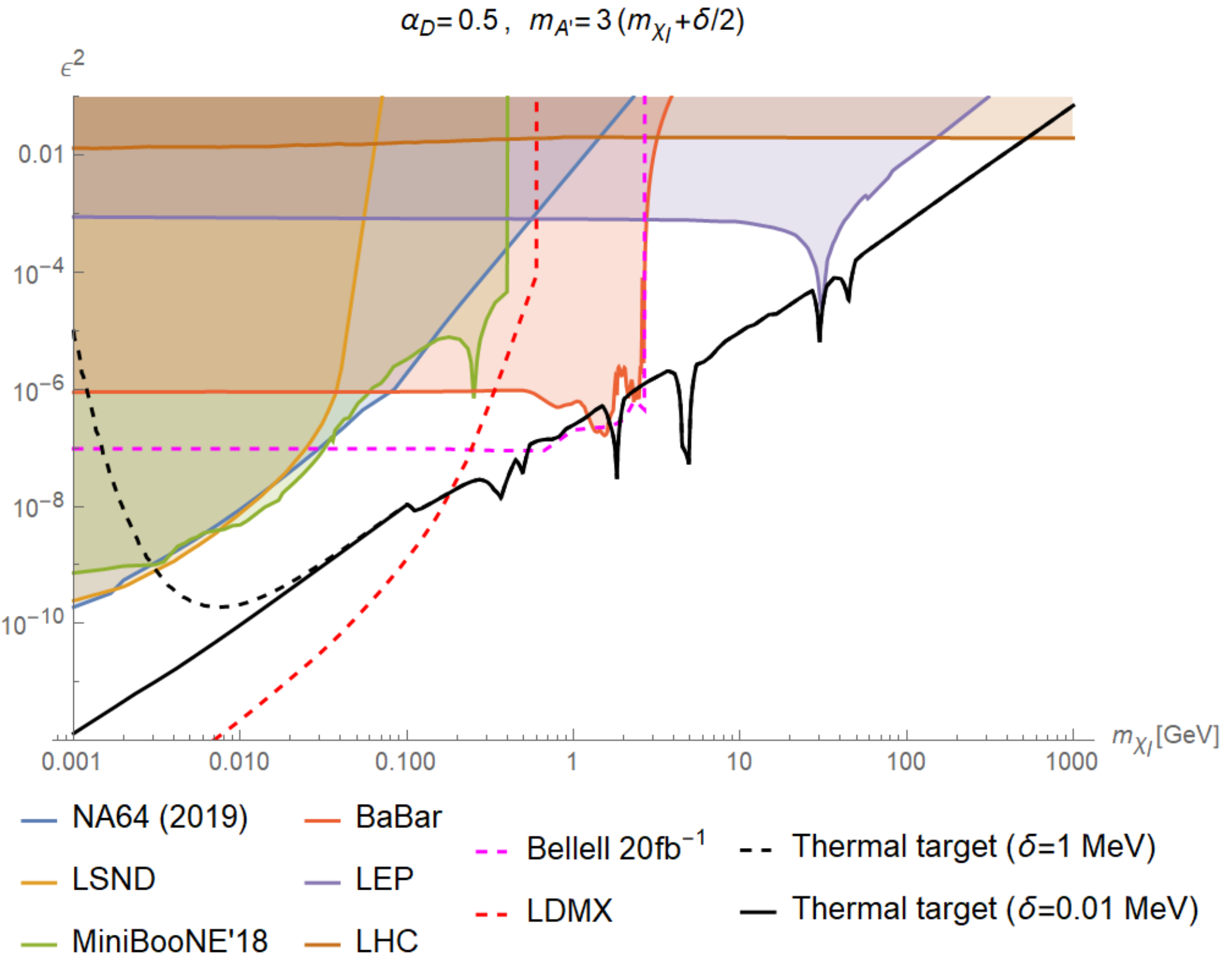}	
 	\end{center}
 	\vspace{-0.5cm}
 	\caption{Constraints from accelerator-based searches on the kinetic mixing parameter $\epsilon^2$ for $m_{A'}=3\left(m_{\chi_l}+\delta/2\right)$ and $\alpha_D=0.5$. Solid color lines correspond to current bounds and dashed color lines to prospective bounds. The black lines denote the $\epsilon^2$ value needed to get the observed dark matter abundance assuming $\chi_l$ and $\chi_h$ constitute all dark matter. For $\delta\lesssim m_{\chi_l}$, the cross section required to obtain the observed thermal relic is large; this leads to a large $\epsilon^2$ value which has already been ruled out. Analogous plots for additional dark-sector benchmarks are shown in Figure \ref{appFig:acceleratorEpsilon}.
 	\label{eps}}
 	\vspace{-0cm}
 \end{figure}

\begin{figure}[!h]
 	\vspace{-0.5cm}
	\begin{center}
		\includegraphics[scale=0.8]{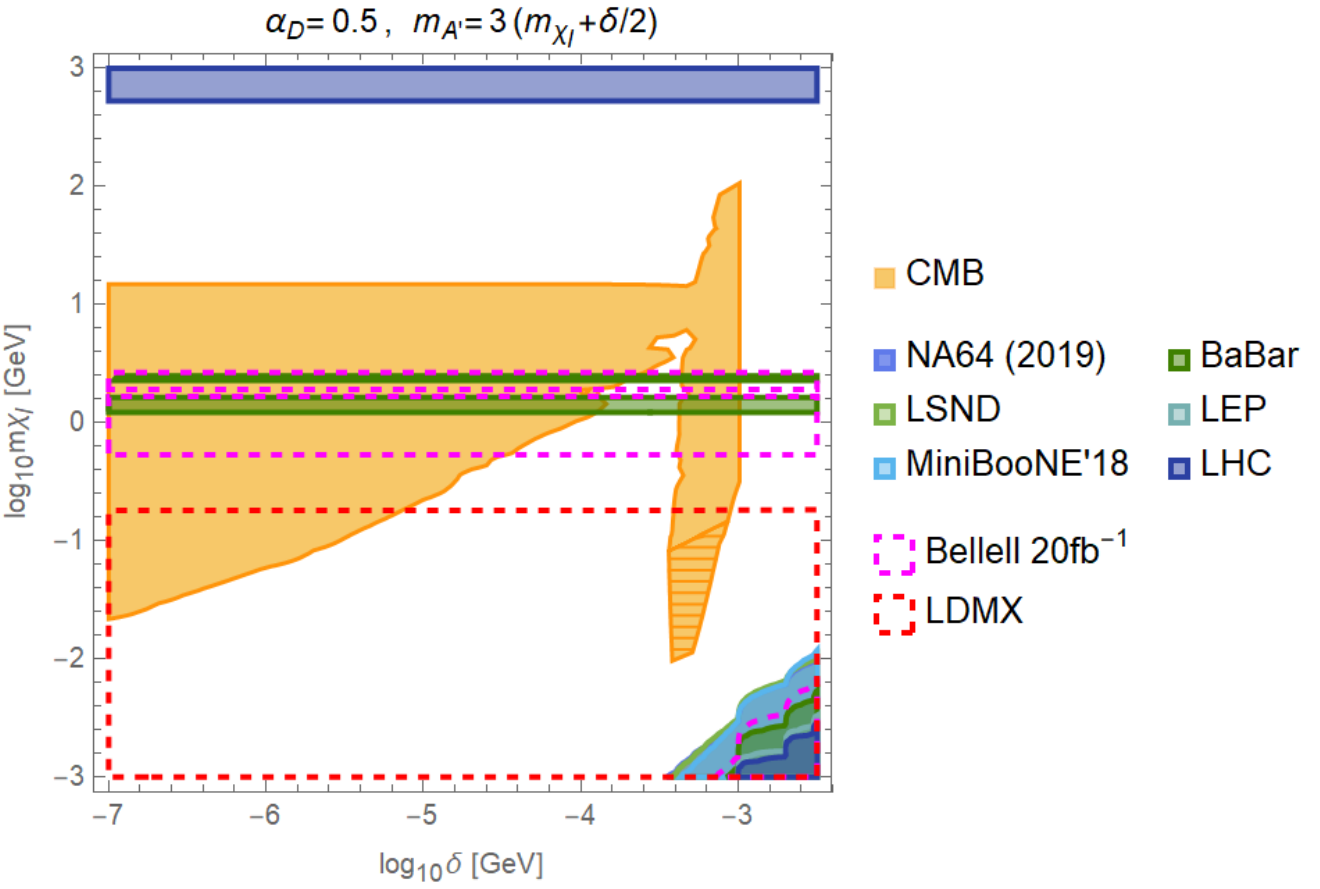}	
	\end{center}
	\vspace{-0.5cm}
	\caption{Regions of parameter space excluded for  $\alpha_D=0.5 $ and $m_{A'}=3 \,\left(m_{\chi_l}+\delta/2\right)$. The gold region is ruled out by CMB observations, for detail see Section \ref{sec:primordial}. The shaded regions are ruled out by different accelerator-based probes while the dashed contours show prospects of future experimental results. Analogous plots for additional dark-sector benchmarks are shown in Figure \ref{appFig:acceleratorCMB}.
	\label{acc}}
	\vspace{-0cm}
\end{figure}

Besides cosmological observations, accelerator-based probes are relevant in constraining dark matter models. These include fixed-target experiments (e.g. beam-dump, missing energy, and missing momentum experiments), which are most powerful as constraints on low-mass dark matter sectors, and collider constraints (missing mass at BaBar, LEP model-independent constraints on kinetic mixing, and LHC missing transverse energy searches), which have lesser coupling sensitivity but explore a much wider range of DM masses. These experiments and their sensitivity are reviewed for example in \cite{Alexander:2016aln,Battaglieri:2017aum,DarkMatterBRNReport,Beacham:2019nyx,Berlin:2020uwy,Lanfranchi:2020crw}. The constraints arising from them are shown in Fig. \ref{eps} for the kinetic mixing $\epsilon$ and in Fig. \ref{acc} for the $m_{\chi}\!-\!\delta$ space. These constraints are complementary to the cosmological and direct detection ones since they do not rely on the survival of dark matter to the present, but depend only on its interaction properties. Currently, accelerator probes only constrain the thermal benchmarks in a small region, but projections for LDMX~\cite{Akesson:2018vlm,Akesson:2019iul,Berlin:2018bsc} and a Belle II mono-photon search with 20 fb$^{-1}$ of data ~\cite{Kou:2018nap}\footnote{The study in \cite{Kou:2018nap} corresponds to 0.04\% of the Belle II goal luminosity, and therefore underestimates the experiment's ultimate sensitivity. Instrumental backgrounds make it challenging to reliably extrapolate Belle II's sensitivity with the full 50 ab${^{-1}}$ dataset, so here we will simply show the low-luminosity projection.}  show that they could constrain the whole sub-GeV region \cite{Battaglieri:2017aum}. In fact, these fixed target constraints have the ability to explore most of the parameter space, see Fig. \ref{acc}.

In Fig. \ref{eps} and \ref{acc}, we observe that LHC\cite{Aaij:2017rft,Izaguirre:2015zva} and LEP\cite{Hook:2010tw} give constraints for large masses, while BaBar~\cite{Lees:2017lec,Izaguirre:2015zva} constraints exclude a sliver of DM masses near 1 GeV.   Likewise, LSND~\cite{Auerbach:2001wg,deNiverville:2011it}, NA64~\cite{NA64:2019imj}, and MiniBooNE~\cite{Izaguirre:2017bqb,Aguilar-Arevalo:2017mqx} constraints are relevant for small masses and large mass splittings.  However, the parameter ranges excluded are small.  On the other hand, future experiments LDMX and Belle II are expected to probe most of the sub-GeV region~\cite{Battaglieri:2017aum}.  We also note that at smaller values of $\alpha_D/m_{A^\prime}^2$ than our primary benchmark, accelerator constraints \emph{increase} in their sensitivity to thermal dark matter, since their sensitivity depends mainly on $\epsilon$.  Thus, for the alternate benchmarks considered in the Appendix (see Figs.~\ref{appFig:acceleratorEpsilon} and \ref{appFig:acceleratorCMB}), the small currently-constrained regions enlarge noticeably.

\section{Direct Detection} \label{directdetection}
\begin{figure}[!htb]
	\begin{center}
		\includegraphics[scale=0.8]{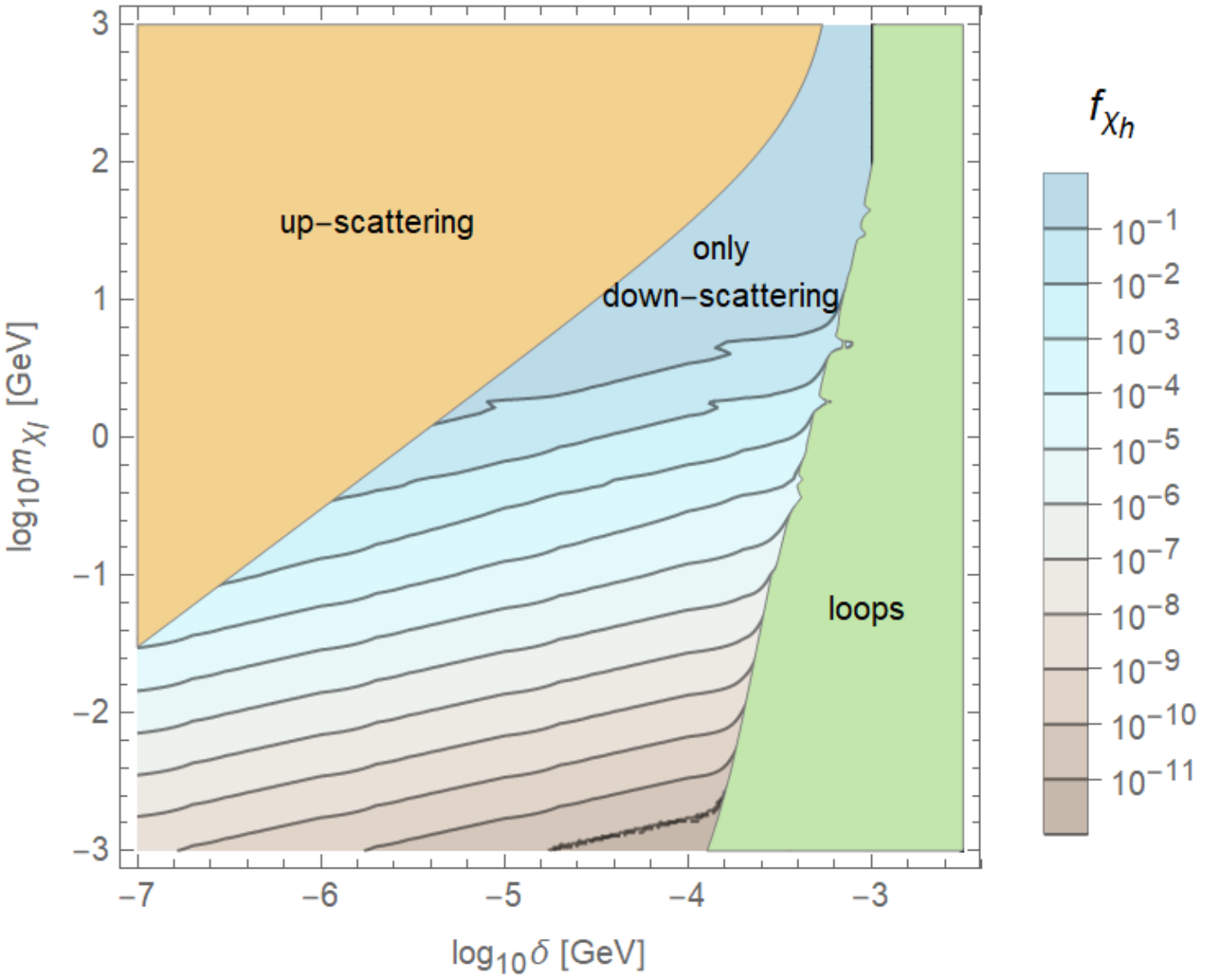}
	\end{center}
	\vspace{-0.3 cm}
	\caption{In the upper left region, up-scatterings are accessible (the kinematic boundary for heavy DM depends on the target nucleus, with improved reach for heavier nuclei; we have assumed A=207 (lead) to illustrate the largest plausible region). The cyan-brown region corresponds to the parameter space where down-scatterings are the only tree-level process that can occur. In this region, we show the contours of the relative abundance of the excited state observed today. Lastly, in the green region decays have heavily depleted the excited state and only loop-level elastic scatterings are expected when the Majorana masses obey $m_\eta = m_\xi$. }
	\label{dd}
\end{figure}

Light, inelastic DM presents a rich variety of signals for direct detection experiments.  The DM parameter space can be loosely subdivided into three regions.  Of these, two are relatively simple and familiar, while the third (the down-scattering region) presents a rich tapestry of new constraints and search opportunities.  After summarizing these three regions, we will discuss in more depth the distinctive phenomenology of the third.  These regions are as follows: 
\begin{itemize}
\item \textbf{Elastic Scattering Only (unconstrained):} At large enough splittings $\delta$, the $\chi_h$ population has decayed by the present era, and up-scattering of $\chi_l$ into $\chi_h$ is not kinematically accessible due to the finite escape velocity of Galactic DM in our local neighborhood.  Therefore, the only possible signals are elastic scattering of the $\chi_l$ DM component off nucleons or electrons.  In our minimal model with equal Majorana masses $m_\xi=m_\eta$, this process occurs only at 1-loop level, via loops involving exchange of two dark photons and an intermediate $\chi_h$ state.  This region is shaded green in Fig.~\ref{dd}.
The resulting cross-sections for both DM-electron and DM-nucleus elastic scattering have been computed in Appendix C of \cite{Berlin:2018jbm}, and are below the sensitivity of current experiments.  Therefore, this region is not presently constrained by direct detection, though parts of it are probed by the CMB and accelerator-based searches discussed in Secs.~\ref{sec:primordial} and \ref{exp}.  Even future SuperCDMS-SNOLAB \cite{CDMSlimitPlotter} and DARWIN \cite{2016arXiv160607001A} projected sensitivities will only probe the loop-level process only in narrow mass ranges (roughly 1--2 and 10--20 GeV); at higher and lower mass ranges, the expected signal is well below the expected neutrino backgrounds \cite{PhysRevD.90.083510,Essig:2018tss}  and will be quite challenging to explore.  Generalizing to unequal Majorana masses $m_\xi \neq m_\eta$ permits tree-level elastic scattering (see Fig.~\ref{Fig:FeynmanGraphs}-h),  but with a cross-section suppressed by $v_{\rm CM}^2 (\delta/m_\chi)^2$ relative to elastic scattering of Dirac DM.  These suppressions make the tree-level scattering process rarer than the loop-level process except for $\delta \gtrsim 10^{-3} m_{\chi}$, and generically undetectable with present experiments. We defer a detailed study of future experiments' sensitivities to these reactions to future work.  
\item \textbf{Up-Scattering/iDM (strongly constrained):} At high DM masses and low splitting $\delta$, the up-scattering/inelastic dark matter (iDM) reaction \cite{TuckerSmith:2001hy} is kinematically allowed.  This process, wherein dominant DM component $\chi_l$ have enough kinetic energy to scatter into $\chi_h$, has distinct kinematics from elastic scattering. Nonetheless, because the expected cross-sections are quite large (at or above the $10^{-39}$ cm$^{2}$ per nucleon scale throughout our mass range of interest, with larger cross-sections at lower masses), most of this region (and in particular, most of the region above CMB constraints) is excluded by many orders of magnitude based on the null results of elastic DM scattering searches.  The precise kinematic boundary for up-scattering depends on the DM escape velocity, which we have taken to be 553 $\km/\s$. At high DM masses, this boundary also depends somewhat on the mass of the target nucleus, with higher-mass target nuclei allowing up-scattering for larger $\delta$.  Therefore, to illustrate inclusively where up-scattering can occur, the orange shading in Fig.~\ref{dd} depicts parameter space where DM up-scattering is possible off lead --- a heavier nucleus than any used in direct detection.  All up-scattering exclusions from terrestrial experiments should be contained in this region, but the actual exclusion from a given experiment is smaller. For example, scattering off xenon overlaps the 4.9-40.9 keV$_{\rm nr}$ search band of the Xenon1T DM search \cite{Aprile:2018dbl} in the fair pink shaded region in Fig.~\ref{DDbounds}.  This entire region is excluded by the Xenon1T search, because the up-scattering cross-section in our model (when kinematically allowed) is much larger than the Xenon1T sensitivity.  It is likely that this exclusion could be extended to the bottom-left (filling out more of the upper left region of Fig.~\ref{dd}) by including lower-threshold searches such as those by CRESST \cite{Angloher:2015ewa,Angloher:2014myn,Angloher:2017zkf,Abdelhameed:2019hmk,Abdelhameed:2019mac} and SuperCDMS \cite{Agnese:2018gze,Agnese:2017jvy}, but these are of limited interest given the overlapping CMB constraints. 
\item \textbf{Mono-energetic Down-Scattering}: The final --- and most novel  and intricate --- channel for direct detection is tree-level down-scattering of the sub-dominant $\chi_h$ population into the lighter $\chi_l$.  While down-scattering of DM off nuclear targets was first proposed as a DM detection signal in \cite{Graham:2010ca}, most of the literature has focused on the regime of small splitting relative to the DM kinetic energy and assumed that the excited state is a dominant or $O(1)$ component of the DM population.  In contrast, we have seen already that the present abundance of $\chi_h$ is quite small in much of our parameter space --- this is illustrated by the density contours in Fig.~\ref{dd}.  Moreover, in most of this region the DM mass splitting $\delta$ can be much larger than the typical DM kinetic energy $\sim 10^{-6}\, m_\chi $. This hierarchy of scales leads to an approximately mono-energetic recoil signal that can be constrained very efficiently even in searches with residual background. 
\footnote{During completion of this work, several papers appeared that explored line-like signals in connection with the Xenon1T electron-recoil excess \cite{Aprile:2020tmw}.  These analyses generally recognized the line-like kinematics achievable via down-scattering, but most did not carefully consider the cosmological abundance of $\chi_h$.}.  
\end{itemize}
The large expected cross-section for DM down-scattering and the striking mono-energetic signal make this reaction a powerful search channel for light DM with moderate inelastic splittings, \emph{even in parameter regions where the $\chi_h$ abundance is quite small}.  We therefore focus on the discussion of the down-scattering signal in this section.  We begin by summarizing the  signal kinematics and yield in Sec.~\ref{ssec:yield}.  In Sec.~\ref{ssec:earthshadow} we discuss passage of $\chi_h$ through the Earth, noting that their expected down-scattering cross-section is sufficiently high to suppress the flux of up-going $\chi_h$.  This ``Earth shadowing'' effect modestly reduces the overall down-scattering signal, while also introducing a (sidereal) daily modulation of this scattering.  In Sec.~\ref{ssec:expConstraints} we derive constraints on line-like DM down-scattering off nuclei from detailed data published by CRESST-II and CRESST-III, and recast other experiments' searches for other line-like signals, such as dark-photon and axion DM, as down-scattering constraints.  We discuss the impact of these experiments' constraints on our parameter space.  In Sec.~\ref{ssec:SensitivityPlots}, we reframe the question, discussing the viability of our model as an explanation of the Xenon1T electron-recoil excess and introducing a parameter space through which current and future experimental results can be compared to the range of predicted signals in our model. We close with a word of caution in Sec.~\ref{ssec:fragility}: while powerful, the down-scattering signal is ``fragile'' in that any additional scattering or decay process that further depletes the $\chi_h$ population relative to our estimates would nullify or at least substantially weaken these constraints. Examples of such processes include the semi-elastic process discussed in Sec.~\ref{sssec:hlll} and dipole transitions from $\chi_h$ to $\chi_l$ (which are, however, subject to other constraints).  

\subsection{Down-Scattering Kinematics and Yield}\label{ssec:yield}
We summarize here the kinematics and yield for DM down-scattering off nuclear targets, following  \cite{Graham:2010ca,Essig:2010ye}, then extend these results to electron targets.  
The nuclear recoil energies in DM-nuclear down-scattering interactions are given by $E_R=\bar{E}_R+\Delta E_R \cos{\theta}$ where $\theta$ is the scattering angle in the CM frame and 
\begin{equation}
	\bar{E}_R=\frac{\mu_{\chi_{h} N}\delta}{m_N} \, \quad 
	\Delta E_R=\frac{ \mu_{\chi_{h} N}^2 v^2}{m_N}\sqrt{1 + \frac{2 \delta}{\mu_{\chi_{h} N} v^2}} \ . \label{erbar}
\end{equation}
Here, $\mu_{\chi_{h}N}$ is the reduced mass of the $\chi_h$-nucleus system and $m_N$ is the mass of the target nucleus $N$.  
The central value $\bar{E}_R$ corresponds to momentum-balanced sharing of the mass energy $\delta$ between the outgoing DM and nucleus in the zero-velocity limit; the width $\Delta E_R$ is parametrically suppressed by $\mu v^2/\delta$ which, in most of the region of interest, is $\ll 1$ (regions where this ratio is $>1$ are subject to stringent up-scattering constraints noted above).
This means that the nuclear recoils from down-scattering are almost mono-energetic and have a recoil energy given by $E_\text{recoil}=\frac{\mu_{\chi_{h}N}}{m_N}\delta $. This will generate a signal sharply peaked at $E_\text{recoil}$. 

The differential rate per unit target mass per unit time is
\begin{equation}
	\frac{dR}{dE_R}=N_T f_{\chi_h}\frac{\rho_\chi}{m_{\chi_h}} \int_{v_\text{min}}^\infty
	 \frac{d\sigma_{\chi_hN}}{dE_R} v \, f({\bf{v}}) d^3 {\bf{v}} \ ,\label{eq:nuckin}
\end{equation}
where $N_T$ is the number of nuclei per unit target mass, $f({\bf{v}})$ is the DM velocity distribution in the lab frame, $\sigma_{\chi_hN}$ is the cross section for $\chi_h$-$N$ scattering, and the minimum $\chi_h$ velocity compatible with recoil energy $E_{R}$ is
\begin{equation}
	v_\text{min}(E_R)\simeq\left| \sqrt{\frac{E_R m_N}{2 \mu _{N \chi _h}^2}}-\frac{\delta }{\sqrt{2 E_R m_N}}\right| \ .
\end{equation}
We assume a DM velocity distribution given by the Standard Halo Model with a escape velocity $v_\text{esc}=553$ km/s and velocity dispersion equal to the Sun's circular rotation velocity, $v_0=220 \km/\s$. When the DM kinetic energy dispersion is much smaller than $\delta$, i.e. $\sim v_0 \ll \sqrt{\delta/\mu_{\chi_{h}N}}$, 
the expected rate can be approximated as 
\begin{equation}
R_N\simeq \epsilon_{\rm det} f^\text{det.}_{\chi_h} \frac{\rho_\chi}{m_{\chi_h}} \frac{N_A}{A_N}\,\sigma_{\chi_hN} v \ , \label{RchiN}
\end{equation} 
where $N_A$ is the Avogadro number,  $A_N$ is the atomic mass of the target nucleus $N$, $\epsilon_{\rm det}$ accounts for detector efficiency, and
\begin{equation}
	\sigma_{\chi_hN} \ v=\frac{\sqrt{2} \ 16 \ \pi \alpha \alpha_D \epsilon^2 \mu_{\chi_hN}^{3/2}\delta^{1/2} Z_N^2}{m_{A'}^4}\, |F(E_R)|^2 \ , \label{eq:sigmachiN}
\end{equation}
 is the cross-section for $\chi$-nucleus scattering, 
with $Z_N$ the atomic number of the nucleus $N$, and $|F(E_R)|^2$ is the nuclear form factor. The form factor is $O(1)$ as long as the momentum transfer $q = \sqrt{2 m_\text{Nuc} E_R} = \sqrt{2 \mu_{\chi N} \delta}$  is small compared to the size of the nucleus --- as is the case in most of our parameter space.  However, for large enough $\delta$ and DM masses above $30$ GeV down-scattering begins to be affected by the structure of the recoiling nucleus.  We use a Helm form factor \cite{Helm:1956zz} with parameters defined in (28) of \cite{Bramante:2016rdh}. The factor of $f^\text{det.}_{\chi_h}$ accounts for the suppressed abundance of $\chi_h$ relative to the total DM density at the detector and is given by
\begin{equation}
f^\text{det.}_{\chi_h}=\epsilon_{\rm Earth} f_{\chi_h} \ .	
\end{equation}
Here, $\epsilon_{\rm Earth}$ accounts for the {\it Earth's Shadow} suppression discussed in the next section, which is approximately 1 for DM masses above a GeV, and never falls below $50\%$.   

Replacing $m_N \rightarrow m_e$ in \eqref{eq:nuckin}, one obtains the recoil energy distribution for DM down-scattering off electrons in the free-electron approximation.   These scatters typically have an even narrower energy distribution than the nuclear scatters, since $\mu_{\chi_h e} \ll \mu_{\chi_h N}$ for most DM masses we consider, centered on $\bar{E}_{R,e}=\frac{\mu_{\chi_{h} e}\delta}{m_e}$. The expected rate can again be written as 
\begin{equation}
	R_e\simeq \epsilon_{\rm det} f^\text{det.}_{\chi_h}\frac{\rho_\chi}{m_{\chi_h}}\ \frac{N_A}{A_N}\,Z_\text{exc.} \, \sigma_{\chi_he} \ v. \label{Refree}
\end{equation} 
Here $Z_\text{exc.}$ is the number of electrons associated with each atom that can be excited by this transition, discussed below. The cross-section for DM down-scattering off an electron is given by
\begin{equation}
\sigma_{\chi_he} \ v=\frac{\sqrt{2} \ 16 \ \pi \alpha \alpha_D \epsilon^2 \mu_{\chi_he}^{3/2}\delta^{1/2}}{m_{A'}^4}	\ .\label{eq:sigma_chie_v}
\end{equation}
which can be approximately related to the cross-section for DM annihilation into electrons as
\begin{equation}
\sigma_{\chi_{he}} \ v \approx \sqrt{2}\sigann_{ee}  \frac{m_e^{3/2}\delta^{1/2}}{\bar m_{\chi}^2} \, \label{eq:sigma_chie_v_approx}
\end{equation}
where we have used \eqref{cann}  and approximated $\langle m_{A'}^2- s \rangle \approx m_{A'}^2$ and $\mu_{\chi e} \approx m_e$.

It is well-known that the precise kinematics of DM-electron scattering depends on the electrons' initial-state wavefunctions (see e.g.~\cite{Essig:2017kqs} for the case of xenon targets).  This problem is simplified in our kinematic regime, where the recoil energy is dictated mainly by the mass difference $\delta$  and the relative momentum of the initial-state electron and DM particle merely spreads this signal in energy.  Roughly speaking, electrons with binding energy $E_b> \bar{E}_{R,e}$ cannot be excited. Electrons with binding energy $E_b < \bar{E}_{R,e}$ will recoil with central energy $\bar{E}_{R,e}-E_b$ (with some enhanced broadening for states with $E_b\sim \bar{E}_{R,e}$), but valence electrons' de-excitation to fill the resulting vacancy will emit additional photons, so that the total energy deposited in each case is approximately $\bar{E}_{R,e}-E_{b,v}$, where $E_{b,v}$ is the binding energy scale of valence electrons (on the scale of 10s of eV, well below our energy scales of interest).  Noting in addition that the broadening effect from sizable bound-state electron momenta only applies to a relatively small fraction of deeply bound energy levels, we simply estimate the signal by using the free-electron rate \eqref{Refree} and kinematics, accounting for the fact that only a subset of electrons in each atom --- those with binding energy less than $\bar{E}_{R,e}$ --- can be excited.  The number of such electrons is denoted by $Z_\text{exc.}$ above.  For example, in our energy range of interest $Z_\text{exc}$ for xenon ranges from 18 at recoil energies of 100 eV to 54 at energies above 33 keV.  We will also consider bounds from experiments using semiconductor targets, such as germanium. Because we are simply counting available electrons that can be excited by an energy deposition \emph{well above} the valence electrons' binding energies, a similar count of excitable electrons per atom based on the energy levels of isolated atoms should apply for semiconductors. In particular, given experimental resolutions, for our kinematics the effects of interatomic interactions in crystalline germanium should not appreciably affect these signals.  We note, however, that an extension of our analysis to smaller splittings $\delta$ of ${\cal O}(10 \mbox{ eV})$ would begin to encounter more significant corrections from bound-state physics and interatomic interactions.

\subsection{Down-Scattering in the Earth and Shadowing of Up-Going Particles}\label{ssec:earthshadow}
The down-scattering cross-section \eqref{eq:sigmachiN} can be quite large compared to those typically considered in direct detection.  For example, for 5 MeV dark matter with $\delta = 3$ keV, with $\alpha_D=0.5$, $m_A'=3 m_{\chi}$, and $\epsilon$ chosen to reproduce a thermal abundance, the down-scattering cross-section on iron is $\sigma_{\chi \text{Fe}}\approx 1.5 \,10^{-32} \rm{cm}^2$ at typical DM velocities of $10^{-3}\, c$.  This large cross-section is due not only to the sizable coupling required for thermal freeze-out, but also to the much larger phase-space for the outgoing particles compared to the incoming ones, $ |{\mathbf p}_{\rm out}|/|{\mathbf p}_{\rm in}| \sim \frac{\delta^{1/2}}{\mu_{\chi N}^{1/2} v}$, which at the benchmark point above enhances the cross-section by a factor of $\sim 25$. Crudely modeling iron as having a uniform mass density of $1.8\, \g/\cm^3$ throughout the volume of the Earth (i.e. 32\% of the Earth's average mass density, corresponding to uniformly distributing all the iron within the earth), and considering iron as the only target on which it can scatter, would give such DM a mean free path of $2\, 10^9 \cm = 0.5 R_{\Earth}$.  This rough estimate suggests that ``down-going'' DM particles $\chi_h$ can easily reach the detector, while ``up-going'' particles, which must pass through ${\cal O}(1)$ Earth radii before reaching the detector, have a significant chance of scattering off nuclei into the lighter $\chi_l$ state.  (We have focus on iron here for simplicity; the mean free path is roughly halved in a more complete but still artificially uniform model of the Earth's composition discussed below).

The outgoing $\chi_l$ produced from this down-scattering will typically be kinematically able to up-scatter again (its kinetic energy is sufficient to offset the recoil energy $\frac{\mu}{m_N} \delta$ carried by the recoiling nucleus).  However, the up-scattering cross-section is \emph{suppressed}, rather than enhanced, by the phase-space factor discussed above and in  \eqref{eq:sigmachiN}. Therefore, the particle is very unlikely to up-scatter within the size of the Earth.  As such, the Earth creates a ``shadow'' that depletes the flux of $\chi_h$'s, converting them into less detectable $\chi_l$'s.  

This suppression of the up-going $\chi_h$ component has two notable physical effects.  The first and most obvious is that it reduces the overall DM flux on the detector --- for mean free paths $\lesssim R_{\Earth}$ but much greater than the detector depth, the suppression is at most $50\%$.  We incorporate this through a suppression factor $\epsilon_{\rm Earth}$ in Eq.~\eqref{RchiN} and Eq.~\eqref{Refree}, which we estimate as described in the paragraphs below. The second effect is more interesting: the DM ``wind'' associated with the Sun's motion through the galaxy defines a preferred direction for DM impinging on the Earth.  As the Earth rotates about its axis, this preferred direction shifts between up- to down-going with a period of one sidereal day. The flux of $\chi_h$ impinging on the detector is highest when the DM wind is more down-going, and reduced  when it is more up-going, leading to a daily modulation of the DM flux.  If a clear line-like signal is observed in direct detection experiments, this modulation feature could allow discrimination between the down-scattering signal discussed here and other physics with monochromatic energy deposition, such as absorption of ultra-light DM (negligible modulation) or of ejecta from the Sun (negligible or Solar-day modulation, depending on particle properties).  

Averaged over a whole sidereal day, the Earth-shadowing effect never suppresses the DM flux below $\sim 50\%$ of a naive estimate, because its mean free path always exceeds the overburden through which down-going DM must penetrate.   Even this modest suppression is relevant to the interpretation of direct detection results, so we account for the Earth-shadowing with the following approximations. We model the Earth as a sphere of uniform density and uniform composition, namely mass fractions of 32.1\% Fe, 30.1\% O, 15.1\% Si, 13.9\% Mg, 2.9\% S, 1.8\% Ni, 1.5\% Ca, and 1.4\% Al (we neglect the contributions from trace elements, which together comprise the remaining 1.2\% of the Earth's mass density) \cite{enwiki:1025011893,Morgan6973}. Given these considerations, we assume the following simplified angular distribution on the detector
\begin{equation}
	f_\text{E.S.}(\theta,|\mathbf{v}_\text{lab}|)=\frac{1}{2}e^{-\frac{L(\theta)}{\lambda_{\chi_h}(|\mathbf{v}_\text{lab}|)}} \ ,
\end{equation}
where $L$ is the Earth-crossing distance and $\lambda_{\chi_h}$ is the mean free path of the DM:
\begin{align}
	&L=\left(R_{\oplus}-l_{D}\right) \cos \theta+\sqrt{ R_{\oplus}^{2}-\left(R_{\oplus}-l_{D}\right)^{2} \sin ^{2} \theta} \ , \\
	&\lambda_{\chi_h}^{-1}(|\mathbf{v}_\text{lab}|)=\sum_{i \in\text{ Earth}}\frac{\rho_E N_A M_{N_i}}{A_{N_i}}\sigma_{\chi_{h}N_i}(|\mathbf{v}_\text{lab}|) \ ,
\end{align}
with $R_{\oplus} \approx 6371 \mathrm{~km}$ the Earth's radius, $l_{D}$ the detector's depth, $\sigma_{\chi_h N}(v)=(\sigma_{\chi_h N} v)/v$ the DM-nucleus cross section with $\sigma_{\chi_h N} v$ given by Eq.\eqref{eq:sigmachiN}, and $M_{N_i}$ is the mass fraction of the element as quoted above. We consider the Earth's shadowing suppression in the direct detection rates by including a factor $\epsilon_{\rm Earth}$ given by
\begin{equation}
	\epsilon_{\rm Earth}=\int_{0}^{\pi}\ud\theta f_\text{E.S.}(\theta,|\mathbf{v}_\text{mean}|) \ , \quad \mathbf{v}_\text{mean}=\int \ud|\mathbf{v}_\text{lab}| |\mathbf{v}_\text{lab}| f(|\mathbf{v}_\text{lab}|) \ ,
\end{equation}
where $f(|\mathbf{v}_\text{lab}|)$ is the DM's velocity distribution on the detector given by the Standard Halo Model. We have checked that the approximations above are accurate within 5$\%$ with respect to a more complete velocity treatment which, instead of evaluating at the mean velocity, includes the integration over a time-independent velocity distribution at each zenith angle $\theta$; corrections from the time- and angle-dependence of the velocity distribution are expected to be smaller. 

\subsection{Down-scattering constraints} \label{ssec:expConstraints}
Down-scattering off both nuclear and electron targets constrains our inelastic DM model.   Nuclear recoils are most constraining for $\sim \GeV$ to multi-GeV DM, while electron scattering provides the greatest sensitivity to lower-mass DM parameter space. 

In nuclear recoils, we have focused on the constraints from CRESST-II \cite{Angloher:2015ewa,Angloher:2014myn,Angloher:2017zkf} and CRESST-III \cite{Abdelhameed:2019hmk,Abdelhameed:2019mac}.  These data sets are well-suited to re-analysis for several reasons: First, although line-like signals are not commonly searched for in nuclear recoils, the CRESST collaboration has published detailed energy spectra from which bounds on line-like signals can be extracted.  Second, these detectors have been optimized to have low energy thresholds, at the level of $30-300$ eV.  Third, the multi-elemental CaWO$_4$ crystal contains both low- and high-mass nuclei.  As a result of the mass-dependence in \ref{erbar}, DM down-scattering should produce 3 distinct ``spikes'' in the energy spectrum.  Constraining the absence of any one of these spikes leads to powerful constraints over a wide region of DM mass and splitting.  For DM below 10 GeV, the characteristic recoil energies $\bar E_{R} \approx \frac{m_{\chi_l}}{m_N}\,\delta$ for oxygen and tungsten nuclei differ by a factor of 11.5, so that the range of $\delta$ explored by this analysis at each DM mass is roughly 10x larger than the range of energies spanned by CRESST data.

For each detector and target nucleus, and for each value of $m$ and $\delta$, we first determine the central recoil energy $\bar E_{R}$ and the signal width $\sigma = \sigma_\text{Det}^2+\Delta E_R^2$.  For simplicity, we use a simple binned analysis, counting observed events within a bin around $\bar E_{R}$  of full-width of $2.8\sigma=2.8\sqrt{\sigma_\text{Det}^2+\Delta E_R^2}$. In the absence of a background model, we determine the maximum expected yield as the value for which Poisson fluctuations to the observed count or lower have probability less than $10\%$.   To convert this into a limit on model parameters, we use the expected rate in Eq.~\eqref{RchiN}, where the detection efficiency includes both the fraction of events expected to fall within our chosen energy bin and the cut-survival probability as given in \cite{Angloher:2015ewa,Angloher:2014myn,Angloher:2017zkf} for CRESST-II and \cite{Abdelhameed:2019hmk,Abdelhameed:2019mac} for CRESST-III.  

\begin{figure}[!hbt]
	\begin{center}
		\includegraphics[width=0.8\textwidth]{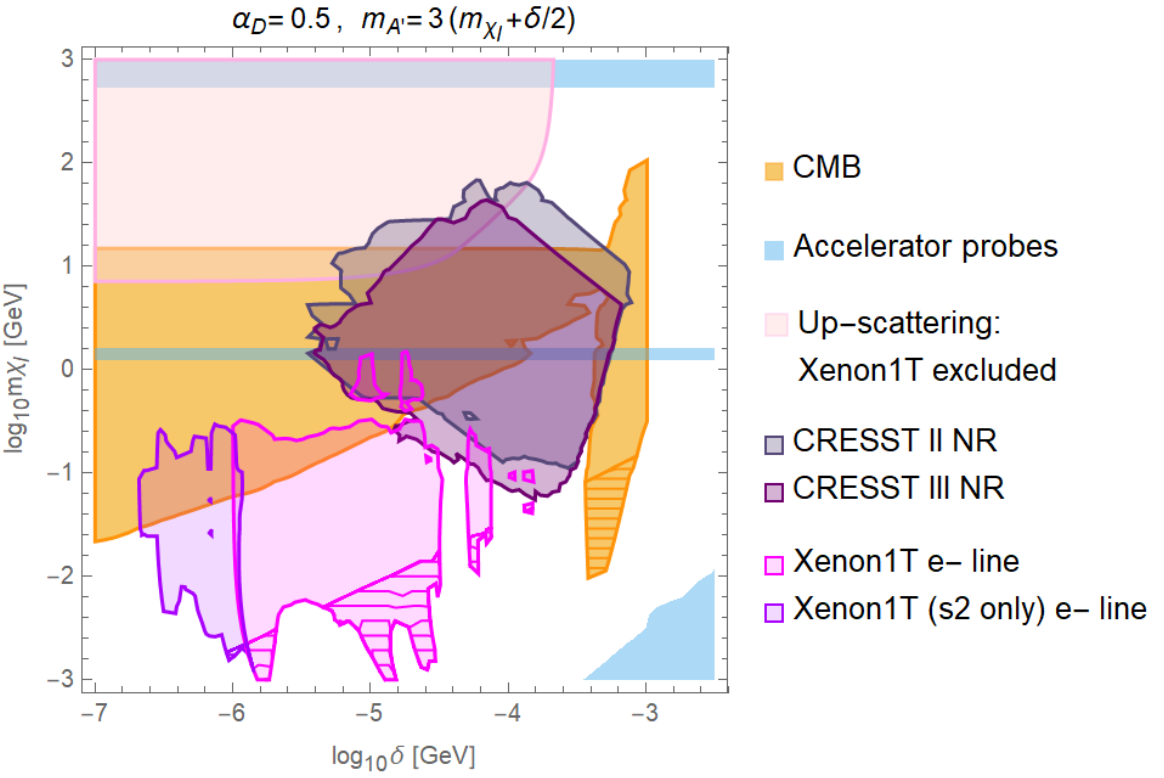}
	\end{center}
	\caption{Detailed exclusion regions for direct detection probes, with summary of CMB and accelerator-based constraints for $\alpha_D=0.5 $ and $m_{A'}=3 \,\left(m_{\chi_l}+\delta/2\right)$. In mustard and light blue, we observe the regions of parameter space ruled out by CMB observations and collider probes respectively. The fair pink region is ruled out by the absence of an up-scattering signal in Xenon1T search window from \cite{Aprile:2018dbl}. The dark purple diagonal regions correspond to the constraints due to the lack of nuclear recoil (NR) signals in CRESST II and CRESST III. Lastly, the sub-GeV regions in magenta shades are ruled out by the absence of the electron recoil signals in Xenon1T. Note that the striated regions for both the CMB and direct detection exclusions correspond to areas where semi-elastic interactions could further deplete the excited state abundance and weaken the constraints. Analogous plots for additional dark-sector benchmarks are shown in Figure \ref{AllBoundsDifBenchmarks}.
	\label{DDbounds}}
\end{figure}

The resulting constraints are shown in Fig. \ref{DDbounds}. We can see that nuclear recoil data from CRESST II and CRESST III is able to rule out a broad diagonal band in our parameter space that reaches DM masses of order $100$ MeV for large mass splittings, and is complementary to collider searches and cosmological constraints. The upper-right and lower-left boundaries of these regions correspond to recoil energies at the upper and lower bounds, respectively, of the CRESST analyses (at the lower bounds, near-threshold increases in background also limit sensitivity).  The lower-right boundary corresponds to efficient depletion of the excited state by decays.  The upper-left boundary is also rate-limited: we note that $f_{\chi_h} \frac{\rho_\chi}{m_{\chi}} \sigma_{\chi_hN} v\propto \delta^{1/2} / m_{\chi}^{5/2}$, which along contours of fixed $E_R$ scales like $1/m_{\chi}^2$.  

Our limit-setting procedure could be further optimized, for example by estimating the background using sidebands at higher and lower energy and subtracting this estimate and/or using it to pick a more optimal bin-width.  An optimized bin width would not affect sensitivity by more than $\sim 25\%$.  Subtraction could significantly improve sensitivity above the left corner of the exclusion region (where high event rates near threshold are limiting sensitivity), but this region is already well constrained by other searches.  Thus, the bounds from our simplified treatment are only modestly conservative, and are likely quite close to the bounds that a more complete analysis could achieve. 

For the bounds arising from DM-electron scattering, we recast dark-photon line-like signal searches as down-scattering constraints. We do so for Xenon1T~\cite{Aprile:2019xxb,Aprile:2020yad, Aprile:2020tmw}, SuperCDMS~\cite{Aralis:2019nfa}, and GERDA~\cite{GERDA:2020emj}. We will also look at projections for SuperCDMS (Ge)~\cite{Bloch:2016sjj} and Lux-Zeplin (LZ, Xe)~\cite{Akerib:2021qbs}. For each experiment, we infer a count-rate limit on monoenergetic signals at a given energy from the constraint on dark photons of a corresponding mass, using the rate formula
\begin{equation}
	R=4 \times 10^{23} \mathrm{keV} \cdot \frac{\epsilon^{2}}{E_{R}} \frac{\sigma_{\mathrm{pe}}}{A_N} \mathrm{~kg}^{-1} \mathrm{day}^{-1} \ ,
\end{equation}
where $\sigma_{\mathrm{pe}}$ is xenon's photoelectric cross-section at energy $E_R$ in barns. For the photoelectric cross-section, we use \cite{Henke:1993eda} for low-energy data, and \cite{NISTpe} for higher-energy data. The resulting count-rate limits can be reinterpreted as constraints on down-scattering via Eq.~\ref{Refree}, so long as the down-scattering signal is indeed line-like, i.e.~ the intrinsic signal width $\Delta E_R$ is less than the detector's energy resolution. We use the resolution models from \cite{Aprile:2020yad} (Xenon1T),  \cite{Agnese:2018gze,Agnese:2017jvy} (SuperCDMS), and \cite{Agostini:2019hzm} (GERDA) to verify that this criterion is satisfied at each parameter point.
 
As in the nuclear recoil case, the electron recoil bounds are weaker for heavier DM due to the smaller dark matter number density which leads to a lower rate. On the other hand,
down-scattering off electrons scales more favorably at low masses than the scattering off nuclei, since since $\sigma_{e \chi_h}\propto\mu_{e \chi_h}^{3/2} \sim m_e^{3/2}$ is approximately constant, while the corresponding cross-section for nuclei falls off as $m_{\chi}^{3/2}$ at low masses.  Therefore, even with the efficient depletion of the heavy DM state $\chi_h$ for light DM, electron scattering remains a viable detection channel. The electron recoil searches are especially relevant for sub-GeV masses as can be seen in Fig.~\ref{DDbounds}.

Two different Xenon1T searches for line-like electron signals are shown in Figure \ref{DDbounds}: the most sensitive search \cite{Aprile:2020tmw} uses a combination of scintillation (s1) and ionization (s2) signals.  A previous search \cite{Aprile:2019xxb} using only the ionization (s2) signal achieved a lower recoil energy threshold (0.186 $\keV$) and therefore offers the best sensitivity to lower $\delta$.  The  SuperCDMS and GERDA line searches do not give rise to any bounds for our main benchmark choice of $\alpha_D$ and $m_{A'}/m_{\chi}$, but do have some sensitivity for other choices of these parameters as illustrated in Fig.~\ref{AllBoundsDifBenchmarks}.

\subsection{Exploring DM Down-Scattering with Current and Future Experiments}\label{ssec:SensitivityPlots}
As we have seen in the previous section, direct detection experiments searching for both electron and nuclear recoil signals are able to explore a region of parameter space complementary to the one which is constrained by CMB data and current accelerator probes. Here, we will analyze in more detail the possible reach of current and future experiments.

\begin{figure}[!htb]
	\begin{center}
		\includegraphics[scale=0.55]{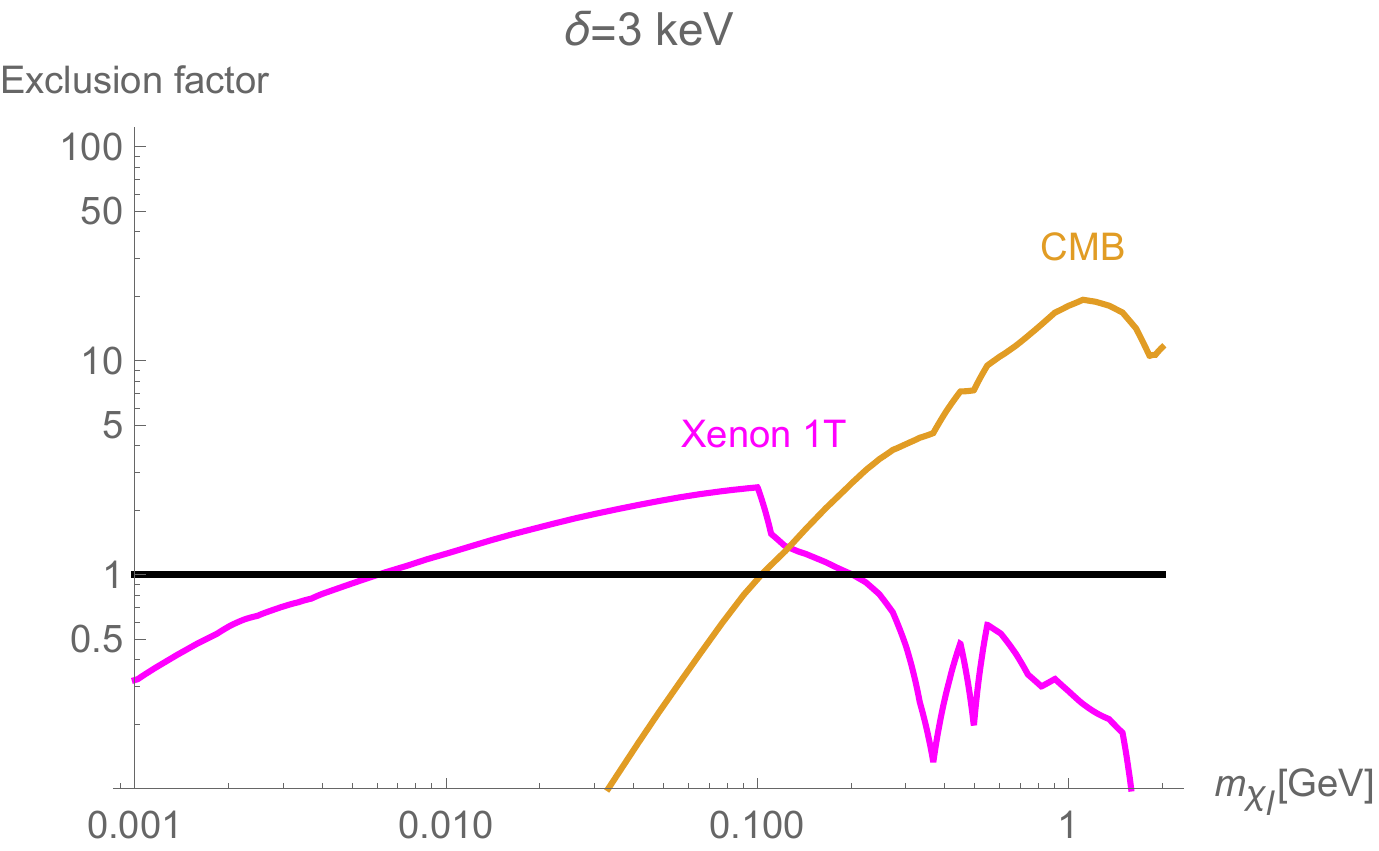}\includegraphics[scale=0.55]{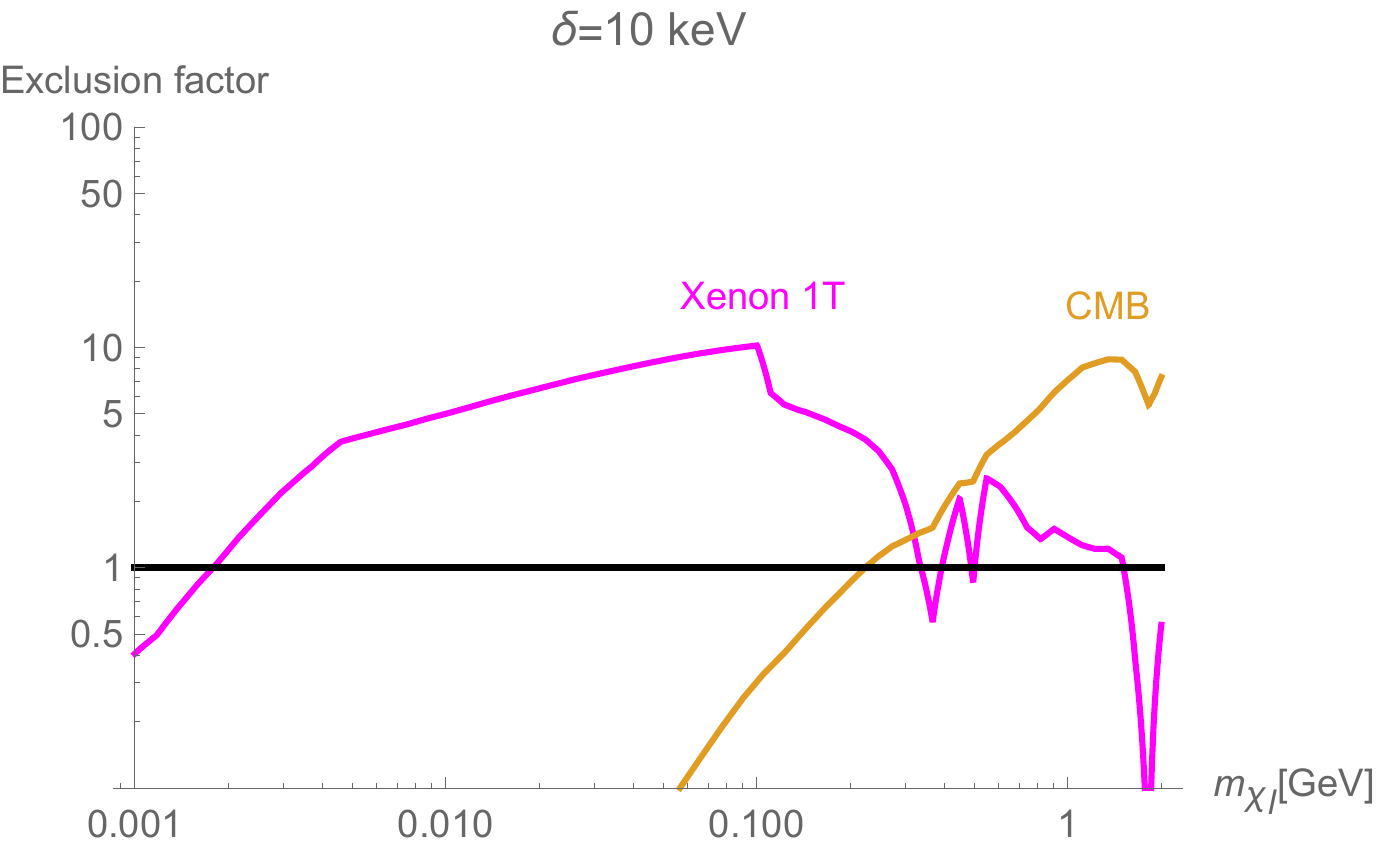}
	\end{center}
	\vspace{-.5cm}
	\caption{Exclusion factors for regions near a mass splitting of order keV. The regions where the Xenon 1T and CMB lines are above 1 are excluded. The exclusion factor for Xenon 1T is given by the expected counts, Eq.~\eqref{Refree}, over the maximum counts that could have occurred in the detector, while the CMB one is given by the inverse of Eq.~\eqref{md}. Note that the exclusion factor is smaller than 10 for a large range of DM masses.}
	\label{fig:exclusionFactors}
\end{figure}

A noteworthy analysis from current experiments is the reported excess of electron recoils peaked at 2-3 keV in the data from the Xenon1T detector \cite{Aprile:2020tmw}. Such a recoil signal could arise from the DM excited state down-scattering off electrons, with a mass splitting is of order keV. We can observe in Fig.~\ref{DDbounds} that the combination of CMB observations and up-scattering in direct detection rule out keV-split thermal dark matter with mass above $\sim 50$ MeV  (with tighter constraints away from our benchmark of $\alpha_D=0.5$ and $m_{A'}=3 m_{\chi}$). In addition, the down-scattering signal expected from $\sim 10-50$ MeV thermal DM is excluded by Xenon1T's observed limit on line-like signals, notwithstanding the excess. Nevertheless, it is important to note that a large part of these regions are excluded only by a factor of order one, see Fig.~\ref{fig:exclusionFactors}. This means that small changes in the cosmology could render these parameter regions allowed. Similarly, semi-elastic interactions (potentially relevant in the striated region of Fig.~\ref{DDbounds}) could further deplete the excited state abundance, thus relaxing the constraints. With these caveats in mind, the Xenon1T excess could be compatible with a down-scattering signal for thermal DM models in the $\sim 1-100$ MeV mass range with splitting $\delta\approx 3 \,\keV$.

\begin{figure}[!hbt]
	\begin{center}
		\includegraphics[width=0.8\textwidth]{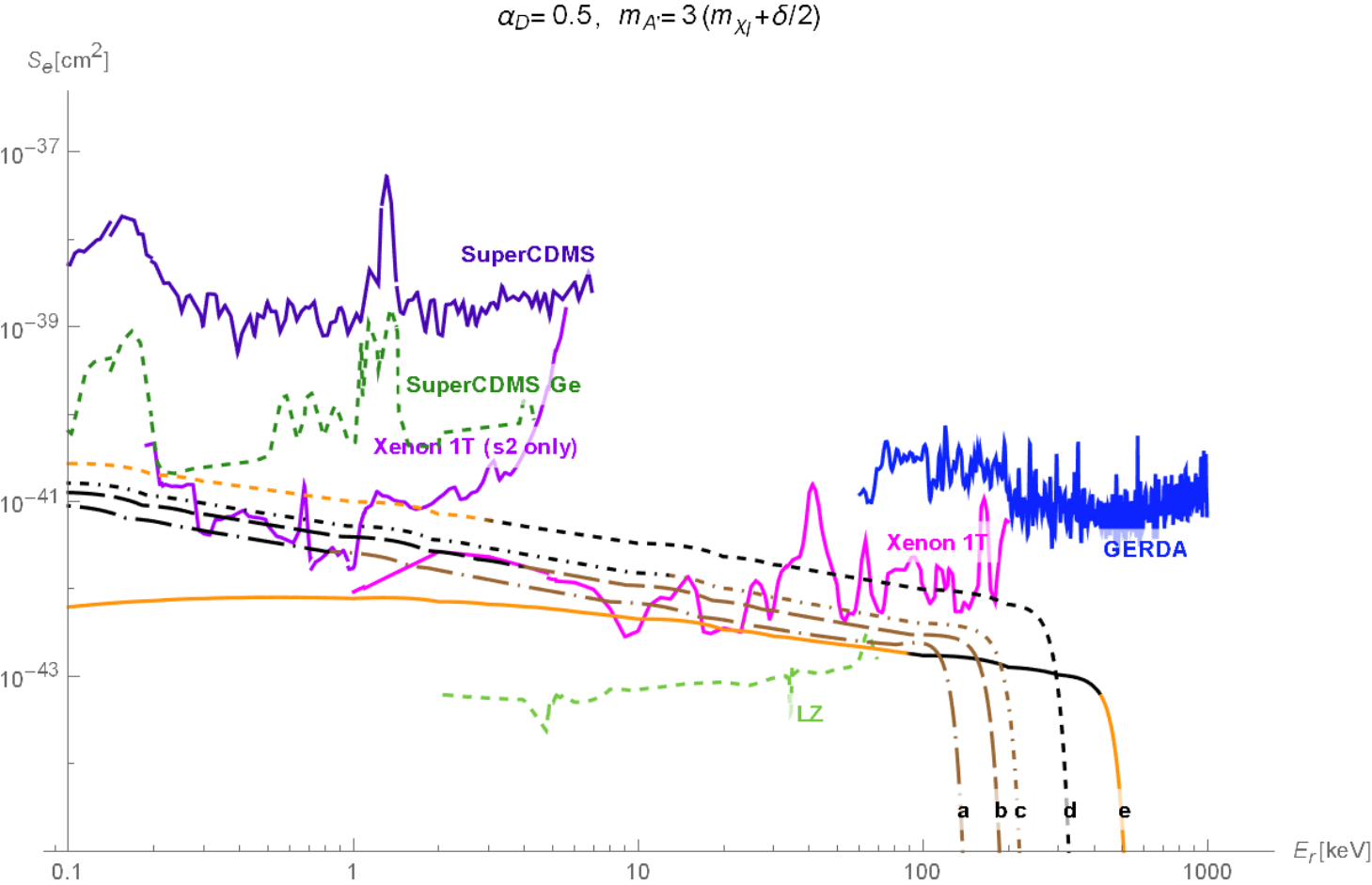}
	\end{center}
	\caption{Sensitivity plot for electron recoil searches. The black/brown/orange lines corresponds to the excited DM fraction at the detector times the DM-electron cross section per GeV DM mass for a thermal target. The lines labeled a, b, c, d, e correspond to $m_{\chi_l}=0.002 \  \text{GeV},  \ 0.005\ \text{GeV}, \ 0.01\ \text{GeV},\  0.1 \ \text{GeV}, \ \text{and} \ 1 \ \text{GeV}$ respectively. The thermal target is excluded by the CMB when the line is orange. When it turns brown the semi-elastic process of the Subsection \ref{sssec:hlll} becomes relevant. The solid color lines show the sensitivity limits for Xenon1T, Xenon1T s2-only, SuperCDMS, and GERDA, obtained by recasting these experiments' bounds on another line-like signal (dark-photon DM), as described in the previous section. The dashed lines show the projections for SuperCDMS with a germanium target and Lux-Zeplin, obtained by the same method from experimental projections. Analogous plots for additional dark-sector benchmarks are shown in Figure \ref{appfig:e-sensitivity}.
	\label{fig:e-sensitivity}}
\end{figure}

To understand the reach of current and future experiments, it is helpful to introduce a parameter space more directly connected to the experimental observables, where our expectations for thermal DM signals at different DM masses can be represented as contours and viewed as sensitivity targets.  We do this starting with electron-recoil searches, where the procedure is simplest and most general.
The central energy for any given model, $\bar E_{R,e} = \mu_{\chi e} \delta/m_e,$ is independent of detector material.   
We can factorize the overall signal yield \eqref{Refree} into a signal strength $S_e$ that depends only on the dark matter model and cosmological history, but not on detector materials, times an exposure factor that depends on the detector material and properties but not on the DM particle properties, except for $\bar E_{R,e}$:

\begin{equation}
S_e = \left(\frac{1\,\GeV}{m_\chi}\right) f_{\chi_h}^{det} \frac{\langle \sigma_{\chi e} v\rangle}{v_\text{ref.}}, \qquad  R_e = \epsilon_{\rm det} \frac{\rho_\chi}{1\,\GeV}\ \frac{N_A}{A_N}\,Z_\text{exc.} \, v S_e, \label{eq:SeDef}
\end{equation}
where $v$ is the average velocity of halo DM particles in the Earth frame (modeling of which does not affect $S_e$ because $\sigma_{\chi e}$ scales as $1/v$), and we have introduced a reference DM mass of $1\,\GeV$ and reference velocity $v_\text{ref.}=v_0=220 \km/\s$ to give $S_e$ units of cross-section. It is therefore straightforward to recast experimental results and projections for mono-energetic line signals in the plane of $S_e$ vs.~$\bar E_{R,e}$, which we have done in Fig.~\ref{fig:e-sensitivity} for the experiments considered above as well as projections for SuperCDMS and LZ, which cover new ground at sub-$\keV$ recoil energies and in the $2-80\,\keV$ range, respectively.   
We have also indicated the expected signals from down-scattering of thermal inelastic DM to indicate the sensitivity ranges that are particularly motivated by this model.  We note that, for DM masses between 2 MeV and 1 GeV, these are clustered within two decades of $S_e$ sensitivity but span a broader range of recoil energies than is explored by any one experiment. 
The theoretical targets are color-coded to indicate additional relevant, model-dependent features of these parameter points, as follows: Regions colored in orange are excluded by CMB energy injection constraints discussed in Sec.~\ref{sec:primordial} (though in some regions only by a narrow margin, as illustrated in Fig.~\ref{fig:exclusionFactors}).  Regions colored in brown are subject to additional depletion by semi-elastic scattering in models with unequal Majorana masses $m_\eta,\xi$ (see Sec.~\ref{sssec:hlll}), so that in these generalized models, signal strengths below the theoretical curve are also consistent with the model.  Regions where neither of the above apply are colored in black. The theory curves shown in Fig.~\ref{fig:e-sensitivity} all assume our primary benchmark choice $\alpha_D =0.5$, $m_{A^\prime} = 3 m_\chi$, to avoid crowding the plot.  Analogous curves for other benchmarks are shown in Fig.~\ref{appfig:e-sensitivity}.

\begin{figure}[!ht]
	\begin{center}
		\includegraphics[width=0.8\textwidth]{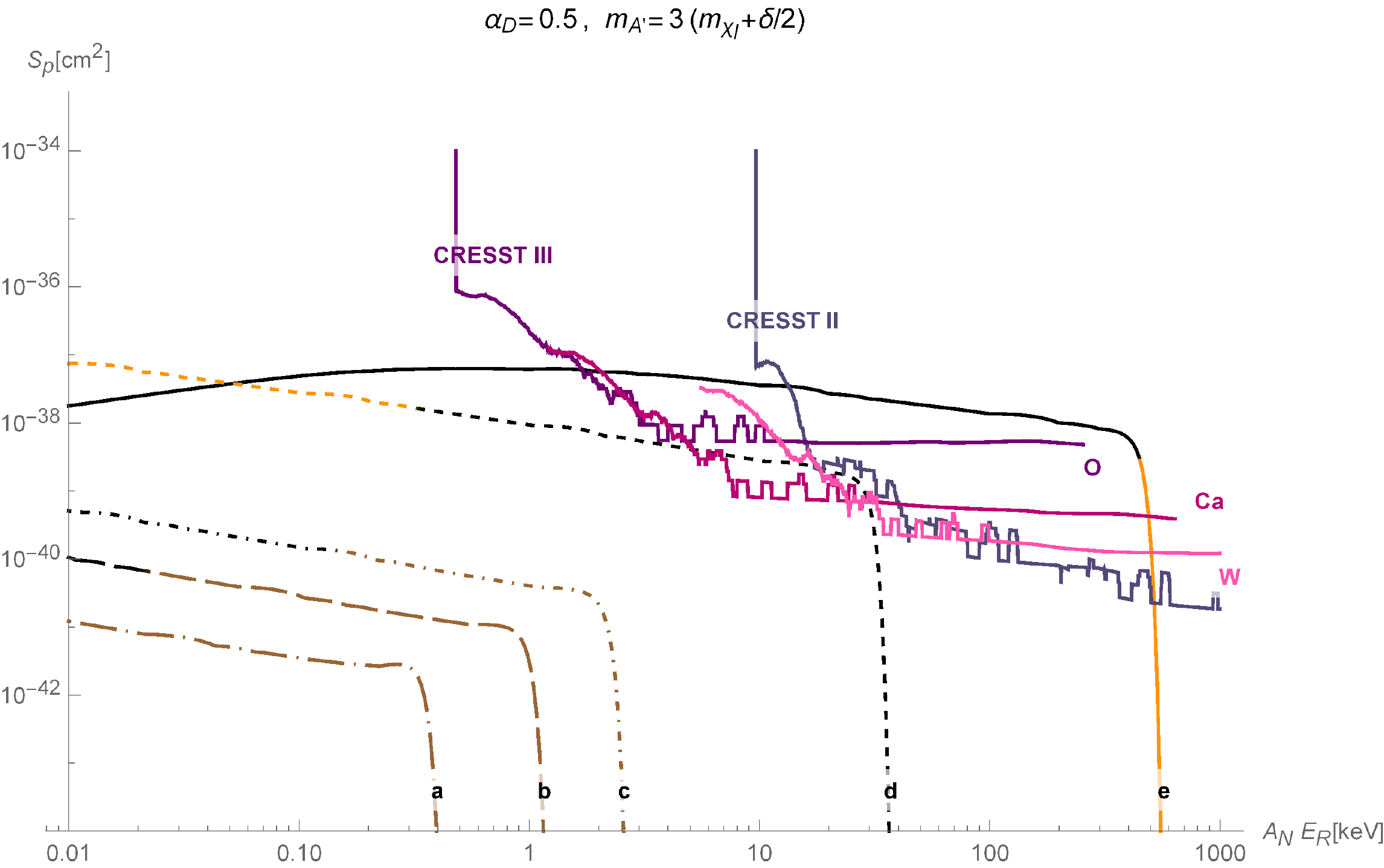}
	\end{center}
	\caption{Sensitivity plot for nuclear recoil searches. The black/brown/orange lines corresponds to the excited DM fraction at the detector times the DM-nucleon cross section over GeV DM for a thermal target. The color scheme and labels of these lines matches that of Fig.~\ref{fig:e-sensitivity}. Meanwhile, the solid purple-shaded lines show the sensitivity limits extracted from our own analysis of published datasets from CRESST II and CRESST III, as described in the previous section. For CRESST III, we have separately plotted the sensitivities to scattering off each nucleus, while for CRESST II, to simplify the figure, we show the strongest of the three sensitivities at each $x$-axis value. Analogous plots for additional dark-sector benchmarks are shown in Figure \ref{appFig:nuc-sensitivity}.}
	\label{sensitivityN}
\end{figure}

In the case of DM down-scattering off nuclei, the recoil energy depends non-trivially on both the DM mass and the target nucleus' mass through the reduced mass $\mu_{\chi N}$.  This prevents a fully general  analysis of the type presented for down-scattering off electrons.  However, a similar approach can be taken if we restrict to the case of DM much lighter than the nucleus where $\mu_{\chi N} \approx m_\chi$.  
In this case,  $\bar E_R \approx m_{\chi} \delta/m_N$, so that $A_N \bar E_R = m_\chi \delta/m_p$ is approximately material-independent\footnote{We note that, because $m_\chi$ is approximating $\mu_{\chi N}$ in the above, it is possible that a line-like DM signal would show up in distinct but nearby bins in this plot for experiments using different materials.  The difference in energies would then offer some indication of the DM mass. For now, in the absence of compelling line-like nuclear-recoil signals, the approximate mapping seems sufficient for comparing different experiments' exclusion sensitivities.}.  
Similar to the case of electron recoils, we can also factorize the signal yield $R_N$ from \eqref{RchiN} into a material-independent signal strength per proton, $S_p$, and a model-independent exposure as
\begin{equation}
S_p =  \left(\frac{1\,\GeV}{m_\chi}\right) f_{\chi_h}^{det} \frac{\langle\sigma_{\chi p} v \rangle}{v_\text{ref.}}, \qquad 
R_N\simeq \epsilon_{\rm det} \frac{\rho_\chi}{1\,\GeV} \frac{N_A}{A_N}\,S_p  Z_N^2 |F(E_R)|^2 v, 
\end{equation}
where our reference velocity is taken to be $v_\text{ref.}=v_0=220 \km/\s$ as before.  In Fig.~\ref{sensitivityN}, we show the sensitivities for CRESST II and CRESST III, as extracted from our analysis of their published datasets, together with the thermal target line for different DM masses. For each detector, we obtain a separate sensitivity curve for each element in the CaWO$_4$ detector.  We have shown the three curves separately for CRESST III to illustrate their complementarity, while for CRESST II we plot the strongest of the three sensitivities at each $x$-axis value to simplify the figure. The curves indicating model expectations are as in Fig.~\ref{fig:e-sensitivity}.  Both the range of signal energies and the range of signal-strength sensitivities required to explore the inelastic models are broader than in the electron-scattering case, due to the hierarchy $m_e < m_\chi < m_N$.   Analogous plots for different dark-sector benchmarks can be found in Fig.~\ref{appFig:nuc-sensitivity}.

\subsection{Fragility of Down-Scattering Signals} \label{ssec:fragility}
The down-scattering signal discussed in this section is a novel approach to detection of inelastic DM models and offers a distinctive signal over a broad parameter space. It bears emphasis, however, that this signal is somewhat fragile: \emph{minor perturbations of the model, such as by irrelevant operators, can lead to further depletion of $\chi_h$ and eliminate this signal or exponentially diminish its rate}.  

The example of semi-elastic scattering $\chi_h \chi_l \rightarrow \chi_l \chi_l$, which occurs when $m_\eta - m_\xi  \neq 0$, has been discussed in Sec.~\ref{sssec:hlll}.  The region where this process would be important for $m_\eta - m_\xi \sim \delta$ is indicated by hatched regions in Figure \ref{DDbounds}.  Because the rate per $\chi_h$ of this scattering reaction is not Boltzmann-suppressed, this reaction can exponentially deplete the $\chi_h$ abundance below the abundance expected from the inelastic process $\chi_h \chi_h \rightarrow \chi_l \chi_l$ alone.   This would, correspondingly, exponentially suppress the direct detection signals (as well as the CMB bound on $\chi_h$ decay).

Alternately, the $\chi_h$ abundance may be suppressed by any decay mode more rapid than the $\chi_h \rightarrow \chi_l + 3\gamma$  that follows from kinetic mixing.   Two examples that could lead to such decays are an electromagnetic dipole coupling or dark-photon mixing with other bosons that open up decays into neutrinos $\chi_h \rightarrow \chi_l \nu \bar\nu$.  
In the case of electromagnetic dipole interactions, ${\cal L} \supset \tfrac{1}{\Lambda} F_{\mu\nu} \chi^h \sigma^{\mu\nu} \chi^l$, leads to a $\chi_h$ lifetime $\pi\Lambda^2/(4\delta^3)$  for decays $\chi_h \rightarrow \chi_l + \gamma$ \cite{Chang:2010en}.  With $\Lambda \lesssim 10^{10}$ GeV, the dipole decays will occur prior to recombination, and as such eliminate both CMB and direct detection constraints.  These scales correspond to much smaller transition dipole moments than are needed for the dipole coupling itself to be observable in direct detection (which have weak- to TeV-scale suppressions) \cite{Chang:2010en,Feldstein:2010su}.For  $10^{10}\,\GeV \lesssim \Lambda \lesssim 10^{12}\,\GeV$, the $\chi_h$ abundance is depopulated within the age of the Universe but CMB constraints on energy injection (as in Sec.~\ref{ssec:primordial-decays}) may constrain these models depending on the precise lifetime, $\chi_h$ depletion by scattering, and mass parameters.  

The $\chi_h$ population can similarly be depleted by a small coupling of the $A^\prime$ to neutrinos, arising for example from a $B-L$ coupling or $A^\prime-Z$ mass mixing \cite{Davoudiasl:2012ag}.  An effective neutrino coupling at the level of $\sim  10^{-9} \left(\frac{m_\chi}{\MeV}\right)^2 \left(\frac{\delta}{\keV}\right)^{5/2}$ leads to a $\chi_h\rightarrow \chi_l\, \nu\, \bar\nu $ decay within the age of the Universe, for our benchmark couplings $\alpha_D=0.5, m_{A^\prime} = 3 m_{\chi_l}$.  For example, for $\delta=3\ \keV$ and $m_{\chi_l} = 50\ \MeV$ this would require neutrino couplings about 3 orders 
of magnitude weaker than the electron-coupling motivated by thermal freeze-out.  

Considering the above, the down-scattering constraints discussed in this section should be interpreted cautiously, and are better viewed as reflecting the discovery potential of direct detection experiments rather than robust exclusions of model parameter-space.   It would be interesting to consider other complementary signals of the various decay reactions mentioned above, or direct detection signals that do not rely on a primordial $\chi_h$ abundance, as means to more conclusively explore this parameter space through direct detection.

\section{Conclusions} \label{concl}
\begin{figure}[!bh]
	\begin{center}
		\includegraphics[scale=0.8]{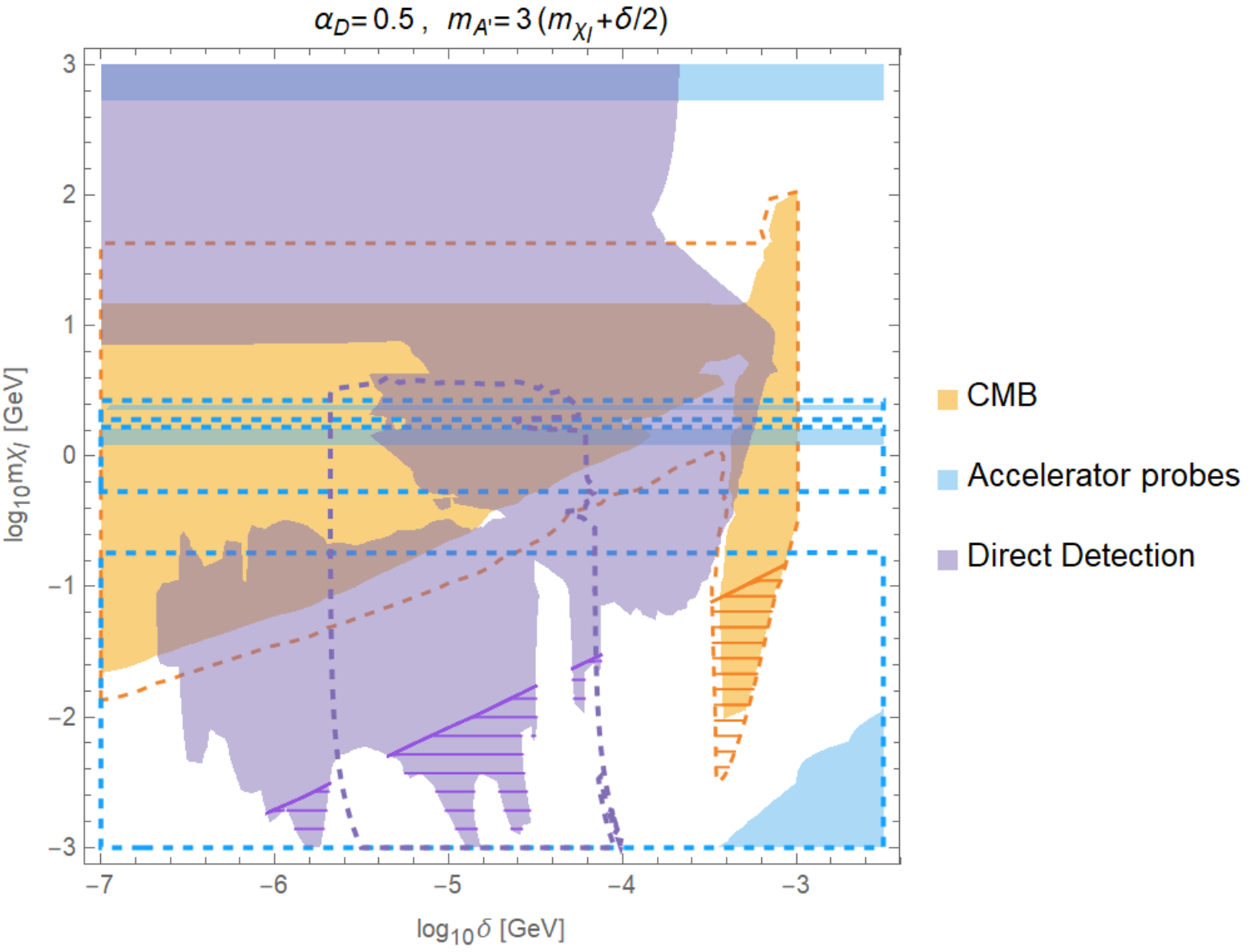}
	\end{center}
	\caption{CMB (mustard), direct detection (purple), and fixed-target (light blue) constraints for $\alpha_D=0.5 $ and $m_{A'}=3 \,\left(m_{\chi_l}+\delta/2\right)$. The projections for each probe are shown with dashed contours of the same color as the excluded regions. In the striated regions, semi-elastic interactions could further deplete the excited state and weaken the exclusions. Details on the CMB, accelerator probes, and direct detection bounds can be found in Figs.~\ref{CMBbounds}, \ref{acc}, and \ref{DDbounds} respectively. The direct detection projections correspond to SuperCDMS and LZ, but for this choice of $\alpha_D$ and $m_{A'}$ only LZ is relevant.Analogous plots for additional dark-sector benchmarks are shown in Figure \ref{AllBoundsDifBenchmarks}.} 
	\label{projections}
\end{figure}

In the search for understanding the physics of the dark sector, it is important to explore the full parameter space for thermal dark matter. 
In this context, pseudo-Dirac or inelastic DM is often discussed as a light DM candidate that is generically allowed by CMB constraints on recombination-era energy injection, as well as being difficult to see in direct detection. This is trivially true in the regime where the heavier DM state decays rapidly into the light state, as is generically the case if their mass splitting $\delta>2m_e$ (or through other non-minimal operators, as discussed in Sec.~\ref{ssec:fragility}).  However, in the parameter regions where the heavier state of pseudo-Dirac DM is long-lived compared to the recombination timescale, the parameter space has not yet been explored quantitatively.   This paper has explored the viability of long-lived pseudo-elastic DM quantitatively and explored several sources of constraints as well as avenues for future detection, using CMB energy injection limits, accelerator-based experiments, and novel direct detection signals.  The current reaches and future projections of each type of experiment are summarized in Fig.~\ref{projections} and described qualitatively below. 

To this end, we have considered the various DM scattering and decay modes that can depopulate the excited state, enabling a quantitative prediction of the DM annihilation rate at recombination and therefore a determination of what parameter space is allowed/disallowed by the CMB energy injection bound on DM annihilation.  This confirms the lore described in the previous paragraph: although the DM annihilation is $s$-wave, the depletion of the excited state can be dramatic enough to decrease the energy injection well below current sensitivity.  For example, for 10 MeV-mass dark matter, the smallest splittings we consider ($\delta=100 eV$) lead to a million-fold suppression of the excited state abundance, resulting in CMB energy injection an order of magnitude below the CMB bound and even beyond the potential sensitivity of a cosmic-variance-limited measurement.  In general, heavier DM requires a larger splitting to be compatible with CMB energy injection limits, but splittings well below $m_e$ are allowed for all sub-GeV DM masses.  
We have also identified the constraints on the model from decays of the DM excited state during recombination, which are a significant constraint on our minimal model for splittings $300 keV\lesssim \delta < 2 m_e$.  These constraints will improve only modestly with future data, as much of the parameter space leads to energy injection beyond the cosmic-variance-limited sensitivity of an ideal CMB measurement. 

We have also discussed accelerator-based constraints. In the benchmark parameter space we consider, they are not yet covering much ground.  This is equivalent to the statement that these experiments have not yet probed the conservative ``thermal target'', but next generation experiments are expected to do so (current experiments are exploring thermal parameter space with weaker dark-sector couplings and/or larger mass ratios than our benchmark, as shown in the Appendix). If these models of DM are correct, it is expected that either LDMX or Belle II should discover hints of invisible new physics at low masses and weak couplings.  An important question then will be the cosmological longevity of the particles being detected in these experiments.  In addition, further experimental avenues that can elucidate different aspects of the DM mass spectrum  will be of great interest. 
 
The canonical direct detection signal of elastic scattering of the light state, is highly suppressed for pseudo-Dirac dark matter, and therefore unobservably small, as well as having characteristic recoil energies below the thresholds of large WIMP detectors. In contrast, we found DM \textbf{down-scattering off electrons or nuclei} to be a surprisingly powerful probe of the low-mass, small-splitting parameter space --- even in parameter regions where the scattering state $\chi_h$ is depleted by many orders of magnitude.  This is possible because the decay energy $\delta$ from a down-scattering reaction, being much larger than the DM kinetic energy, leads to a line-like signal at higher energies than elastic scatters and with parametrically enhanced scattering cross-section.  This enhancement and the large target masses of WIMP detectors can effectively compensate for the small $\chi_h$ abundance.  Searches for other line-like signals, such as absorption of axion or dark-photon DM, can be recast as down-scattering searches.   We found relevant and complementary constraints on both kinds DM-nuclear and DM-electron down-scattering  reactions from current experiments, with CRESST II and III nuclear recoils constraining larger DM masses and splittings, and Xenon1T electron recoils providing constraints on the lower-mass/splitting parameter range.  Intriguingly, mass splittings around 3 keV for light ($\lesssim 10 \,\MeV$) DM are a viable explanation of the Xenon1T electron recoil excess.   Assessing the viability of this scenario requires accounting for effects such as ``Earth shadowing'' --- the screening of up-going $\chi_h$ flux by the Earth --- which suppresses the signal rate by up to 50\% in a parameter range where the expected signal is within a factor of 2 of the size of the Xenon1T electron excess. 

To facilitate future experimental analyses of these line-like direct detection signals, and their comparisons across experiments, we have introduced a signal-strength vs. recoil energy parameter space and plotted the theoretical curves of interest on this space.   We note that both nuclear and electron recoils are of considerable interest.  In nuclear recoils, low thresholds are paramount and light nuclei are beneficial, to explore parameter space at recoil energies of $100 \eV$/$A$ in nuclei of atomic mass $A$.  Low-threshold experiments can probe new parameter space even with relatively low exposure, as illustrated by CRESST-III.  In electron scattering, LZ's expected sensitivity to this process will enable robust sensitivity to the putative 3 keV Xenon1T signal as well as broad sensitivity to similar models with $\delta$ within a few to hundreds of keV.  Fully exploring these models motivates efforts to extend experimental sensitivity to line-like electromagnetic energy deposition to both lower and higher electron-recoil energies. 

It should be noted that all of the non-accelerator-based constraints discussed above and shown in Fig.~\ref{projections} presume that the DM is stable on cosmological timescales (to recombination and/or to the present day), with the leading decay mode being the one induced by kinetic mixing.  We note that additional depletion mechanisms are readily accommodated in this model with minor modifications ---several examples are discussed in Sec.~\ref{ssec:fragility} --- so that accelerator production remains the only \emph{robust} probe fo the small-splitting parameter space.   In addition, in many regions the exclusions are only by ${\cal O}(1)$ factors and so modest changes to the cosmological history could alter the exclusion boundaries.

Our findings point to several directions for future study:  
A clear direction for further analysis is the direct detection signals.  More careful analyses of the experimental data would yield some improvements in sensitivity, and additional detectors may be capable of line-like searches in new parameter space.   In addition, the signal-strength predictions could be improved by a more exact treatment of the freeze-out than our instantaneous-freeze-out approximation.  Moreover, our quantitative treatment of Earth shadowing (the depletion of up-going $\chi_h$ flux at the detector due to down-scatters in the Earth) was limited  to calculating the time-averaged reduction of signal. However, this reduction should be strongly correlated with the (mis)alignment of the ``up-going'' direction with the velocity of the DM wind, giving rise to a signal modulation with period of a sidereal day.  This feature could be used to distinguish a line due to DM down-scattering from other sources, including halo DM and Solar ejecta, which would have negligible sidereal-day modulation.  It may also be possible to improve search sensitivity by taking advantage of the signal's expected modulation to reject background.  These considerations motivate a more thorough  theoretical analysis of Earth-shadowing.  

In the minimal model, our findings favor low DM masses, with upper bounds from $\sim 5$ to 100 MeV over a range of mass splittings from 100 eV to 1 MeV (constraints are essentially eliminated when the splitting $\delta >2 m_e$).  The lower part of this mass range has been argued to be excluded based on measurements of $N_\text{eff}$, the effective number of relativistic species  \cite{Serpico:2004nm,Ho:2012ug,Steigman:2013yua,Steigman:2014uqa,Boehm:2013jpa,Nollett:2013pwa}. Nevertheless, given that the Hubble tension can be partly ameliorated by increasing $N_\text{eff}$, it may be reasonable to consider somewhat larger departures in $N_\text{eff}$ from Standard Model expectations. Simple solutions to the Hubble tension given by having additional light degrees of freedom are excluded since they change the ratio of the sound horizon to the diffusion length which is consistent with $\Lambda$CDM. More involved models with large $N_\text{eff}$ could be consistent with CMB data, but the phenomenology is more complicated and model dependent (see \cite{Knox:2019rjx,DiValentino:2021izs}).   

Finally, as was already emphasized above, the CMB energy injection and down-scattering constraints are substantially relaxed if additional depletion mechanisms are relevant. These can be associated with very high-scale physics (e.g. a magnetic dipole transition suppressed by a scale $\Lambda \lesssim 10^{12}\,\GeV$), or can even arise from non-degenerate Majorana masses within our simple model.  However, in some cases the mechanisms giving rise to these depletions may induce other constraints --- for example, at large $\Lambda$ the dipole transition could give rise to significant energy injection into the CMB. Decays that produce only neutrinos and the light DM state $\chi_l$ are not constrained by CMB energy injection, but may be observable by other means. The interplay of effects in these generalized scenarios therefore deserves a more thorough examination.  Alternately, there may be present-day interactions of $\chi_l$ that re-generate the heavy state, that enable its detection even if the cosmological abundance of $\chi_h$ decays away.  These could help to establish the microphysical properties of DM in more detail.  

\subsection*{Note added}
Most of our results on the early-Universe depletion of inelastic DM and its impact on CMB annihilation constraints was completed as part of MCG's PSI Masters Thesis in 2015, while the main results on accelerator-based constraints and line-like down-scattering nuclear recoil signals were completed in 2015-17.  Other commitments interrupted our progress on this work for several years until the Xenon1T electron excess \cite{Aprile:2020tmw} rekindled our interest in mid-2020.  Several other analyses of the Xenon1T result, published during the extended completion of this paper, overlap the scope of our results (e.g. \cite{Baryakhtar:2020rwy,Bloch:2020uzh}), and are approximately consistent with our results.   Our work offer a complementary perspective: while these works focused mainly on $\delta \sim 3 \keV$ to address the Xenon1T excess, our aim is to assess the viability and detection prospects for light, inelastic DM over a broader range of mass splittings. Conversely, our focus is mainly on models that achieve standard thermal freeze-out via $\chi_l \chi_h \rightarrow \rm{SM}$ final states, i.e. those lying on the ``thermal relic target/line'' of interaction strength vs. mass, while the papers noted above have relaxed this mass-coupling relation.   To this end, our work includes a detailed recasting of CRESST-II and CRESST-III results throughout the sub-GeV and multi-GeV DM mass regions, as well as of electron-recoil constraints on light DM down-scattering which to our knowledge has not been considered elsewhere.  We have also identified several new physical effects of interest for this parameter space, including CMB decay constraints, the potential for Earth's shadowing of heavy DM particles to suppress direct detection signals and induce their daily modulation, and the possibility that semi-elastic scattering could further exponential suppress the DM abundance.  Of these, the semi-elastic scattering and earth-shadow effects are quite relevant to interpretation of the $\sim 3 \keV$ Xenon1T electron-recoil excess, and more broadly to light DM with modest splittings, while CMB decay constraints are most relevant for splittings above a few hundred keV. 

\section{Acknowledgments}
We thank Asher Berlin for useful conceptual discussions prior to his work on \cite{Baryakhtar:2020rwy}, for alerting us to the relevance of this model to the Xenon1T electron excess, and for detailed numerical cross-checks of our results against those in \cite{Baryakhtar:2020rwy}.  MCG is supported by the European Union's Horizon 2020 Research Council grant 724659 MassiveCosmo ERC-016-COG and the STFC grants ST/P000762/1 and ST/T000791/1.  NT is supported in part by the U.S. Department of Energy under Contract DE-AC02-76SF00515.  Significant portions of this work were conducted while MCG and NT were supported in part by Perimeter Institute for Theoretical Physics. Research at Perimeter Institute is supported by the Government of Canada through Industry Canada and by the Province of Ontario through the Ministry of Research and Innovation.

\appendix
\section{Thermally averaged Cross Sections} \label{App:CrossSec}
\subsection{Annihilation ${\chi_l}\;{\chi_h}\rightarrow e\;e$+ heavier fermions + baryons} 
The annihilation cross section for the process ${\chi_l}\;{\chi_h}\rightarrow e\;e$ is given by
\begin{equation}
(\sigma_\text{ann})_{ee}= \frac{2\pi\,\alpha\, \alpha_D\, \epsilon^2 (s+2 m_e^2)\sqrt{s(s-4m_e^2)(s-\delta^2)} (2s+4\bar{m}_\chi^2)}{3s^2\sqrt{s-4\bar{m}_\chi^2}((s-m_{A'}^2)^2+m_{A'}^2\Gamma_{A'}^2)} 
\end{equation}
where $s$ is the usual Mandelstam variable, $\bar{m}_\chi=(m_{\chi_l}+m_{\chi_h})/2$, $m_e$ is the electron mass, $\delta=m_{\chi_h}-m_{\chi_l}$ is the mass splitting, and $\Gamma_{A'}$ is the dark photon decay width given by
\begin{equation}
\Gamma_{A'} \equiv \Gamma\left(A' \rightarrow \chi_{h} \chi_{l}\right)=\frac{\alpha_D m_{A'}}{3}\left(1-\frac{4\bar{m}_\chi^{2}}{m_{A'}^{2}}\right)^{1 / 2}\left(1+\frac{2\bar{m}_\chi^{2}}{ m_{A^{\prime}}^{2}}\right)\left(1-\frac{\delta^{2}}{m_{A'}^{2}}\right)^{3 / 2} \ .
\end{equation} 
Taking into account not only the annihilation to electrons, but also the annihilation to muons and hadrons the total annihilation cross section is given by
\begin{equation}
\sigma_\text{ann}=(\sigma_\text{ann})_{ee}\left(1+\frac{(\sigma_\text{ann})_{\mu\mu}}{(\sigma_\text{ann})_{ee}}\left(\theta
(s-4m_\mu^2)+\theta
(s-4m_\pi^2)R\left(s\right)\right) \right) ,
\end{equation}
where $R(s)$ is the ratio of hadron and muon production in $e^+\,e^-$ annihilation, $R(s) \equiv \sigma(e^+\,e^-\rightarrow \text{hadrons})/\sigma(e^+\,e^-\rightarrow \mu+\,\mu-)$ \cite{10.1093/ptep/ptaa104}. We compute the thermal average at $T=T_{fo}$ numerically by following \cite{GONDOLO1991145}, and performing the following integral
\begin{equation}
\sigann=\frac{1}{8 m_{\chi_l}^{2} m_{\chi_h}^2 T K_{2}^{2}(m_{\chi_l} / T) K_{2}^{2}(m_{\chi_h} / T)} \int_{4 \bar{m}_\chi^{2}}^{\infty} \sigma_\text{ann} \left(s-4 \bar{m}_\chi^{2}\right) \sqrt{s} K_{1}(\sqrt{s} / T) \ d s	\ . \label{sigmaThAv}
\end{equation}
We will use this thermally averaged cross-section to find the $\epsilon^2$ required for a thermal target as given by Eq.~\eqref{epsilon}.
Note that the thermally averaged cross-section for the annihilation to electrons has a simple analytic expression in the non-relativistic limit given by
\begin{equation}
\sigann_{ee}\simeq\frac{16\pi\,\alpha\, \alpha_D\, \epsilon^2 \bar{m}_\chi^2}{(4\bar{m}_\chi^2-m_{A'}^2)^2} \ .\, \label{cann}
\end{equation}
where we have taken the limits $ m_f\ll m_{\chi_l},\, \delta< m_{\chi_l}, 2 m_{\chi_l}<m_A$.
\subsection{Scattering ${\chi_h}\,f\rightarrow {\chi_l}\, f$} \label{csxf} 
For this process, the usual expansion of the cross section in small velocities fails since the expansion grows in negative powers of $\tfrac{\delta}{m_{\chi_l}}$. Instead, we compute the full cross section and then take power law approximations for five different regions in order to obtain analytic functions after the thermal average. The piece-wise cross section times velocity now is given by

{ \begin{equation}
(\sigma \, v)_{\chi_h f} = \begin{cases}
\vspace{0.2 cm}
\frac{m_{A'}^2}{2  \ m_{\chi_l}^2}\sigann_{ee} &\text{for}\; s>m_{A'}^2 \\ 
\frac{1}{3 m_{{\chi_l}}^2}\sigann_{ee}\, s  &\text{for}\; s<m_{A'}^2\; \\ \vspace{0.2 cm}
&\text{and}\; T> m_{\chi_l}+m_f \\ \vspace{0.2 cm}
\frac{(m_{\chi_l}^4-2m_{\chi_l}^2 s +4s^2)(m_{\chi_l}^2-s)^2 }{12\,m_{\chi_l}^2\,s^2\,(m_{\chi_l}^2+s)}\!\sigann_{ee}&\text{for}\;\mu< \!T\!< m_{\chi_l}\!\!+\!m_f\\ \vspace{0.2 cm}
\frac{\,\mu^2}{m_{\chi_l}^2}\sigann_{ee} \,v & \text{for}\; \delta< T< \mu \\ \vspace{0.2 cm}
\frac{\,\sqrt{2}\,\mu^{3/2}\sqrt{\delta}}{m_{\chi_l}^2}\sigann_{ee} & \text{for}\; T<\delta
\end{cases}, \\ \label{sigxf}
\end{equation} }
where $\sigann_{ee}$ is given by Eq. \eqref{cann}, and $\mu$ is the reduced mass of the $\chi_h$-fermion system. If the dark gauge boson mass is comparable to the dark matter mass, the cross section should be multiply by $(m_{A'}^2-4\bar{m}_{\chi}^2)^2/m_{A'}^4$.

\begin{figure}[!h]
	\begin{center}\includegraphics[width=0.6\textwidth]{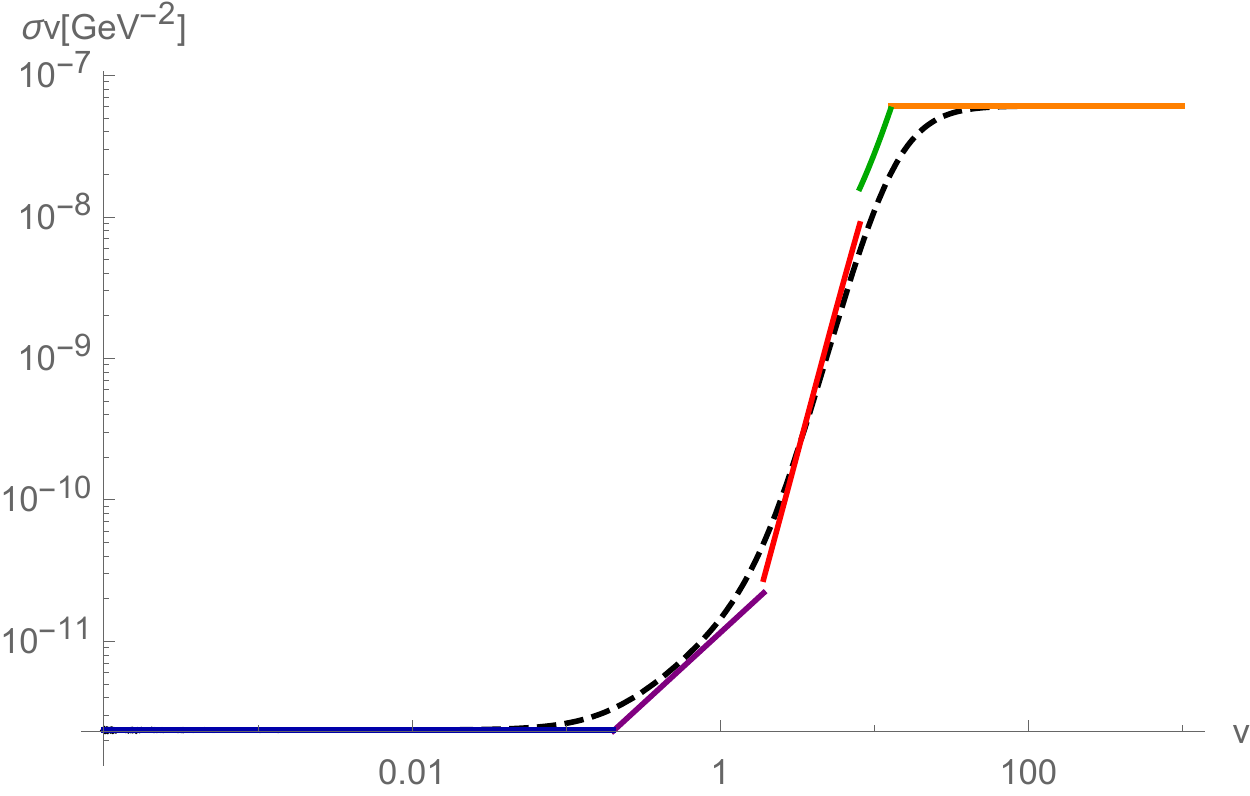}\end{center}
	\caption{Plot of the approximation for the cross section times velocity for the scattering ${\chi_h}\;e^\pm\rightarrow{\chi_l}\;e^\pm$ for masses $m_{\chi_l}=10\,\text{MeV}$, $m_{A'}=5m_{\chi_l}$ and $\delta=10^{-2}\,\text{MeV}$. The dashed black line correspond to full expression for the cross section, the colored lines are the cross section approximations.}
\end{figure}

These five regions consist of three relativistic zones and two non-relativistic. The least  relativistic region (third one above) corresponds to the region where only one particle is relativistic, either the electron (positron) or the DM. From these expressions, we can take the thermal average, with the corresponding approximations in each limit and obtain:

{ \begin{equation}
\langle \sigma \, v\rangle_{\chi_h f}= \begin{cases}\vspace{0.4 cm}
\frac{m_{A'}^2}{2 m_{\chi_l}^2}\sigann_{ee} &\text{for}\; T> \frac{1}{2 \sqrt{2}}\, m_{A'} \\
\frac{4 T^2}{m_{\chi_l}^2}\sigann_{ee} &\text{for}\; \!\frac{1}{8}\left(\!\left(\!\frac{15}{2}\!\right)^2\!\pi\!\right)^{\frac{1}{3}}\! \mu < T \\ \vspace{0.4 cm}
& \hphantom{\text{for}} \leq \frac{1}{2 \sqrt{2}}\, m_{A'} \\
\frac{15\sqrt{\pi}\mu^{3/2}\sqrt{T}}{8\sqrt{2} \,m_{\chi_l}^2} \sigann_{ee}& \text{for}\; \frac{256}{225 \pi}\,\delta< T \\ \vspace{0.4 cm}
&\hphantom{\text{for}} \leq  \!\frac{1}{8}\left(\!\left(\!\frac{15}{2}\!\right)^2\!\pi\!\right)^{\frac{1}{3}}\! \mu\\ \vspace{0.4 cm}
\frac{\,\sqrt{2}\,\mu^{3/2}\sqrt{\delta}}{m_{\chi_l}^2}\sigann_{ee} & \text{for}\; T\leq \frac{256}{225 \pi}\,\delta
\end{cases}\!.\!\!\label{tsigxf}
\end{equation} }

After integrating the cross section for each region, we find that two of the relativistic regions collapse into one. The limits of the piece-wise function were chosen in order to have a continuous function.

\begin{figure}[!h]
	\begin{center}
		\includegraphics[width=0.6\textwidth]{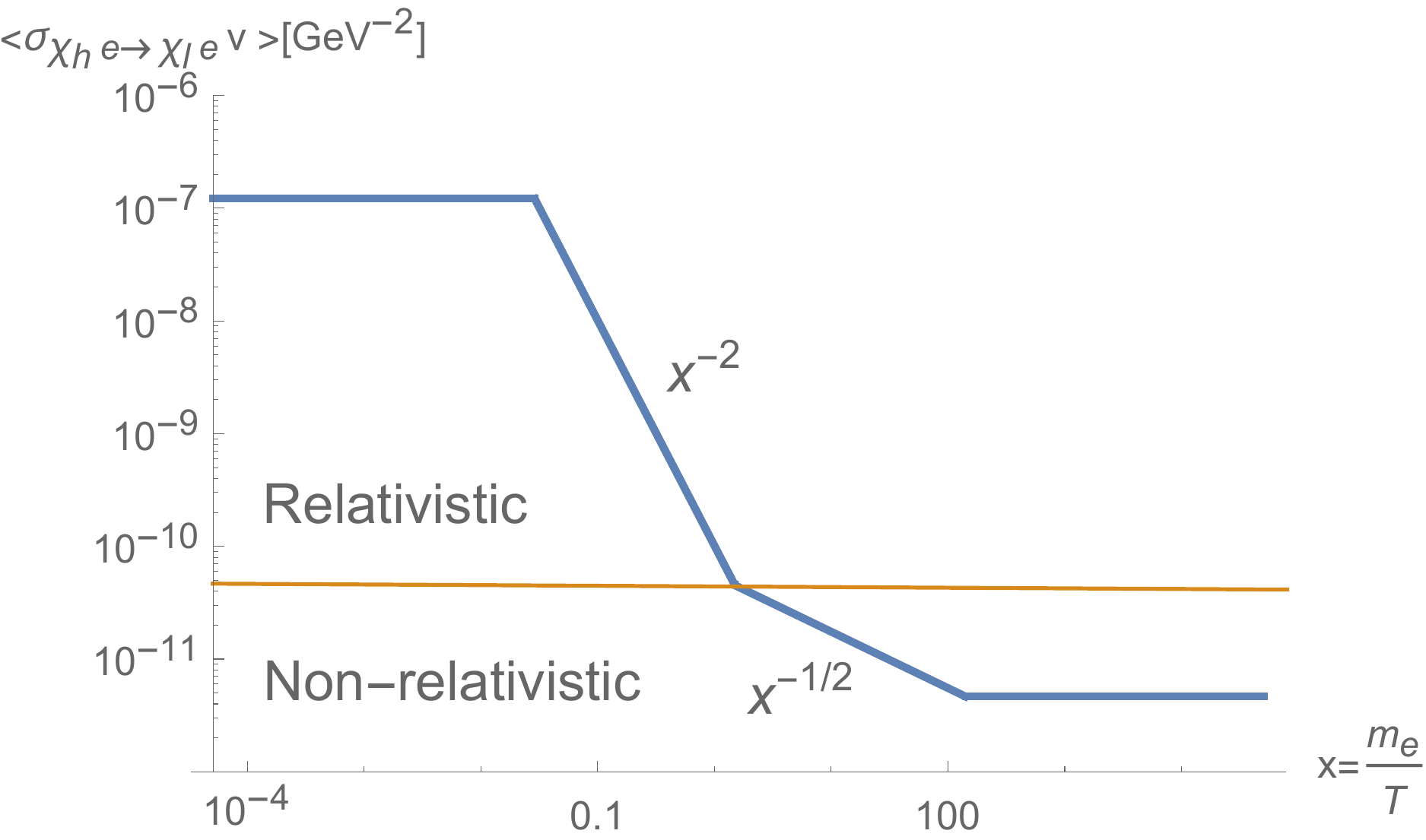}	
	\end{center}
	\caption{Thermally averaged cross section for a DM mass of $10$ MeV, mass splitting of $10^{-2}$ MeV, and $m_{A'}= 5 \,m_{\chi_l}$. We can observe that the cross section is constant for both high and low temperatures, but in the middle it has a non-trivial dependence on the temperature.}
	\label{csxfT}
	\vspace{-0cm}
\end{figure}

\subsection{Scattering ${\chi_h}\,{\chi_h}\rightarrow {\chi_l}\,{\chi_l}$}\label{x} 
For this annihilation process, we find again that the velocity expansion is not possible, thus we follow the same procedure as before and obtain that the cross section times velocity is given by:
\begin{equation}
(\sigma \, v)_{\chi_h\chi_h}= \begin{cases}
\frac{8\, \pi \,\alpha_D^2}{m_{A'}^2} &\text{for}\; s>m_{A'}^2 \\
\frac{28\,\pi\, \alpha_D^2\;s }{3\,m_{A'}^4} &\text{for}\; s<m_{A'}^2\; \text{and}\; T> \mu \\
\frac{2\,\pi\, \alpha_D^2 \,m_{\chi_l}^2}{m_{A'}^4}\,v& \text{for}\; \delta< T< \mu \\
\frac{4\,\sqrt{2}\, \pi\,\alpha_D^2\; m_{\chi_l}^{3/2}\sqrt{\delta}}{m_{A'}^4}& \text{for}\; T<\delta
\end{cases}.\label{sigx}
\end{equation} 

This time the cross section is divided in four different zones, two where the DM is relativistic and two where it is non-relativistic. When the mass splitting is of order of the DM mass, instead of using the approximation of Eq. \eqref{sigx} in the ultra non-relativistic case, we should use
{\small \begin{align}
(\sigma \, v)_{\chi_h\chi_h}&=\frac{4\pi \, \alpha_D^2 \,\sqrt{4\delta  \bar{m}_\chi}(m_{\chi_l}+2 \delta)^2}{ (m_{\chi_l} +\delta )
	\left(m_{A'}^2+4\delta \bar{m}_\chi\right)^2} \ .
\end{align}}

 \begin{figure}[!h]
	\begin{center}\includegraphics[width=0.6\textwidth]{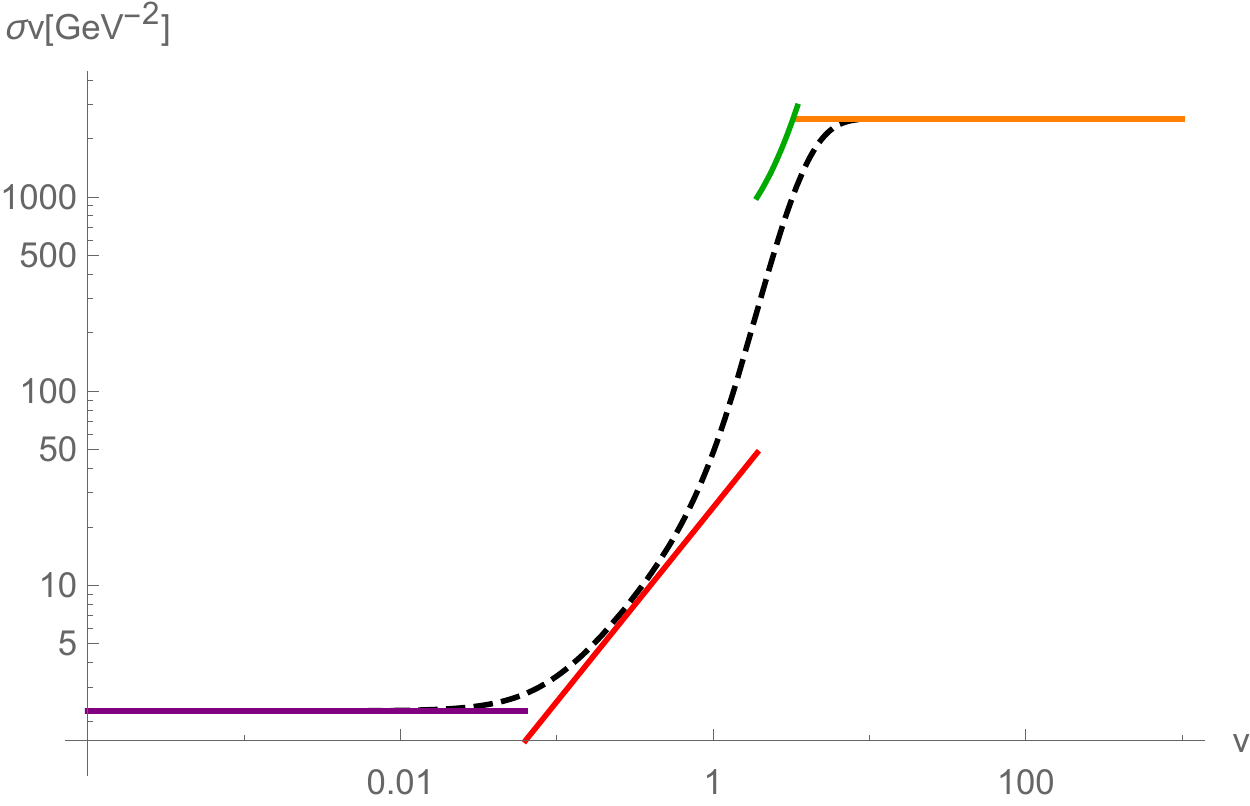}\end{center}
	\caption{Plot of the approximation for the cross section times velocity for the scattering ${\chi_h}\,{\chi_h}\rightarrow{\chi_l}{\chi_l}$ for masses $m_{\chi_l}=10\,\text{MeV}$, $m_{A'}=5m_{\chi_l}$, $\delta=10^{-2}\,\text{MeV}$, and dark coupling $\alpha_D=0.5$. The dashed black line correspond to full expression for the cross section, the colored lines are the cross section approximations.}
\end{figure}

From this equation we obtain the thermal average which is given by:
{ \begin{equation}
\langle\sigma \, v\rangle_{\chi_h\chi_h} =
\begin{cases}
\frac{8\, \pi\, \alpha_D^2}{m_{A'}^2} &\text{for}\; T>\frac{m_{A'}}{\sqrt{14}}\\ \vspace{0.4 cm}
\frac{112 \,\pi \, T^2\, \alpha _D^2}{m_A'^{'4}} &\text{for}\; \frac{\left(\frac{15}{7}\right)^{2/3} \pi^{1/3} m_{\chi_l}^{4/3}}{16
	({m_{\chi_l}+\delta )^{1/3}}}<T\leq\frac{m_{A'}}{\sqrt{14}}\\
\frac{15 \pi ^{3/2} \, \alpha _D^2\, m_{{\chi_l} }^2\,\sqrt{T}}{4 m_{A'}^4 \sqrt{m_{{\chi_l}}+\delta}}& \text{for}\; \frac{512 \left(\delta ^2+\delta  m_{\chi _l}\right)}{225 \pi  m_{\chi _l}}< T\\ \vspace{0.4 cm}
&\hphantom{\text{for}}\leq\frac{\left(\frac{15}{7}\right)^{2/3} \pi^{1/3} m_{\chi_l}^{4/3}}{16
	({m_{\chi_l}+\delta )^{1/3}}} \\\vspace{0.4 cm}
\frac{4\,\sqrt{2}\, \pi\,\alpha_D^2\; m_{\chi_l}^{3/2}\sqrt{\delta}}{m_A^4}& \text{for}\; T\leq \frac{512 \left(\delta ^2+\delta  m_{\chi _l}\right)}{225 \pi  m_{\chi _l}}
\end{cases}.\label{crossecxx}
\end{equation}} 

\begin{figure}[!h]
	\vspace{-0.3 cm}
	\begin{center}
		\includegraphics[width=0.6\textwidth]{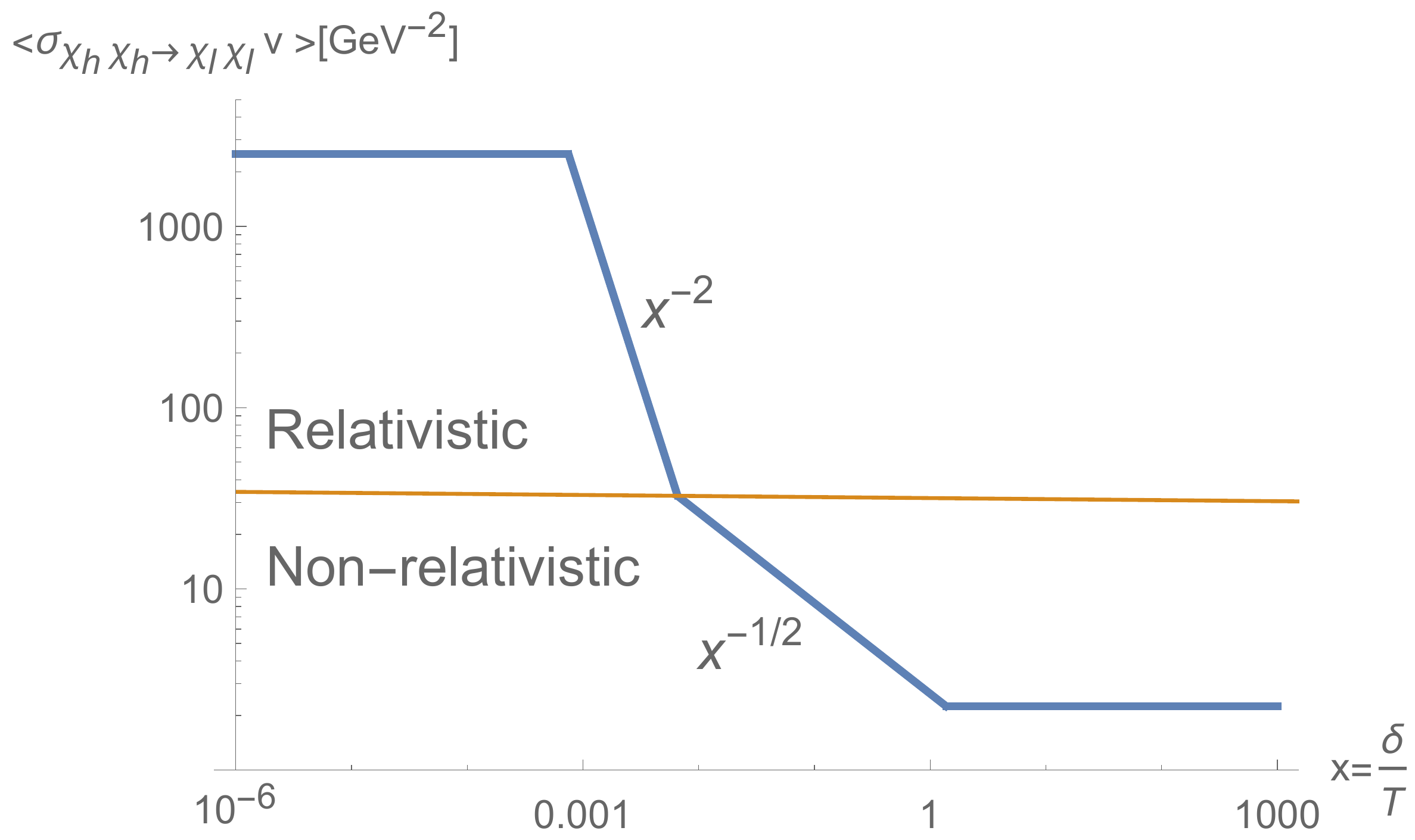}	
	\end{center}
	\vspace{-0.3 cm}
	\caption{Thermally averaged cross section for a DM mass of $10$ MeV, mass splitting of $10^{-2}$ MeV, $\alpha_D=0.5$, and $m_{A'}= 5 \,m_{\chi_l}$. Again, we observe that for both high and low temperatures the cross section is constant and in the middle has a non-trivial dependence on $x$.}
	\label{csxx}
	\vspace{-0.15 cm}
\end{figure}

\section{Annihilation Freeze-out Calculation} \label{apfo}
In this Appendix, we show in detail the calculation for the freeze-out of the total dark matter abundance. The Boltzmann equation for the total dark matter abundance is given by Eq. \eqref{totall}. Using $Y^\text{eq}_\text{tot}\propto e^{-x}(1+e^{-\frac{\delta}{m_{\chi_l}}x})$, we can solve for the freeze-out temperature and abundance:
\begin{align}
x_f&=\frac{1}{1+\frac{\delta}{m_{\chi_l}}}\log{\left[0.076 \frac{g}{ g_*^{1/2}}\, \frac{M_{Pl}\, m_{\chi_l}\, \sigann x_f^{1/2}}{\left(1+e^{\frac{-\delta}{m_{\chi_l}}x_f}\right)}\right]}, \label{xann}\\
Y_{\text{tot}}&=\frac{45}{2\pi^4}\sqrt{\frac{\pi}{8}}\frac{g\, x_f^{3/2}}{g_{*s}}\left( \frac{g_*^{1/2}(1+e^{\frac{-\delta}{m_{\chi_l}}x_f})}{0.076\, g\, M_{Pl}\; m_{\chi_l} \langle \sigma_{\text{ann}} \rangle x_f^{1/2} }\right)^{\!\!\frac{1}{1+\frac{\delta}{m_{\chi_l}}}}\nonumber\\
&\times\left(1+\left( \frac{g_*^{1/2}(1+e^{\frac{-\delta}{m_{\chi_l}}x_f})}{0.076\, g\, M_{Pl}\; m_{\chi_l} \langle \sigma_{\text{ann}} \rangle x_f^{1/2}}\right)^{\frac{\frac{\delta}{m_{\chi_l}}}{1+\frac{\delta}{m_{\chi_l}}}}\right), \label{yinf}
\end{align}
where $g=2$ is the number of internal degrees of freedom, $g_*$ is the effective number of relativistic degrees of freedom and $g_{*s}$ the effective number of relativistic degrees of freedom for entropy. Notice that in the recursive relation, Eq. \eqref{xann}, the terms depending on $x$ give a small contribution since the dependence is logarithmic and we have that $x\sim\mathcal{O}(10)$. This fact implies that the error from solving the equation recursively is $\sim0.5\%$. We now look at two different limits $\delta<m_{\chi_l}$ and $\delta\sim m_{\chi_l}$. For the case $\delta<m_{\chi_l}$ we find
\begin{equation}
Y_{\text{tot}}=\left(1+e^{\frac{-\delta}{m_{\chi_l}}x_f}\right) \frac{3.79 \,x_f}{(g_{*s}/g_*^{1/2})M_{Pl}\; m_{\chi_l} \langle \sigma_{\text{ann}} \rangle }. \label{ytot}
\end{equation}
Meanwhile for $\delta\sim m_{\chi_l}$
\begin{align}
Y_{\text{tot}}&=\frac{45}{2\pi^4}\sqrt{\frac{\pi}{8}}\frac{g\,x_f^{3/2}}{g_{*s}}\left( \frac{g_*^{1/2}(1+e^{\frac{-\delta}{m_{\chi_l}}x_f})}{0.076\, g\, M_{Pl}\; m_{\chi_l} \langle \sigma_{\text{ann}} \rangle x_f^{1/2} }\right)^{\frac{1}{1+\frac{\delta}{m_{\chi_l}}}} \nonumber \\
&\rightarrow\frac{45}{2\pi^4}\sqrt{\frac{\pi}{8}}\frac{g^{1/2}}{g_{*s}}x_f^{5/4}\left( \frac{g_*^{1/2}(1+e^{\frac{-\delta}{m_{\chi_l}}x_f})}{0.076\, M_{Pl}\; m_{\chi_l} \langle \sigma_{\text{ann}} \rangle }\right)^{1/2}.
\end{align}
After the dark matter annihilation process has frozen-out, the total dark matter abundance will only change due to the expansion of the Universe. The only quantity that keeps changing due to particle interactions is the relative abundance of the heavy and light state. Given this, we can compute the value of the dark matter density today in the two previous limits. For $\delta<m_{\chi_l}$
\begin{align} \Omega_\text{tot}\, h^2&=\frac{Y_{\text{tot}}\,s_0\,m_{\chi_l} }{3 H_0^2 M_{Pl}^2}=\left(1+e^{\frac{-\delta}{m_{\chi_l}}x_f}\right)\frac{3.79\,g_*^{1/2}\,x_f s_0}{6g_{*s} H_0^2 M_{Pl}^3\; \langle \sigma_{\text{ann}} \rangle} \nonumber\\
&=8.77\times10^{-11} \text{GeV}^{-2}\left(1+e^{\frac{-\delta}{m_{\chi_l}}x_f}\right)\frac{\,g_*^{1/2}\,x_f }{g_{*s} \,\langle \sigma_{\text{ann}} \rangle},
\label{om}
\end{align}
while for $\delta\sim m_{\chi_l}$ we have
\begin{align} \Omega_\text{tot}\, h^2=\frac{Y_{\text{tot}}\,s_0\,m_{\chi_l} }{3 H_0^2 M_{Pl}^2}=\frac{45}{2\pi^4}\sqrt{\frac{\pi}{8}}\frac{g\,x_f^{3/2}}{g_{*s}}\frac{s_0\,m_{\chi_l} }{3 H_0^2 M_{Pl}^2} \nonumber \\
\times\left( \frac{g_*^{1/2}(1+e^{\frac{-\delta}{m_{\chi_l}}x_f})}{0.076\, g\, M_{Pl}\; m_{\chi_l} \langle \sigma_{\text{ann}} \rangle x_f^{1/2} }\right)^{\frac{1}{1+\frac{\delta}{m_{\chi_l}}}}\nonumber\\
\rightarrow0.30\, \text{GeV}^{-1/2}\left( \frac{g\,g_*^{1/2}\,m_{\chi_l}}{g_{*s}^2\; \langle \sigma_{\text{ann}} \rangle }\right)^{1/2}x_f^{5/4}(1+e^{\frac{-\delta}{m_{\chi_l}}x_f})^{1/2}.
\label{om1}
\end{align}
Assuming that all the contributions  to the dark matter density comes from ${\chi_l}$ and ${\chi_h}$ and using the current observational value of the dark matter density given by Planck 2018 \cite{Aghanim:2018eyx}: $ \Omega_{\text{CDM}} h^2=0.1200\pm0.0012 $, it is possible to find an estimate for the annihilation cross section. First we solve for $\sigann$ in Eq. \eqref{om} or Eq. \eqref{om1} and use it in Eq. \eqref{xann}, for both limits we find
\begin{equation}
x_f=\log{\left[\frac{0.144  s_0}{\Omega_\text{tot}\, h^2 3H_0^2 M_{Pl}^2} \frac{g\,}{ g_{*s}}\, m_{\chi_l}\, x_f^{3/2}\right]}.
\end{equation}
We solve this equation recursively until $|x_i-x_{i+1}|<10^{-5}$. Having this solution, we use it in Eq. \eqref{yinf} and Eq. \eqref{om} or Eq. \eqref{om1} to find the abundance and the annihilation cross section.

\section{Constraints for several benchmark points} \label{app:MoreBenchmarks}
In this appendix, we show the results obtained in the main text for three different benchmark points: $\alpha_D=0.5, \ m_{A'}=3 \,\left(m_{\chi_l}+\delta/2\right)$, $\alpha_D=0.5, \ m_{A'}=10 \,\left(m_{\chi_l}+\delta/2\right)$, $\alpha_D=0.05, \ m_{A'}=3 \,\left(m_{\chi_l}+\delta/2\right)$. This helps us understand the difference in the constraints as the ratio $\alpha_D^2/m_{A'}^4$ varies away from our primary benchmark considered in the main text.  In general, the constraints become stronger at larger $m_{A^\prime}/m_\chi$ or smaller $\alpha_D$.
\begin{sidewaysfigure}[!ht]
	\begin{center}
		\includegraphics[width=0.9\textwidth]{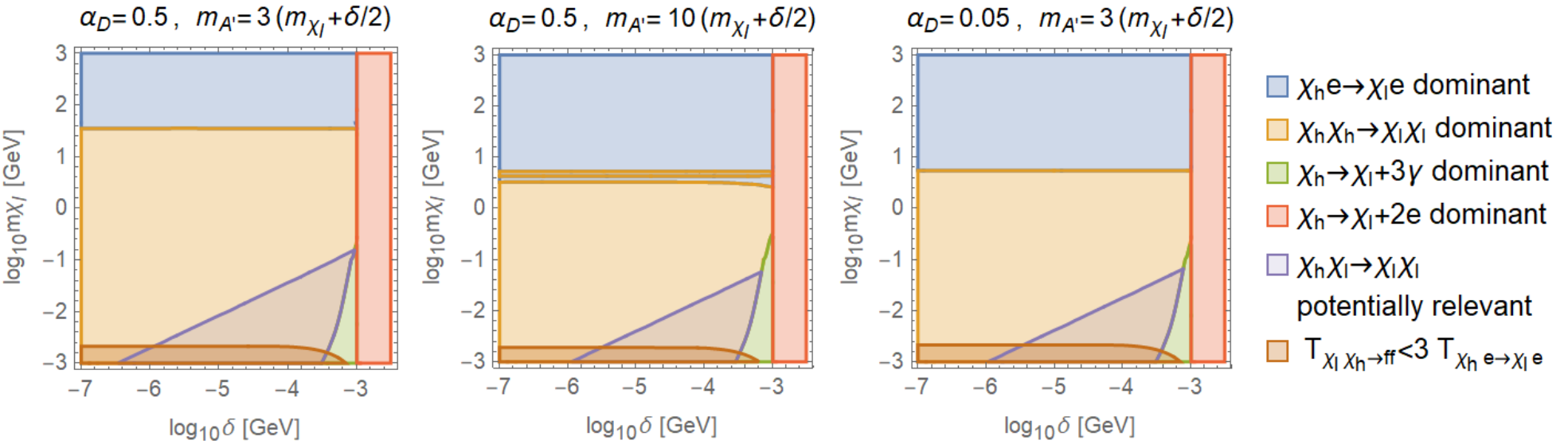}
		\caption{Regions of the parameter space where the $\chi_h$ abundance at recombination is set by the decay to electrons, decay to photons, scattering off fermions, or scattering of dark matter, as in Fig.~\ref{regions}(left) but for different values of $\alpha_D$ and $m_{\chi}/m_{A'}$. In the purple region, there could be an additional depletion of the excited state due to semi-elastic interactions. The brown region corresponds to the case where the chemical decoupling occurs close, but still after, the total dark matter freeze-out. We note that at lower  $\alpha_D^2/m_{A'}^4$ the scattering off DM becomes less relevant for $\sim 10$ GeV masses.\label{appFig:dominantReactions}}
	\end{center}
\end{sidewaysfigure}
\begin{sidewaysfigure}[!ht]
	\begin{center}
		\includegraphics[width=0.9\textwidth]{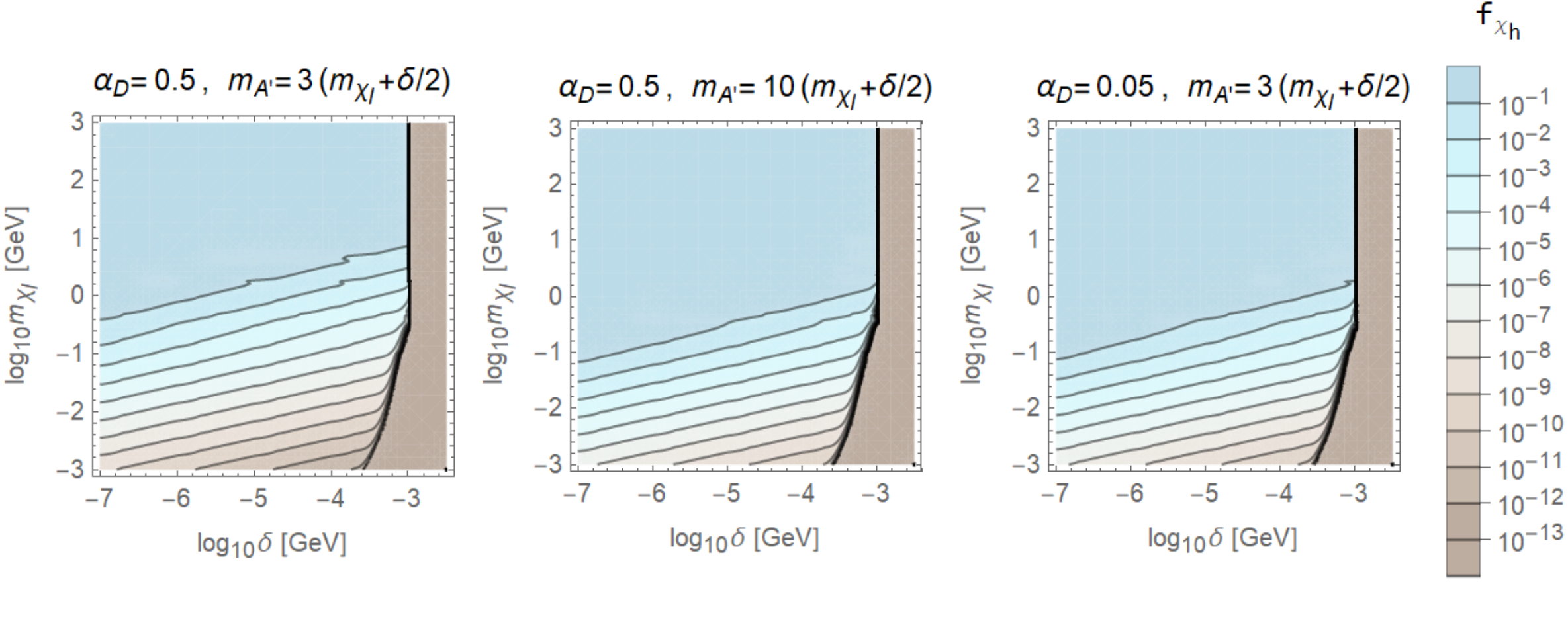}
		\caption{Contours of the relative abundance of the dark matter excited state, as in Fig.~\ref{regions}(right) but for different values of $\alpha_D$ and $m_{\chi}/m_{A'}$. We can observe the exponential depletion in the regions where the decays dominate. For smaller $\alpha_D^2/m_{A'}^4$ the depletion due to scatterings off dark matter dominates at lower masses which leads to a large excited state abundance for GeV DM masses. \label{appFig:abundances}} 
	\end{center}
\end{sidewaysfigure}

\begin{figure}[!ht]
	\begin{center}
		\includegraphics[scale=1]{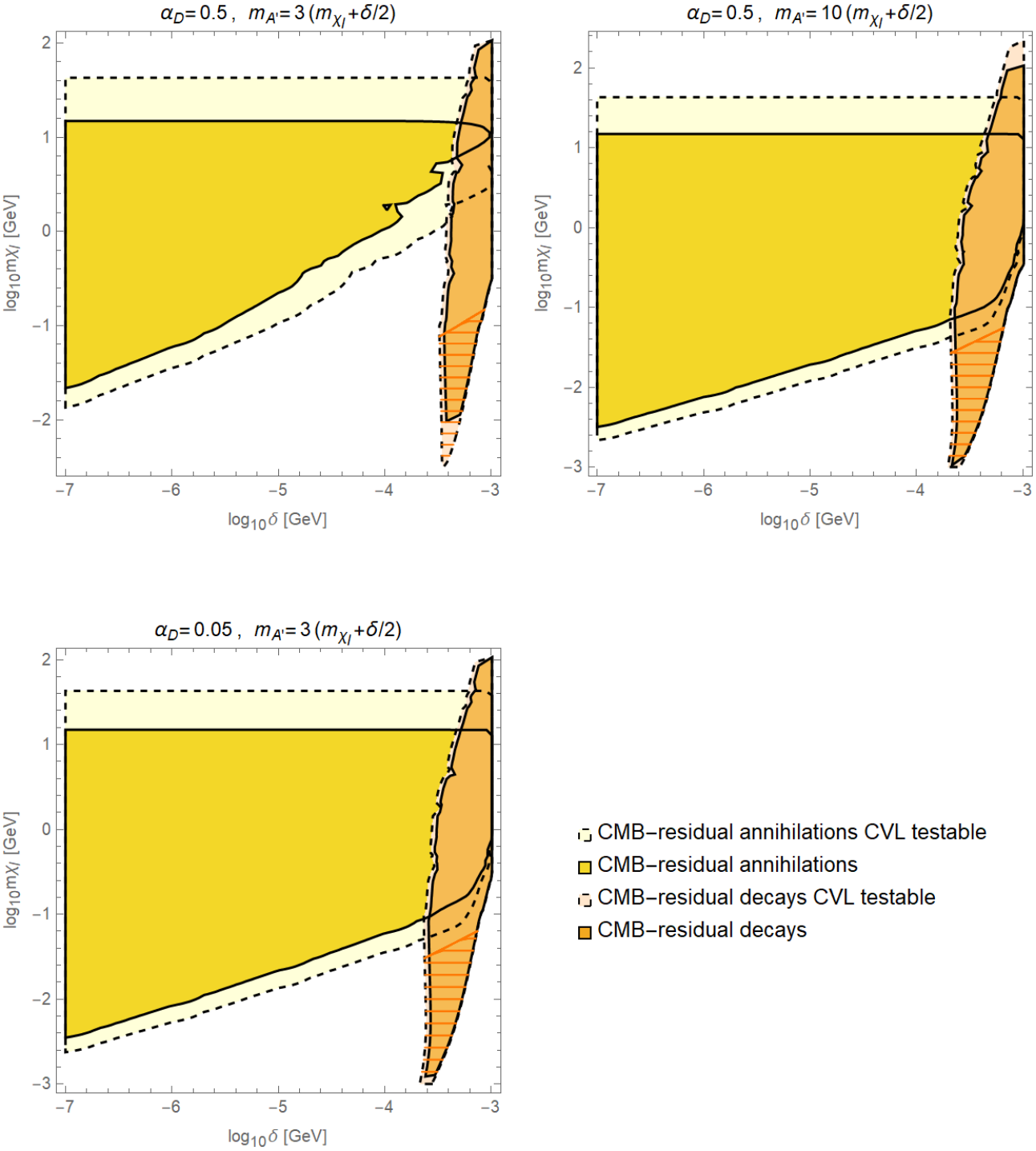} 
	\end{center}
\caption{Constraints arising from residual annihilations (yellow) and decays (orange) that can alter the CMB anisotropies, as in Fig.~\ref{CMBbounds} but for different values of $\alpha_D$ and $m_{\chi}/m_{A'}$. The dashed contours correspond to the regions that a CVL limited experiment could probe. The striated area is the region where semi-elastic interactions may further deplete the excited state and weaken the constraints.\label{appFig:cmb}}
\end{figure}
\begin{figure}[!ht]
	\begin{center}
		\includegraphics[width=1\textwidth]{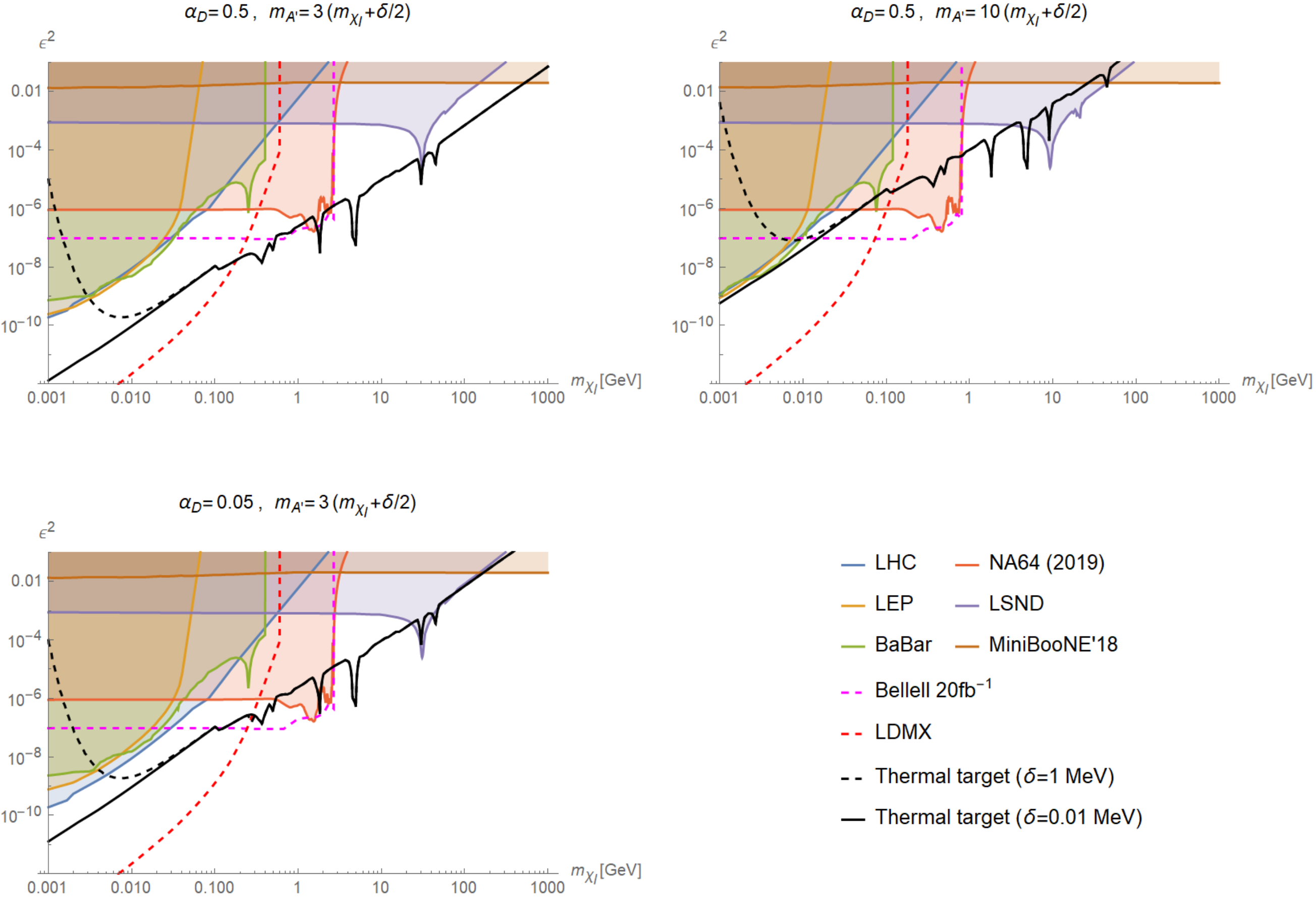} 
	\end{center}
\caption{Current bounds and projections from fixed-target experiments for a thermal target with MeV ans sub-MeV mass splitting,  as in Fig.~\ref{eps} but for different values of $\alpha_D$ and $m_{\chi}/m_{A'}$. We note that the exclusions become stronger for smaller $\alpha_D^2/m_{A'}^4$ ratios, while the projections have the ability to constrain most of the sub-GeV region. \label{appFig:acceleratorEpsilon}}
\end{figure}
 \begin{figure}[!h]
	\begin{center}
		\includegraphics[width=1\textwidth]{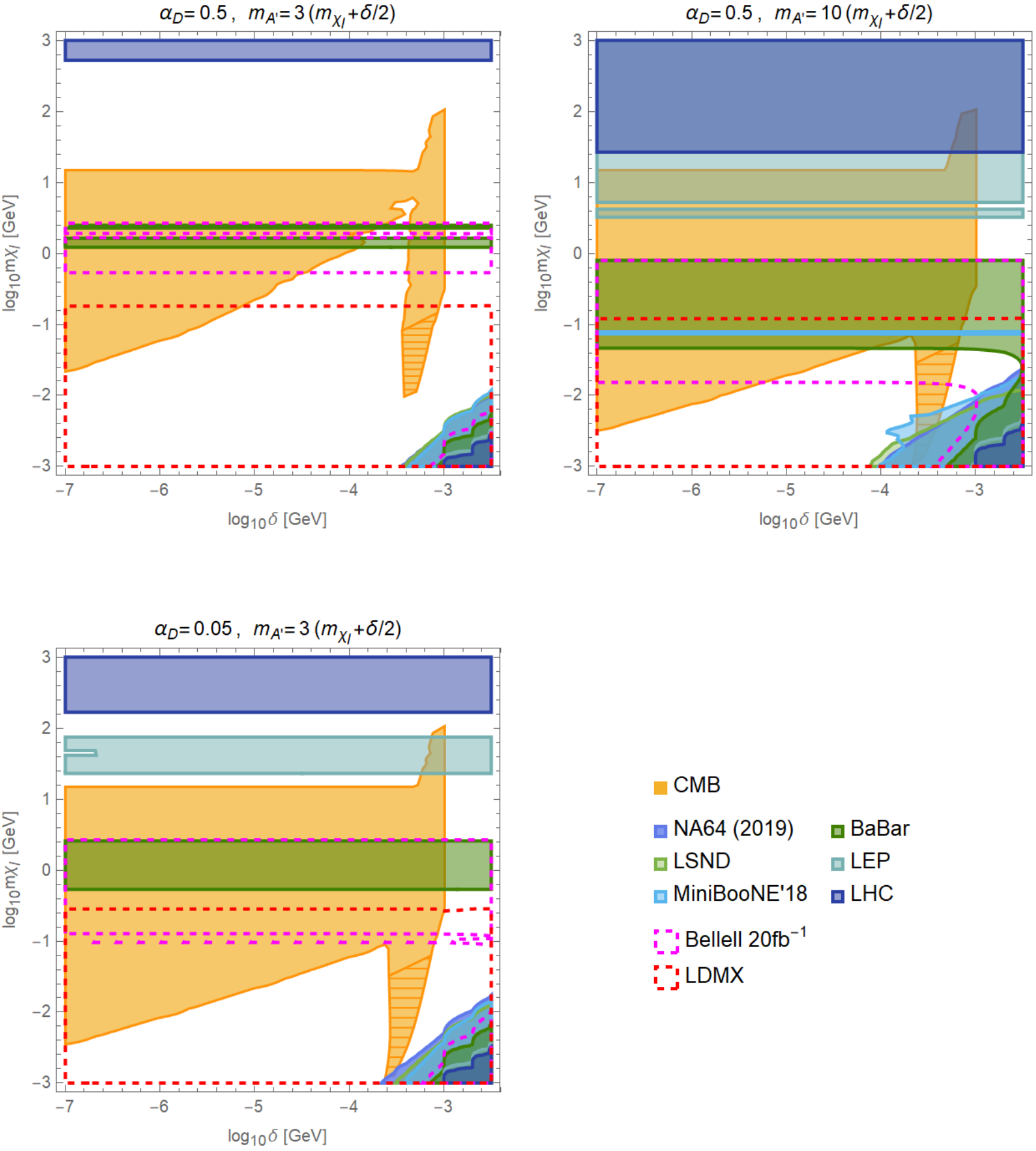}	
	\end{center}
	\caption{Regions of parameter space excluded by CMB observations and accelerator-based probes, as in Fig.~\ref{acc} but for different values of $\alpha_D$ and $m_{\chi}/m_{A'}$. The mustard region is ruled out by CMB observations and the color shaded regions are ruled out by different accelerator-based probes. The dashed contours correspond to Belle II and LDMX projections.\label{appFig:acceleratorCMB}}
	\label{acc3}
	\vspace{-0cm}
\end{figure}
\begin{figure}[!ht] 
	\includegraphics[width=\textwidth]{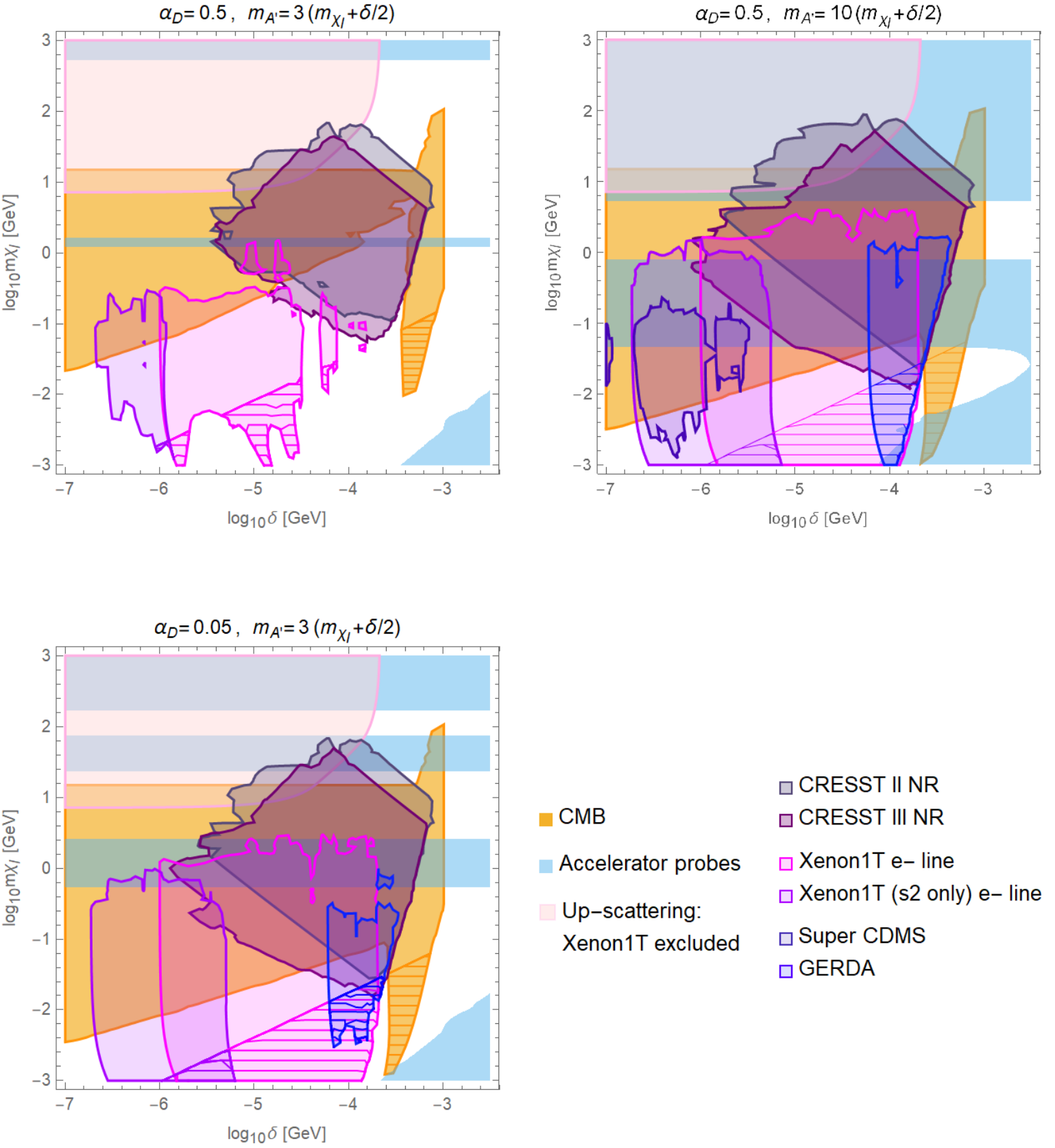} 
	\caption{Detailed exclusion regions for direct detection probes, as in Fig.~\ref{DDbounds} but for different values of $\alpha_D$ and $m_{\chi}/m_{A'}$. We also show the constraints from collider probes in light blue and CMB in mustard. Here, we can clearly see that small $\alpha_D/m_A^2$ ratios are heavily constrained. In the striated areas, semi-elastic interactions can further deplete the $\chi_h$ abundance and weaken the constraints.}
	\label{AllBoundsDifBenchmarks}
\end{figure}
\begin{sidewaysfigure}[!ht]
	\begin{center}
		\includegraphics[width=0.9\textwidth]{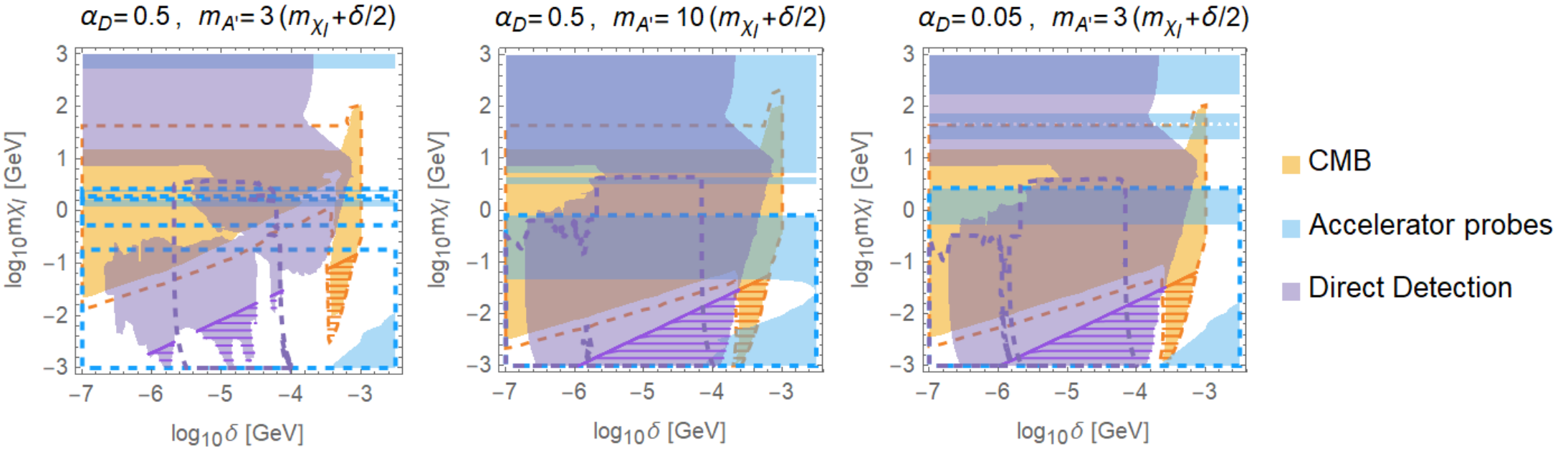} 
		\caption{High-level summary of exclusion regions and search prospects arising from CMB (mustard), collider (light blue), and direct detection (purple) probes, as in Fig.~\ref{projections} but for different values of $\alpha_D$ and $m_{\chi}/m_{A'}$.  The dashed line contours correspond to projections. Meanwhile, the striated area shows the region where semi-elastic interactions can further deplete the excited state and weaken the exclusions. The direct detection projections correspond to SuperCDMS and LZ. \label{appFig:overallSummaryPlots}}
	\end{center}
\end{sidewaysfigure}

\begin{sidewaysfigure}[!ht]
	\begin{center}
	\includegraphics[width=\textwidth]{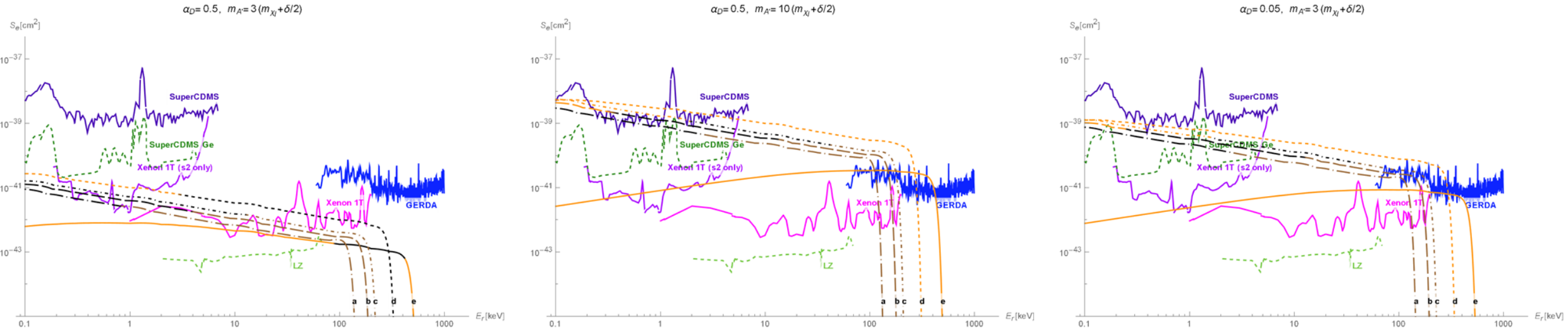}
	\end{center}
	\caption{Sensitivity plots for electron recoil searches, as in Figure \ref{fig:e-sensitivity} but for different values of $\alpha_D$ and $m_{\chi}/m_{A'}$. The black/orange/brown lines corresponds to the excited DM fraction times the DM-electron cross section per GeV DM mass for a thermal target.  The lines labeled a, b, c, d, e correspond to $m_{\chi_l}=0.002 \  \text{GeV},  \ 0.005\ \text{GeV}, \ 0.01\ \text{GeV},\  0.1 \ \text{GeV}, \ \text{and} \ 1 \ \text{GeV}$ respectively. The thermal target is excluded by the CMB when the line is orange. When it turns brown the semi-elastic process of the Subsection \ref{sssec:hlll} becomes relevant. The solid color lines show the sensitivity limits for Xenon1T, Xenon1T s2-only, SuperCDMS, and GERDA. The dashed lines show the projections for SuperCDMS with a germanium target and Lux-Zeplin. \label{appfig:e-sensitivity}}
\end{sidewaysfigure}

\begin{sidewaysfigure}[!ht]
	\begin{center}
	\includegraphics[width=\textwidth]{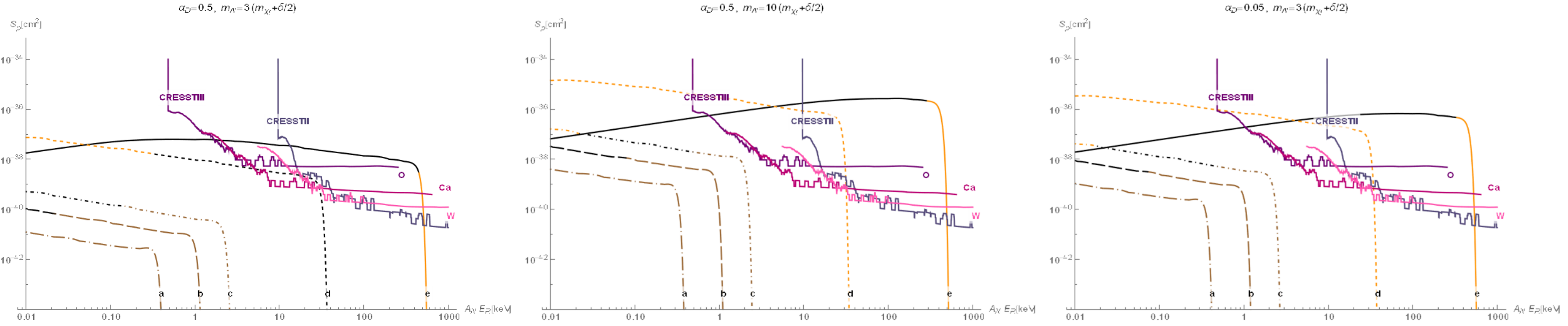}
	\end{center}
	\caption{Sensitivity plots for nuclear recoil searches, as in Fig.~\ref{sensitivityN} but for different  values of $\alpha_D$ and $m_{\chi}/m_{A'}$. The black/brown/orange lines corresponds to the excited DM fraction times the DM-nucleon cross section per GeV DM mass for a thermal target. The color scheme and labels of these lines are the same as in Fig.~\ref{appfig:e-sensitivity}. The purple-shaded lines show the sensitivity limits for CRESST II and CRESST III. For CRESST III we show in different shades the contribution from each nucleus. \label{appFig:nuc-sensitivity}}
\end{sidewaysfigure}

\FloatBarrier

\pagebreak

\bibliography{newbib}

\end{document}